\documentclass{aa}
\usepackage[varg]{txfonts}
\usepackage{dblfloatfix}
\usepackage{afterpage}
\usepackage{graphicx}
\usepackage{natbib}
\bibpunct{(}{)}{;}{a}{}{,}

\def\0{\hspace*{0.5em}}
\begin{document}

\title{Multiple short-lived stellar prominences
on O stars:\\ The O6.5I(n)fp star $\lambda$ Cephei\thanks{
Based on observations obtained the Mercator Telescope on La Palma, operated by the Institute of Astronomy of the University of Leuven, Belgium}}

\author{N.P. Sudnik\inst{1,2}  \and H.F. Henrichs\inst{1}
}
\institute{Anton Pannekoek Institute for Astronomy, University of Amsterdam,
Science Park 904, 1098 XH Amsterdam, Netherlands
\and
Saint-Petersburg State University, Universitetskii pr. 28,
Saint-Petersburg 198504, Russia
}

\offprints{H.F. Henrichs,
\email{h.f.henrichs@uva.nl}}

\date{Received date / Accepted date}

\keywords{Stars: early-type -- Stars: individual: $\lambda$ Cephei --  Stars: magnetic field -- Stars: winds, outflows -- Stars: rotation}

\abstract
{Most O-type stars and many B stars show unexplained cyclical variability in their spectral lines, i.e.,\ modulation on the rotational timescale, but not strictly periodic. The variability occurs in the so-called discrete absorption components (DACs) that accelerate through the UV-wind line profiles and also in many optical lines. For such OB stars no dipolar magnetic fields have been detected with upper limits of $\sim300$ G.
}
{We investigate whether multiple magnetic loops on the surface rather than non-radial pulsations (NRPs) or a dipolar magnetic field can explain the observed cyclical UV and optical spectral line variability.}
{We present time-resolved, high-resolution optical spectroscopy of the O6.5I(n)fp star $\lambda$ Cephei. We apply a simplified phenomenological model in which multiple spherical blobs attached to the surface represent magnetic-loop structures, which we call stellar prominences, by analogy with solar prominences. We compare  the calculated line profiles as a function of rotational phase, adopting a rotation  period of 4.1 d, with observed relative changes in subsequent quotient spectra.}
{We identify many periodicities in spectral lines, almost none of which is stable over timescales from months to years.  We show that the relative changes in various optical absorption and emission lines are often very similar. Our proposed model applied to the \ion{He}{ii} $\lambda$4686 line can typically be fitted with  2--5 equatorial blobs with lifetimes between $\sim$ 1 and 24 h.
}
{Given the irregular timescales involved, we propose that the azimuthal distribution of DACs correspond to the locations of stellar prominences attached to the surface. This could explain the observed variability of optical and UV lines, and put constraints on the strength and lifetime of these structures, which can be compared  with recent theoretical predictions, in which bright magnetic surface spots are formed by the action of the subsurface convection zone.}

\titlerunning{Stellar prominences
on O stars: The O6.5I(n)fp star $\lambda$ Cephei}

\authorrunning{N.\ Sudnik, H.~F.\ Henrichs}

\maketitle

\section{Introduction}
Mass loss of OB-type stars occurs in the form of a stellar wind, which is driven and accelerated by the radiation pressure exerted by the luminosity in numerous spectral lines, resulting in a terminal velocity that exceeds the local escape velocity. In the absence of significant short-term luminosity changes, the wind is expected to be constant in time. Such winds, however, are  known to be intrinsically unstable and clumped; they  are also  likely affected by the presence of rotation, pulsation, and/or magnetic fields, which obviously makes the wind structure  far more complex. Observationally, two types of wind variability can be distinguished using  ultraviolet (UV) resonance lines.  The first type shows variability that is strictly periodic, with the wind line profile varying from absorption to emission and back again, symmetrically in velocity range around the rest wavelength. This type is encountered among all studied magnetic oblique rotators, with the periodicity being the rotation period of the star. This class includes many helium peculiar stars with spectral type from $\sim$B7 for the He-weak stars, $\sim$B1/B2 for the He-strong stars, up to the Of?p stars.  For instance, the gallery presented by \cite{henrichs:2012} contains He-weak, He-strong, $\beta$ Cephei variables, and a slowly pulsating B star, all with very similar profile variations. Early discovered magnetic O-star examples include the O7Vp star $\theta^1$~Ori~C \citep{donati:2002} and the Of?p star HD~191612 \citep{donati:2006}. All discovered magnetic OB stars until 2012 are listed in \cite{petit:2013}. In the galactic magnetic OB star survey MiMeS \citep{wade:2014} $7\pm 1\%$ out of over 550 OB stars have a magnetic field, likely of the type producing this first type of wind variability (although only a minority of this sample have time-resolved UV spectra). All these magnetic stars are in the hydrogen core burning phase, including the supergiant $\zeta$\,OriAa O9.5Ia \citep{bouret:2012, blazere:2015}, the star with the weakest polar field strength.

The second type of wind variability is called cyclical, i.e.,\ without strict periodicity (like sunspots) as seen in variable UV wind lines, which is mostly in the form of discrete absorption components (DACs) in unsaturated parts of P Cygni profiles. Discrete absorption components  accelerate from low velocity (at about $v$sin$i$) towards negative velocity and approach the terminal wind velocity as measured at the bottom of the blue edge of saturated UV \mbox{P Cygni} wind profiles.  Whereas the formation of these wind structures are ascribed to corotating interacting regions \citep[CIRs,][]{mullan:1984, mullan:1986}, as in the solar wind, their nature is not known.  \cite{cranmer:1996} showed that such wind structures indeed give rise to DACs in UV profiles. The velocity contrast arises from differences in wind driving between places with and without surface inhomogeneities.
So far all investigations point towards a near-photospheric origin of the CIRs, e.g.,\ \cite{henrichs:1994}, \cite{kaper:1997}, and have been recently substantiated by \cite{massa:2015}.
The cyclical timescale scales with the estimated stellar rotation period \citep{henrichs:1988,prinja:1988}. Proposed mechanisms include non-radial pulsations (NRPs), a dipolar magnetic field, and/or multiple magnetic loops on the surface, which are all latitude-dependent features causing azimuthal changes in the wind, and make them likely candidates.
Direct evidence of a periodic DAC originating from a magnetic pole was shown for the magnetic  O star $\theta^1$~Ori~C by \cite{henrichs:2005}, by noting that 13 years of rotationally phase-folded \ion{C}{IV} UV profiles of this star show a progressing absorption feature very similar to a single (cyclical) DAC, whose  starting epoch could be traced back to the phase when the only observable magnetic pole passes through the line of sight.
\cite{david-uraz:2014} closely investigated the magnetic dipolar field hypothesis as the origin of DACs in the best available sample of OB stars, and concluded that such fields are highly unlikely to be responsible for these wind structures in all massive stars. This would leave NRPs and/or multiple magnetic loops on the surface as a possible cause for the cyclical wind variability, likely present in the remaining $93\%$ of non-bipolar magnetic OB stars, which is the subject of this investigation.

In our search for the origin of the cyclical wind behavior, we focussed on the enigmatic star \object{$\lambda$~Cephei} (\object{HD~210839}, \object{HIP~109556}), classified as O6I(n)fp by \cite{walborn:1972}, but later revised to O6.5I(n)fp by \cite{sota:2011}. Thanks to its brightness, it has been studied in great detail in many wavelength regions. The \ion{He}{II} $\lambda$4686 line is always in emission (hence its spectral type), but it is double peaked and variable without any stable periodicity. Night to night variations in this spectral line were reported by \cite{leep:1979} and \cite{grady:1983}, the latter simultaneously with UV spectra obtained with the {\it IUE} satellite. The shortest reported timescale of significant variations in the \ion{He}{II} $\lambda$4686 line is about 15 minutes \citep{henrichs:1991}. In these data, simultaneously acquired with UV spectroscopy in 1989, the overall equivalent width of this line within $-$400 and +400 km~s$^{-1}$  curiously varied in concert with the blue edge of the \ion{C}{IV} UV resonance line around $-$2300 km\,s$^{-1}$. This parallel behavior has not been explained. Simultaneous H$\alpha$ and {\it IUE} spectra were presented by \cite{kaper:1997} also with  possibly correlated behavior.  The star shows nearly saturated UV resonance P Cygni lines with strong variability in the absorption part, both in the form of DACs in the \ion{Si}{IV} line, and in the blue edge of the more saturated \ion{C}{IV} and \ion{N}{V} lines \citep{kaper:1996, kaper:1999} with typical timescales varying from $1.4\pm 0.2 $  to $2.2\pm 0.7$ and $4.3 \pm 1.4$\,d for different datasets. Two of these time series were recently reanalyzed by \cite{massa:2015}. \cite{marchenko:1998} reported a 0.6326 d photometric period in {\it Hipparcos} data.  Two modes with periods of 6.6 h and 12.3 h of small-amplitude non-radial pulsations of $\lambda$~Cep were discovered by \cite{dejong:1999}. This could not be confirmed, however, by \cite{uuh-sonda:2014}, which will be discussed below. Simultaneous X-ray and optical H$\alpha$ and \ion{He}{II} $\lambda$4686 spectroscopy was presented by \citep[hereafter RH15]{rauw:2015}, who found a 4.1 d periodicity in H$\alpha$ and an X-ray emitting region between 1.1 and 2 stellar radii. 

The rotation period of the star is not known; it is on the order of a few days, as estimated from the stellar parameters obtained from spectral model fits \citep[e.g.][]{markova:2004, bouret:2012}, and is similar to the value found in  the UV time series, which reflect corotating wind structures.   \cite{david-uraz:2014} determined an upper limit of 136 G for the dipolar magnetic field strength.

In our effort to understand the cause of the cyclic variability of optical lines we analyzed four time-resolved high-resolution series of spectra obtained with various instruments in 1989, 2007, 2012, and 2103, and particularly studied the behavior of the strongly variable \ion{He}{II} $\lambda$4686 double-peaked emission profile. This line probes the
base of the stellar wind, which is therefore particularly suited to studying the
azimuthal structure of the wind close to the surface. Because the behavior of this spectral line is  cyclic, and not periodic, on the same daily rotational timescale, we consider it to be strong evidence of a photospheric -- wind connection. In our attempt to model the variability we circumvented the well-known
difficulty of modeling the asymmetric double-peaked emission line itself (see, e.g.,~\cite{sundqvist:2011}, although see \cite{hillier:2012} for $\zeta$ Pup) by considering {quotient} spectra, obtained by dividing subsequent spectra,
such that only relative profile changes are monitored. These quotient spectra display a large diversity
in shape, from zero amplitude (within the noise) to more than $5\%$ of the relative flux. The main aim of this paper is to develop a first approximation model to describe the shape and time evolution of these quotient spectra by means of introducing a variable number of corotating spherical blobs of emitting gas attached to the surface of the star with adjustable optical depth, size, and velocity dispersion. These blobs are envisaged to represent ``stellar prominences'' presumably caused by local transient magnetic structures. We develop a formalism to calculate the rational phase dependent effect of such a blob on the line profile, which would be superposed on the spectral line of the photosphere plus ambient wind. Support for this picture comes from a comparison of quotient spectra from several \ion{H}, \ion{He}{I}, and \ion{He}{II} lines, which often look similar and display similar behavior, although differences do occur.  Within this limited description we aim to derive the typical characteristics of these transient features, in particular the lifetime of configurations and of individual blobs (hours, days), and the long-term timescale (years) of this phenomenon. Indirect support for the presence of transient small-scale phenomena very close to the star may come from space-based high-precision photometry of the mid O-type giant $\xi$ Per, presented by \cite{ramiaramanantsoa:2014}, who conclude that the observed modulation is consistent with several co-rotating bright spots on the stellar surface.

We emphasize that the importance of these exploratory calculations is the relevance to the behavior of many other, if not all, massive O- and early B-supergiant stars. Preliminary comparable pilot studies of a handful other O stars (e.g.,\ $\xi$ Per, 19\,Cep, $\alpha$\,Cam, $\lambda$\,Ori, $\zeta$\,Ori) showed that their short- and long-term optical line profile variability is very similar to what we observed in $\lambda$\,Cep. Since these stars also show cyclic variability in their winds, we take this as further support that the presence of what we call stellar prominences could well be a common phenomenon of massive stars.

This paper is organized as follows. We first review the stellar parameters  used in the paper. In Sect.\,\ref{section:wind} we summarize the UV wind properties. 
In Sect.\,\ref{section:optical} the optical spectroscopic observations are presented. Section\,\ref{section:model} describes the properties of the model. The results of the model fits are given in Sect.\,\ref{section:fits}. Section\,\ref{section:discussion} contains the discussion, and  Sect. 8 summarizes the conclusions.


\section{Stellar properties and the distance problem}
\label{star}

\begin{table}[b!]
\caption[Stellar parameters]{Adopted parameters for \object{$\lambda$ Cep  used in this paper. }}
\label{param}
\centering
\begin{tabular}{llc}
\hline
\hline
& & Reference\\
\hline
Spectral Type                         & O6.5\,I\,(n)fp         & 1 \\
$V$                                   & 5.043                  & 2 \\
$d$ (pc)                              & 649$^{+112}_{-83}$     & 1 \\
$T_{\rm eff}$ (K)                     & $36000 \pm 1000$       & 3 \\
$R/R_{\odot}$                         & 17.5                   & 4 \\
$v_{\rm rad}$ (km\,s$^{-1}$)          & $-$75.1 $\pm$ 0.9        & 5 \\
$v$sin$i$ (km\,s$^{-1}$)              & 204                    & 6 \\
$P_{\rm rot}$ (d)                     & 4.1                    & Sect.\,\ref{star}\\
$i$                                   & $68^{\circ}$           & Sect.\,\ref{star}\\
\hline
\vspace{-0.6cm}
\label{tab:parameters}
\end{tabular}\\
\tablebib{(1) \citet{sota:2011}; (2) GOS catalogue \citep{maiz-apellaniz:2004}; (3) \citet{bouret:2012}; (4) \citet{markova:2004}; (5) \citet{gontcharov:2006};  (6) \citet{simon-diaz:2014}.}
\end{table}

The projected equatorial rotational velocity, $v$sin$i$, is firmly established at 204 km\,s$^{-1}$ \citep{simon-diaz:2014}. Since the description presented in this paper hinges on an assumed period of rotation, $P_{\rm rot}$, consistent values for the radius, $R$, and angle of inclination, $i$, are required. Since no interferometric diameter is known, $R$ can only be estimated from the stellar effective temperature and luminosity and  therefore depends critically on the distance, which turns out to be problematic \citep[see, e.g.,][]{schroeder:2004}, and is therefore discussed here. $\lambda$~Cep is a runaway star \citep{blaauw:1961} with a radial velocity of $-75$ km s$^{-1}$ \citep{gontcharov:2006} and is located in the direction of the Cep OB2 association. According to dynamical calculations by \cite{hoogerwerf:2001}, the star cannot be a member of Cep OB2 but instead belonged originally to Cep~ OB3 (subgroup $a$) located $\sim$6.6$^{\circ}$ to the east  of its current position. It escaped through a supernova explosion of the presumed binary companion star $\sim$4.5 $\times 10^6$~yr ago, and traveled at 75 km s$^{-1}$ to end up at a distance of $\sim$450~pc from the Sun in its current direction. After an analysis of its space motion, these authors regard $\lambda$ Cep as a blue straggler member of Cep OB3, which is consistent with the shape and orientation of the bowshock discovered with {\it IRAS} by \cite{vanburen:1988} and confirmed with a 24 $\mu$m {\it Spitzer} image \citep{gvaramadze:2011}. The latter authors, however, concluded that the star escaped only $\sim$2.5$\times 10^6$~yr ago with a peculiar velocity of 60 km s$^{-1}$, which was based on a more recent distance determination of Cep OB3 of 700 pc \citep{melnik:2009}, not very different from the distance reported by \cite{crawford:1970} and used by \cite{hoogerwerf:2001}, although \cite{moreno:1993} found $0.9\pm0.1$ kpc, which was revised to 800 pc by \cite{peter:2012}. In short, in both dynamical calculations $\lambda$ Cep should be 140 to 280 pc closer to the Sun than Cep OB3, which is in the same range as the most recent Hipparcos distance of 649$^{+112}_{-83}$~pc \citep{sota:2011}. However, this distance is inconsistent with its spectral class, energy distribution, luminosity, and radius;  the effective temperature is rather strongly constrained at $36000\pm 1000$~K \citep{bouret:2012}, which would put the star farther away.  Reported distances include 0.83 kpc \citep{markova:2004}, 0.9 kpc \citep{fullerton:2006}, 1.08 kpc \citep{puls:2006}, $0.95\pm 0.1$ kpc \citep{bouret:2012}, and even $1.6 \pm 0.3$ kpc from interstellar Ca~H and K line strengths \citep{megier:2009}, implying an uncertainty of about a factor of 2 in $R$. We have simply adopted $R = 17.5~R_{\odot}$, which is consistent with the values from different papers containing a spectral line analysis, and have ignored the distance problem. Future improvement may come from apparent diameter measurements by optical interferometry \citep{vanbelle:2012} as well as from a superior parallax determination by the forthcoming {\it GAIA} mission.

To arrive at a value for the rotation period we considered both the estimated recurrence time of DACs \citep{kaper:1999,david-uraz:2014} and the phase coherence between the equivalent widths of the \ion{He}{ii} $\lambda$4686 line and the blue edge of the UV \ion{C}{IV} wind line. A periodicity of
$P_{\rm rot}$~$\sim2$~d would be consistent with the cyclic behavior, but with the adopted radius such a short period would imply $i <30^{\circ}$, which would not be compatible with the always blue-to-red marching absorption features in the \ion{He}{I} $\lambda$4713 line  \citep{dejong:1999} and in the \ion{He}{I} $\lambda$4471 and \ion{He}{II} $\lambda$4542 lines  \citep{uuh-sonda:2014}. With such a near pole-on orientation,  features moving in the opposite direction that originate at the far side of the star are also expected,  which is not the case (see also Sect.\,\ref{quotient}). We therefore adopted twice this value, $P_{\rm rot} =4.1$ d, similar to the conclusion by RH15 based on a Fourier-analysis of the H$\alpha$ variability.  With $R = 17.5 R_{\sun}$, this implies $i = 68^{\circ}$ for the inclination angle.
With this radius the rotation period cannot exceed 4.4 d, and should be more than 3.4 d for $i \approx 50^{\circ}$.
The adopted stellar parameters and other relevant observational aspects are presented in Table~\ref{tab:parameters}.


\begin{table*}[!b]
\caption{Optical spectral observations of \object{$\lambda$ Cep} used in the present work.
}
\centering
\begin{tabular}{llllllll}
\hline
\hline
Dates            & Telescope   & Spectrograph & Resolution & Coverage (\AA )& Exp.\ time (min) & N & S/N \\
\hline
1989 Oct 17--21  & Calar Alto, 2.2\,m & 90 cm f/3 camera & 25000 & 4600--4730 & 7--10  & 109\tablefootmark{a}& 500 \\
                 & Kitt Peak, 0.9\,m  &camera $\#5$ & 22000 & 4650--4735 & 10--15 &  71\tablefootmark{a}& 300 \\
2007 Nov 20--22, & BOAO, 1.8\,m       & BOES   & 45000 & 3830--8260 & 3--5   & 30 & 300 \\
\hspace{0.78cm}Dec 16   &                    &        &       &            &        &    &     \\
2012 Oct 8--14   & Mercator, 1.2 m    & HERMES & 85000 & 3770--9000 & 3.3    & 23 & 150 \\
2013 Oct 9--15   & Mercator, 1.2 m    & HERMES & 85000 & 3770--9000 & 3.3 & 42 & 150    \\
\hline
\label{table:instruments}
\end{tabular}
\vspace{-0.6cm}
\tablefoot{
The columns list respectively the dates, telescope, name of spectrograph, spectral resolution $\lambda/\Delta\lambda$, wavelength coverage in \AA, exposure time in min,\ number of spectra acquired (N), and average signal-to-noise ratio (S/N) of the unbinned data.
\tablefoottext{a}{After rebinning all 180 spectra in the time domain, in total 33 spectra were finally utilized for the analysis, see Sect.\,\ref{sub:1989analysis}.}
}
\end{table*}

\section{Stellar wind observations}
\label{section:wind}

\begin{figure}[!hb]
\centering
\includegraphics[width=0.85\columnwidth]{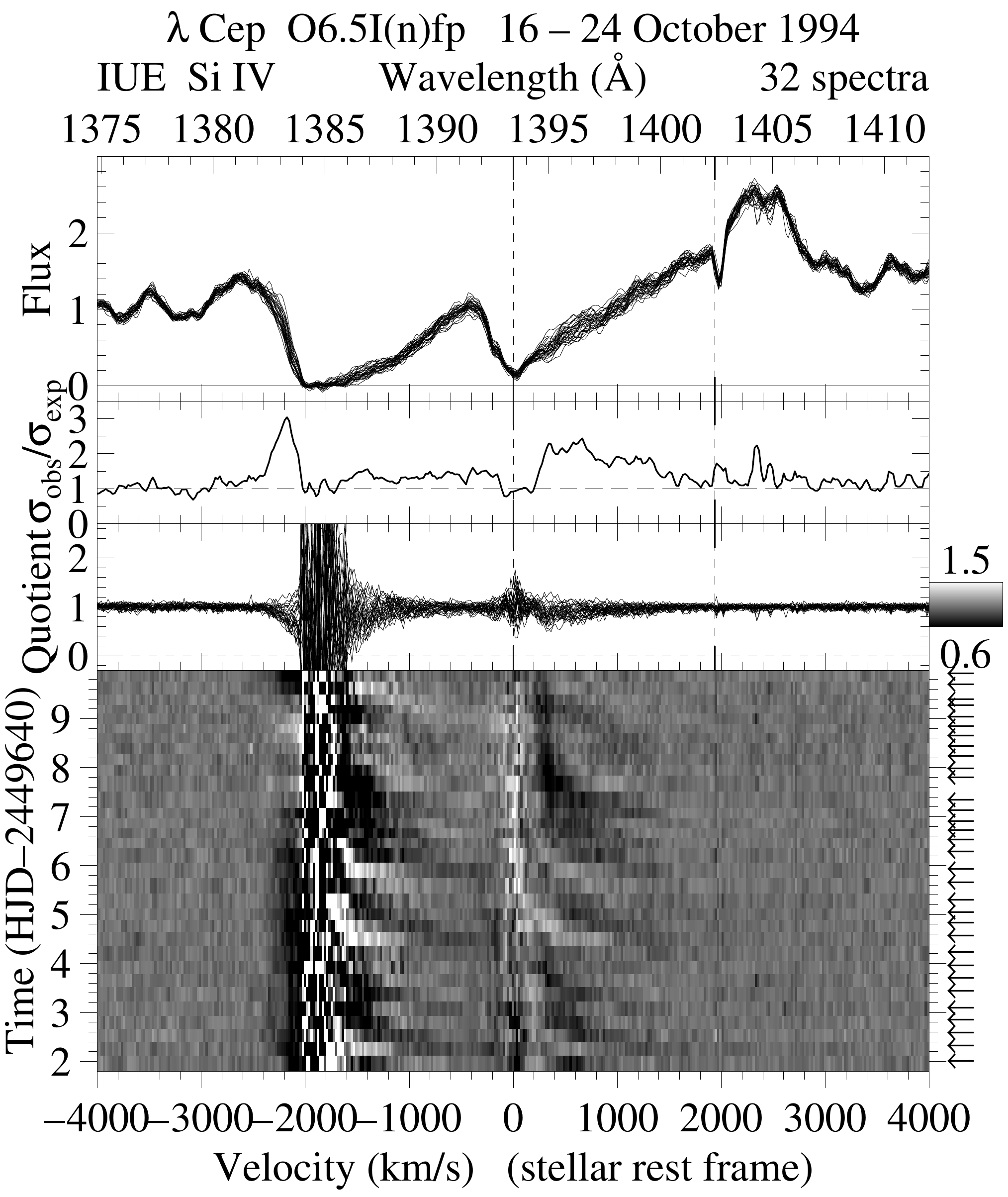}
\caption{Variability in the \ion{Si}{iv} $\lambda$1394--$\lambda$1403 doublet of $\lambda$~Cep in October 1994 showing the cyclic appearance of DACs. Vertical dotted lines denote the rest wavelengths. {\it Lower panel:} Grayscale representation of quotient spectra in the panel above it, obtained with a template spectrum constructed from the spectra in the top panel (flux units in $10^{-10}$~erg~cm$^{-2}$s$^{-1}$\AA $^{-1}$) using the method by \cite{kaper:1999}. The template represents the undisturbed, underlying wind profile, which makes the DACs appear as absorption features. The second panel from the top gives the temporal variance spectrum (TVS$^{1/2}$) expressed as the ratio of the observed to the expected standard deviation (see Sect.\,\ref{section:optical}). }
\label{fig:si4}
\end{figure}

The star $\lambda$ Cep is the fourth most frequently observed O star with the {\it IUE} satellite; there have been 179 good quality spectra in six groups.  The evolution of the stellar wind lines of \ion{C}{iv}, \ion{Si}{iv}, \ion{N}{V}, and \ion{N}{IV} has been described by \cite{kaper:1996, kaper:1997, kaper:1999}. Two series with the best time resolution were reanalyzed by \cite{massa:2015}. Figure\,\ref{fig:si4} shows the evolution of the \ion{Si}{iv} $\lambda$1394--$\lambda$1403 doublet from the 1994 dataset;   cyclical DACs are clearly present.

\begin{figure*}[!ht]
\includegraphics[width=0.85\linewidth]{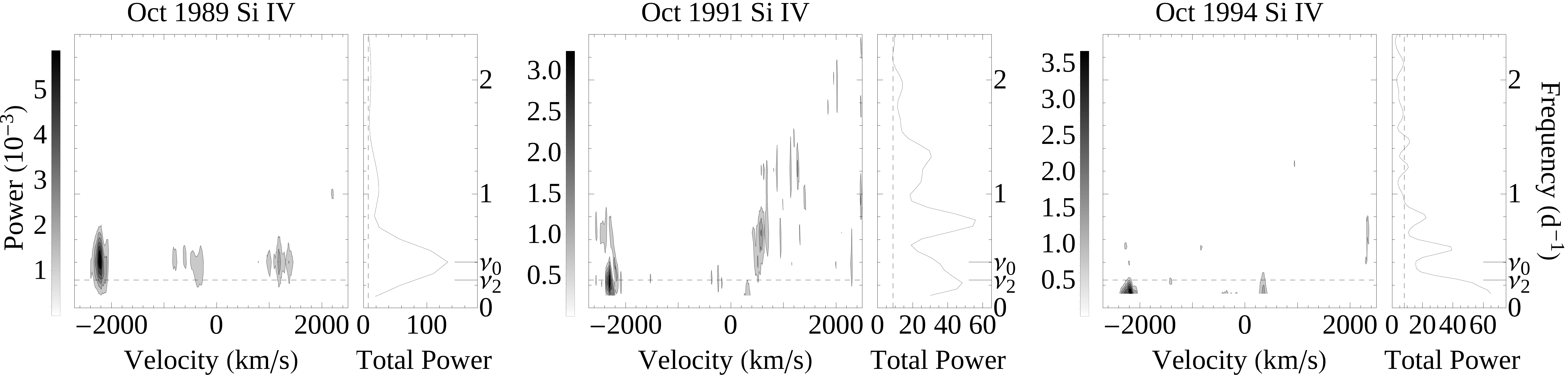}
\caption{From left to right: grayscale representation of the CLEANed periodograms as a function of velocity in the stellar rest frame around the \ion{Si}{iv} $\lambda$1394 line for the {\sl IUE} datasets of October 1989, 1991, and 1994. The profiles of the 1994 data are shown in Fig.\,\ref{fig:si4}. The legend bars give the corresponding grayscale codings. The horizontal dashed line and tickmark at $\nu_2 = 0.245$~d$^{-1}$ indicate our adopted rotation frequency of the star. The three side panels display the total power over the velocity range of the main panel in the same units. The vertical dashed line indicates the estimated noise level in the power (see text). The tickmark at $\nu_0 = 0.403$~d$^{-1}$ (2.48 d) shows the dominant frequency in the 1989 data.}
\label{fig:powerSi4}
\end{figure*}

To search for possible periodicities in the wind lines we performed a CLEAN analysis \citep{roberts:1987} as a function of velocity with respect to the $\lambda$1394 line for the three best  {\it IUE} datasets in 1989, 1991, and 1994. We used  1000 iterations with gain 0.1. The lowest frequency is dictated by the Nyquist frequency of the dataset. Figure \ref{fig:powerSi4} presents the grayscale representations of the resulting periodograms along with the total power, summed over the full width of the doublet. We attempted to calculate the power above which the peaks are significant at a given level, but this is a complicated issue in the case of unequally spaced data like ours (see, e.g.,\ \cite{frescura:2008}). We have followed the empirical approach by \cite{dejong:1999} in which the noise is calculated as the level below which 98$\%$ of the points lie in the frequency range between 5 and 10 d$^{-1}$, where there is likely not any significant power. We note  that this noise level is only a lower limit to any useful significance level, but the lack of a proper treatment prohibits a more specific approach.

In the total power diagrams the prominent peak at $\nu_0 = 0.403$~d$^{-1}$ (2.48~d) does not show up in the  simultaneous \ion{He}{ii} $\lambda$4686 data of 1989 (see Figs.\,\ref{fig:powerHe2_4686} and \ref{fig:EWblobph1989}), and has  not been seen in any other dataset. There is no obvious shared frequency with significant power in these three datasets, which underlines the cyclical behavior of the wind.
Our adopted rotation frequency at $\nu_2 = 0.245$~d$^{-1}$ (4.1~d) appears -- perhaps marginally -- at low velocity in 1989 and 1991, but is absent in 1994, which confirms that there is no sign of permanent corotating wind structures close to the star, as is well known.

\section{Optical observations}
\label{section:optical}

\begin{figure*}[!t]
\includegraphics[width=0.95\linewidth]{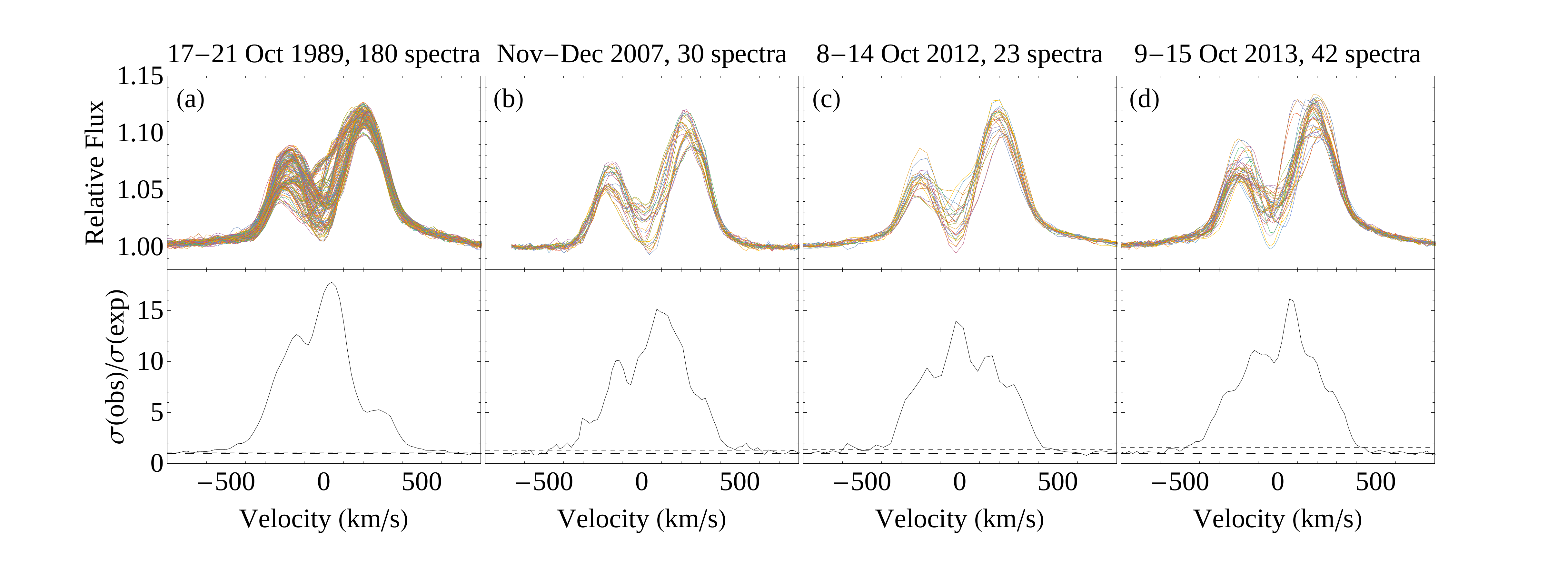}
\caption{{\it Top panels:} Overplots of \ion{He}{ii} $\lambda$4686 spectra, showing dramatic changes on short and long timescales.  The spectra have been rebinned such that the S/N equals 810 in all four panels. Vertical dashed lines are drawn at $\pm v$sin$i$. {\it Lower panels:} Temporal variance spectra expressed as the ratio of the observed to the expected standard deviation, $\sigma_{\rm obs}/\sigma_{\rm exp}$ (similar to TVS$^{1/2}$). The lower long-dashed horizontal line is drawn at unity. The upper horizontal short-dashed line indicates the 1$\%$ significance level. From left to right: {\it (a)} acquired at Calar Alto and Kitt Peak in 1989 with shortest time resolution of about 15 min. The corresponding periodogram is  in the left-hand panel of Fig.\,\ref{fig:powerHe2_4686}; {\it (b)} acquired at BOAO in 2007, spanning 27 days; {\it (c)} acquired at the Mercator telescope in 2012 with its corresponding periodogram shown in the middle panel of Fig.\,\ref{fig:powerHe2_4686};  {\it (d)}  acquired at the Mercator telescope in 2013.  The corresponding periodogram  is in the right-hand panel of Fig.\,\ref{fig:powerHe2_4686}. }
\label{fig:TVS4686}
\end{figure*}

\begin{table}[!b]
\caption{Spectral lines considered in the analysis.}
\centering
\begin{tabular}{ll}
\hline
\hline
Dataset & Spectral lines with rest wavelength in \AA .\\
\hline
1989    & \ion{He}{i}  4713.145 \\
        & \ion{He}{ii} 4685.804\\
2007    & \ion{H}{i}   4101.734, 4340.472, 4861.332, 6562.852 \\
        & \ion{He}{i}  4026.191, 4471.480, 5875.621 \\
        & \ion{He}{ii} 4199.87,  4541.591, 4685.804, 5411.521\\
2012    & \ion{H}{i}   4101.734, 4340.472, 4861.332, 6562.852 \\
        & \ion{He}{i}  4026.191, 4471.480, 4713.145, 5875.621 \\
        & \ion{He}{ii} 4199.87,  4541.591, 4685.804, 5411.521\\
2013    & \ion{H}{i}   4101.734, 4340.472, 4861.332, 6562.852 \\
        & \ion{He}{i}  4026.191, 4471.480, 4713.145, 5875.621 \\
        & \ion{He}{ii} 4199.87,  4541.591, 4685.804, 5411.521\\
\hline
\label{table:lines}
\end{tabular}
\vspace{-0.6cm}
\tablefoot{ The years marked in the first column correspond to the first column of Table~\ref{table:instruments}.}
\end{table}

Four datasets obtained at several observatories were analyzed for this investigation (see Table \ref{table:instruments}).
The spectra taken in 1989 were part of an international campaign including {\it IUE} (see Fig.\,\ref{fig:powerSi4}) and ground-based observations \citep{henrichs:1991,dejong:1999}. This same dataset of 180 high-resolution spectra with time resolution down to 15 min over 4 days has led to the discovery of the non-radial pulsations in $\lambda$\,Cep by the analysis of the \ion{He}{i} $\lambda$4713 line. The 2007 dataset was collected at the Bohyunsan Optical Astronomy Observatory in South Korea and was described by \cite{kholtygin:2011}, and included a first analysis.
The observations in 2012 and 2013 were carried out with the fiber-fed HERMES spectrograph mounted on the 1.2 m Mercator telescope on the island of La Palma, Spain, and covered a wavelength
range from 3770 to 9000 \AA\ with a resolution of $R = 85000$ \citep{raskin:2011}.
The spectral lines that were analyzed are listed in Table \ref{table:lines}. The wavelength scale of each line was transformed into velocity scale in the stellar rest frame after correction for the barycentric velocity of the observatory and the radial velocity of the star. All lines were normalized to a local continuum, which was approximated by a linear fit through two carefully chosen segments adjacent to the line center.  Atmospheric lines around H$\alpha$ in the Mercator spectra were removed with an empirical telluric template spectrum.

\begin{figure*}[!t]
\includegraphics[width=0.935\linewidth]{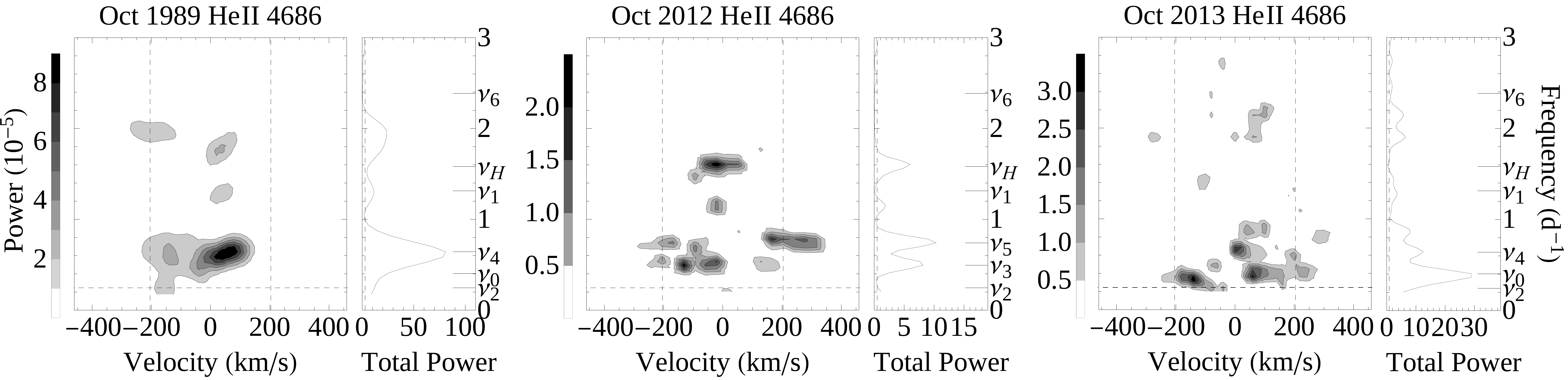}
\caption{From left to right: Grayscale representation of the CLEANed periodograms as a function of velocity of \ion{He}{ii} $\lambda$4686 spectra for the   1989, 2012, and 2013 datasets. For the layout see Fig.\,\ref{fig:powerSi4}. Vertical dashed lines in the main panels are drawn  at $\pm v$sin$i$. For the full profiles see Fig.\,\ref{fig:TVS4686}.  The tickmark at $\nu_1= 1.315$ d$^{-1}$ indicates the frequency with the strongest power in this line in August 2013 (see RH15), which is absent in our three datasets. The tickmarks at $\nu_2= 0.245$ d$^{-1}$ (= 4.1 d, our adopted rotation period) and $\nu_3 = 0.495$ d$^{-1}$ indicate the frequencies found in H$\alpha$ in August 2013 by RH15. The dominant frequencies at $\nu_0$ in the \ion{Si}{iv} in 1989 and at $\nu_4= 0.642$ d$^{-1}$ (1.56 d) are drawn at the peaks of the   1989 and 2013 dataset, respectively. The dominant peak in 2012 is at $\nu_5 = 0.741$ d$^{-1}$ (1.35 d). The tickmark at $\nu_{\rm H} = 1.5808$ d$^{-1}$ (= 0.6326 d) denotes the periodicity derived from the Hipparcos lightcurve  by \cite{marchenko:1998}, which is remarkably present in 2012. The indicated frequency  $\nu_6= 2.386$ d$^{-1}$ is absent in this line, but present in nearly all other helium lines in 2013, see Table\,\ref{table:frequencies}.}
\label{fig:powerHe2_4686}
\end{figure*}

\begin{figure*}[!t]
\includegraphics[width=0.65\linewidth]{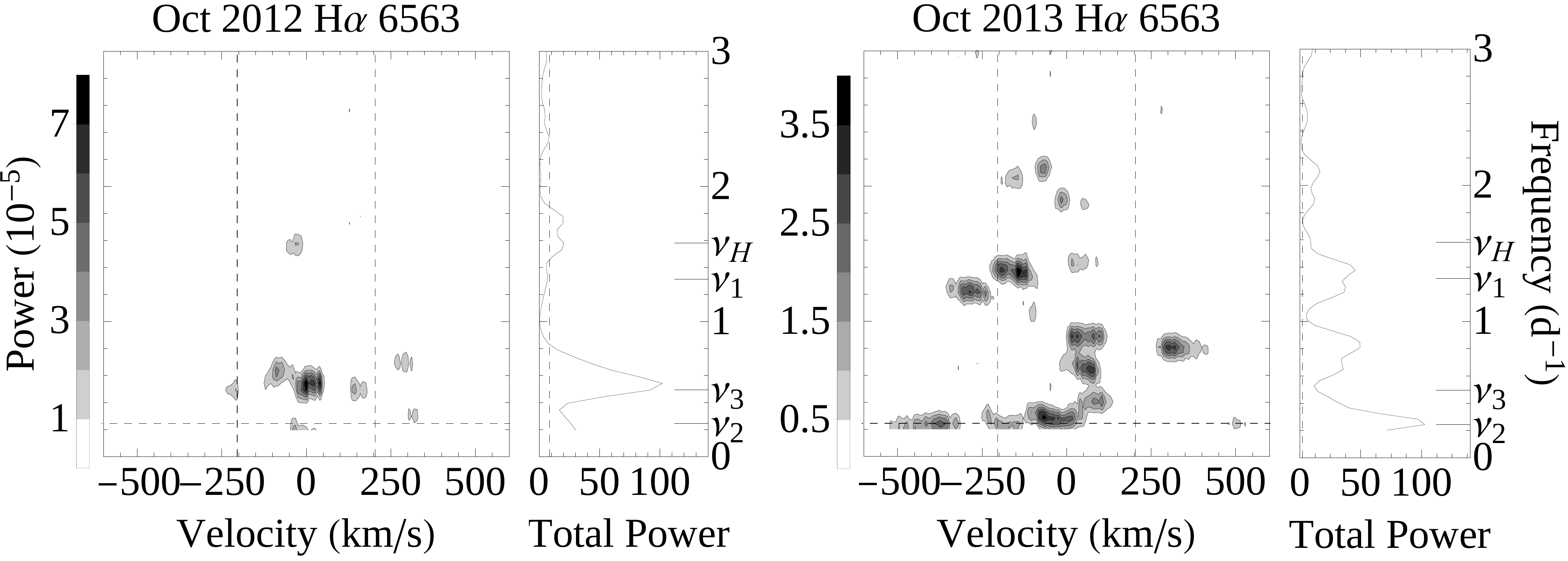}
\caption{Same as Fig.\,\ref{fig:powerHe2_4686}, but for H$\alpha$ spectra for the  2012 and 2013 datasets. The full profiles are displayed in Fig.\,\ref{fig:TVSha}. Of the two frequencies found in H$\alpha$ in August 2013 by RH15 we find $\nu_3$ only in 2012, and $\nu_2$ only in 2013. The 2012 data may contain some power at the Hipparcos frequency $\nu_{\rm H}$.}
\label{fig:powerHa}
\end{figure*}

We applied the model described below to spectra of $\lambda$ Cep of all four datasets.  Although the available spectral range included several hydrogen and helium lines, we restricted our analysis to the \ion{He}{II} $\lambda$4686 line simply because the amplitude of variations in this line is the most conspicuous of the available lines.  Overplots of all profiles are presented in Fig.\,\ref{fig:TVS4686}, which demonstrate the dramatic changes encountered. To allow a comparison, the profiles have been rebinned onto a velocity grid such that the S/N equals 810 for all four datasets. In the lower panels of the figure we show the ratio of the observed to the expected standard deviation, which is similar to TVS$^{1/2}$ with TVS the temporal variance spectrum \citep{fullerton:1996}. This ratio equals unity at the absence of statistically significant variation. As a reference we plotted the $1\%$ significance level, which shows that the range of variability extends from $-500$ to $+500$ km s$^{-1}$ in the \ion{He}{II} $\lambda$4686 line in the 1989 dataset, which has the highest quality (Fig.\,\ref{fig:TVS4686}a).

Overplots of all other studied lines are presented in Figs.\,\ref{fig:TVS4713}--\ref{fig:TVS4542a} and in Appendix A.  We now summarize the main characteristics of these datasets; in particular, we describe the temporal behavior of the profiles and of the subsequent quotient spectra of the different lines.

\subsection{Power spectra}

In order to search for periodic signals in the profiles we constructed grayscale representations of periodograms as a function of velocity in the frequency range from 0.2 to 10 d$^{-1}$ (period 0.1 to 5 d) for all datasets except for 2007, which is undersampled. This range includes both the assumed rotation period and the data coverage. We used the same CLEAN algorithm  described in Sect.\,\ref{section:wind}, with 1000 iterations and gain 0.1.  Using more iterations and/or different gains did not affect  the results very much. The significant part of the periodograms in the frequency range 0.2 to 3 d$^{-1}$ for \ion{He}{ii} $\lambda$4686 and H$\alpha$ are displayed in Figs.\,\ref{fig:powerHe2_4686} and \ref{fig:powerHa}, respectively. The periodograms of all other lines are presented in Fig.\,\ref{fig:powerHe1_4713} and in Appendix B. In these figures --  in the accompanying diagrams with the total power -- we have indicated the frequencies that are relevant for a comparison, in particular $\nu_2$, our adopted rotation frequency. All frequencies with total power likely above the noise and whose the origin can be identified in the periodogram are listed in Table\,\ref{table:nu}. Given the relatively short coverage we did not attempt to determine their accuracy, but at least two decimal places are significant. The occurrence of the frequencies in the different datasets is tabulated in Table\,\ref{table:frequencies}. We summarize in turn the behavior of the different lines.

Our most important dataset is for \ion{He}{II} $\lambda$4686 (Fig.\,\ref{fig:powerHe2_4686}). We note that the power distribution among the three datasets is very different. The strong power in 1989 at $\nu_4$ is entirely absent in 2012, but  is present in 2013 (although much weaker and also concentrated at line center); it is not   present two months earlier  (RH15) and  is also absent in all other spectral lines. We also marked $\nu_6$, which is found in nearly all other helium lines in 2013 only, but entirely absent in this \ion{He}{ii} line. The only common frequency with all hydrogen lines is $\nu_3$, but only in 2012. Such behavior in the various lines signifies different formation regions as shown by \cite{martins:2015} and can be used as a diagnostic for the stratification \citep[see also][]{hillier:2012}, which is, however, beyond the scope of this paper. The absence of significant power at $\nu_1$, the frequency with the highest power in this line August 2013 found by RH15, also illustrates the strong variable nature of this line. It may be significant that $\nu_{\rm H}$, found in the Hipparcos lightcurve by \cite{marchenko:1998}, is clearly present in the center of the line in 2012, but in no other spectral line. The dominant frequency in \ion{Si}{iv} in 1989 was $\nu_0$; it  is absent in the simultaneous \ion{He}{II} $\lambda$4686 data, but appeared as the dominant frequency in the center of the line in October 2013, although it is absent in RH15. See also Fig.\,\ref{fig:EWblobph1989} for a comparison with the UV \ion{C}{iv} edge  behavior.

\begin{table}[!b]
\caption{Main frequencies (rounded values) and corresponding periods referred to in this paper.
}
\centering
\begin{tabular}{lll|lll}
\hline
\hline
Name     & Frequency & Period & Name     & Frequency & Period\\
   & (d$^{-1}$)& (d) &      & (d$^{-1}$)  & (d)\\
\hline
 $\nu_0$ & 0.403 & 2.48  & $\nu_7$       &  1.74  & 0.574  \\
 $\nu_1$ & 1.315 & 0.760 & $\nu_8$       &  1.19  & 0.840  \\
 $\nu_2$ & 0.245 & 4.1   & $\nu_9$       &  0.620 & 1.613  \\
 $\nu_3$ & 0.495 & 2.02  & $\nu_{10}$    &  1.41  & 0.709  \\
 $\nu_4$ & 0.642 & 1.56  & $\nu_{\rm H}$ &  1.581 & 0.6326 \\
 $\nu_5$ & 0.741 & 1.35  & NRP1          &  1.96  & 0.510  \\
 $\nu_6$ & 2.386 & 0.419 & NRP2          &  3.64  & 0.275  \\
\hline
\label{table:nu}
\end{tabular}
\vspace{-0.5cm}
\tablefoot{
The labeling $\nu_{1, .., 3}$ is the same as in RH15. Our adopted rotation frequency is $\nu_2$. The frequency reported by \cite{marchenko:1998} from the Hipparcos lightcurve is denoted  $\nu_{\rm H}$. The NRP modes reported by \cite{dejong:1999} are marked as NRP1 and NRP2. See Table \ref{table:frequencies} for the occurrence of these frequencies.
}
\end{table}

The time behavior of the hydrogen lines is  similar for a given dataset, but not from year to year (see Table\,\ref{table:frequencies}), except for H$\delta$ which is partly blended with \ion{N}{iii} $\lambda$4097 and  \ion{Si}{iv} $\lambda$4089. \cite{sundqvist:2011} argued that the photospheric contribution of the \ion{He}{ii} blend with H$\alpha$ is negligible. \cite{rauw:2015} found the dominant frequency to be $\nu_2$ (our adopted rotation frequency) in this line; it is most likely also present in our data two months later, but this is difficult to confirm as $\nu_2$ is close to the lower frequency limit (set by the temporal coverage) in our Fourier analysis.

\begin{figure*}[!b]
\includegraphics[width=2\columnwidth]{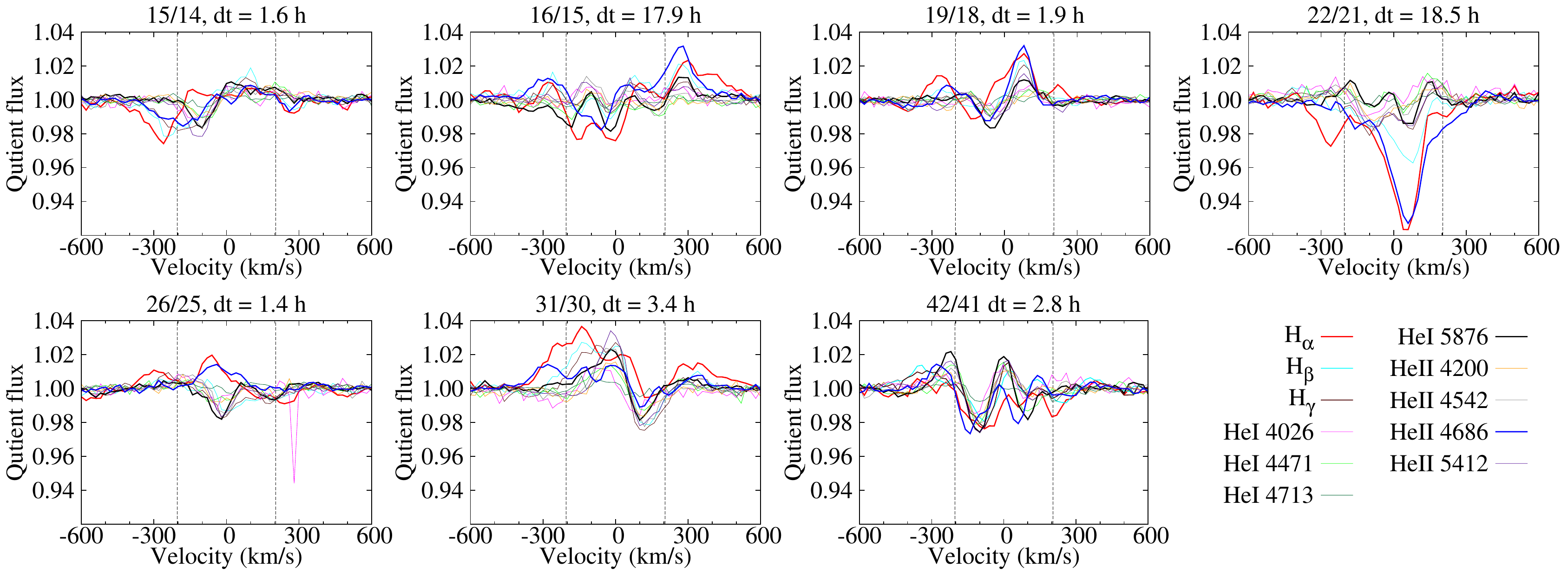}
\caption{Selected sample of subsequent quotient spectra of different lines from the 2013 dataset. The top label denotes the sequence numbers used for the quotient and the time difference between them. The legend gives the color codes used. The description of the individual panels is given in Sect.\ref{quotient}.}
\label{fig:quotients}
\end{figure*}

\begin{table}[!b]
\caption{Occurrence of frequencies in the diagrams with total power of the datasets, organized per year. }
\centering
\begin{tabular}{lccc}
\hline
\hline
Spectral line     & Oct 1989    & Oct 2012                       & Oct 2013 \\
\hline
H$\alpha$ 6563    &         & $\nu_3$                    & $\nu_2$, $\nu_1$ \\
H$\beta$  4861    &         & $\nu_3$                    & $\nu_2$, $\nu_1$ \\
H$\gamma$ 4341    &         & $\nu_3$                    & $\nu_2$, $\nu_1$, $\nu_8$ \\
H$\delta$ 4102    &         & $\nu_3$, $\nu_7$           & $\nu_8$, $\nu_2$ \\
\ion{He}{i} 5876  &         & $\nu_3$, $\nu_8$, $\nu_7$  & $\nu_3$, $\nu_1$, $\nu_6$, $\nu_2$\\
\ion{He}{i} 4713  & NRP1, NRP2 & -                       &  - \\
\ion{He}{i} 4471  &         & $\nu_3$                    & $\nu_6$, $\nu_1$, $\nu_2$ \\
\ion{He}{i} 4026  &         & -                          & $\nu_6$          \\
\ion{He}{ii} 4686 & $\nu_4$ & $\nu_5$, $\nu_3$, $\nu_{\rm H}$ &  $\nu_0$, $\nu_4$   \\
\ion{He}{ii} 5412 &         & $\nu_9$, $\nu_7$           & $\nu_{10}$, $\nu_9$, $\nu_6$\\
\ion{He}{ii} 4542 &         & -                          & $\nu_{10}$, $\nu_6$ \\
\ion{He}{ii} 4200 &         & -                          & $\nu_6$\\
\ion{Si}{iv} 1394 & $\nu_0$ &                            &                       \\
\hline
\label{table:frequencies}
\end{tabular}
\vspace{-0.5cm}
\tablefoot{The frequencies are given in Table \ref{table:nu}, and marked where appropriate in Figs.\,\ref{fig:powerSi4}, \ref{fig:powerHe2_4686}, \ref{fig:powerHa}, \ref{fig:powerHe1_4713}, and Appendix\,B. A horizontal dash indicates the absence of power above the noise level.  In each column the frequency with the strongest power is listed first.
}
\end{table}

In all the remaining helium lines, except  \ion{He}{i} $\lambda$4713 (see Sect.\,\ref{nrpdiscussion}), we find $\nu_6$ near line center in 2013, but not in any other dataset. We note  the occurrence of $\nu_3$, which dominates in 2012 in all hydrogen lines and in \ion{He}{I} $\lambda$5876 in both 2012 and 2013, but which is only present in \ion{He}{I} $\lambda$4471 in 2012. We consider this behavior as also being due to a significant change in the lower wind structure. Of the many frequencies listed by \cite{uuh-sonda:2014}  only their $\nu_2$ in \ion{He}{ii} $\lambda$4542 corresponds to our $\nu_{10}$ in this line, and no other match is found.

The fact that the periodograms of the spectral lines differ so much, also from year to year, strongly confirms the previous findings \citep[e.g.,][and RH15]{uuh-sonda:2014}  that there is no stable periodicity, certainly not over the whole width of the line profile (as in the case of pulsations), and also not near the adopted rotation period of 4.1 d. It appears, though, that a number of the same frequencies recur over the course of time.

\subsection{Subsequent quotient spectra}
\label{quotient}

We  modeled the \ion{He}{ii} $\lambda$4686 line not only because of its strongest variability, but also because the rapid changes (hourly, or shorter)  in this line often (but not always) occur simultaneously in several other optical lines. This is most prominently seen in quotient spectra of subsequent exposures. Selected typical examples of the 2013 dataset are displayed in Fig.\,\ref{fig:quotients}. We describe here the behavior of the subsequent quotients of 11 different spectral lines; H$\delta$ was left out because of partial blending (see Fig.\,\ref{fig:quotients}). The panels are labeled with the ratio of the sequence numbers used, and the time difference (in hours) between them. We note that none of the \ion{He}{i} $\lambda$4713 quotients in 2012 and 2013 shows significant amplitude. Only in the 1989 data, with superior S/N, does this line vary (see Sect.\,\ref{nrpdiscussion}). A good example of parallel behavior is seen in quotient 19/18, which occurred in 1.9 h. All quotient profiles have a similar structure with increasing amplitude towards higher ionization; \ion{He}{ii} $\lambda$4686 and H$\alpha$ are about equally the strongest. Only H$\alpha$ shows some extra features at $-300$ km s$^{-1}$. About 60$\%$ of the subsequent quotient spectra display such behavior, which partly validates our choice of only fitting the \ion{He}{ii} $\lambda$4686 line. The strongest change, $\sim7\%$, is seen in quotient 22/21, taken 18.5 h apart;  the highest amplitudes are in \ion{He}{ii} and H$\alpha$, followed by H$\beta$.  In about 30$\%$ of the cases a trend of stratification is observed, as in quotients 15/14 and 31/30 with intervals 1.6 and 3.4 h, respectively. Both extend far beyond $v$sin$i$. The structure and the ratio of amplitudes is similar to the previous parallel case, but the peak velocity shifts among the different lines. Quotient 16/15 show night to night variations with a strong amplitude far beyond $v$sin$i$. In less than a few percent of the cases the different lines behave in the opposite sense to quotient 26/25, in 1.4 h. More complex behavior also occurs, as in quotient 42/41, where some parts move in parallel and other parts move in opposite directions.
We note that in general the subsequent quotient spectra of H$\alpha$ and \ion{He}{II} $\lambda$4686 mostly resemble each other (mainly shape and to a lesser extend amplitude), but because the H$\alpha$ line is difficult to normalize owing to partial blending, this line was discarded for line fitting.


\section{Modeling line profile variations}
\label{section:model}
Our approach that explains the cyclical optical line profile variability is to assume that  relatively short-lived, multiple stellar prominences emerge on the stellar surface, presumably owing to localized magnetic activity, more or less in analogy to what is observed on the solar surface (although sunspots are dark, whereas magnetic spots on massive stars are brighter than their surroundings). Reasonings behind this approach include that rapid changes (approximately hourly)  occur simultaneously in several optical lines (Sect.\,\ref{section:optical}), and that the rapid timescale of such changes is much shorter than the rotation timescale. In our exploratory calculations of describing the properties of a stellar prominence, we assume as a first approximation the geometry of a spherical blob with the size of a fraction of the stellar radius that  stays at a fixed point above the surface, and hence corotates.

\begin{figure}[!t]
\centering
\includegraphics[width=0.7\columnwidth]{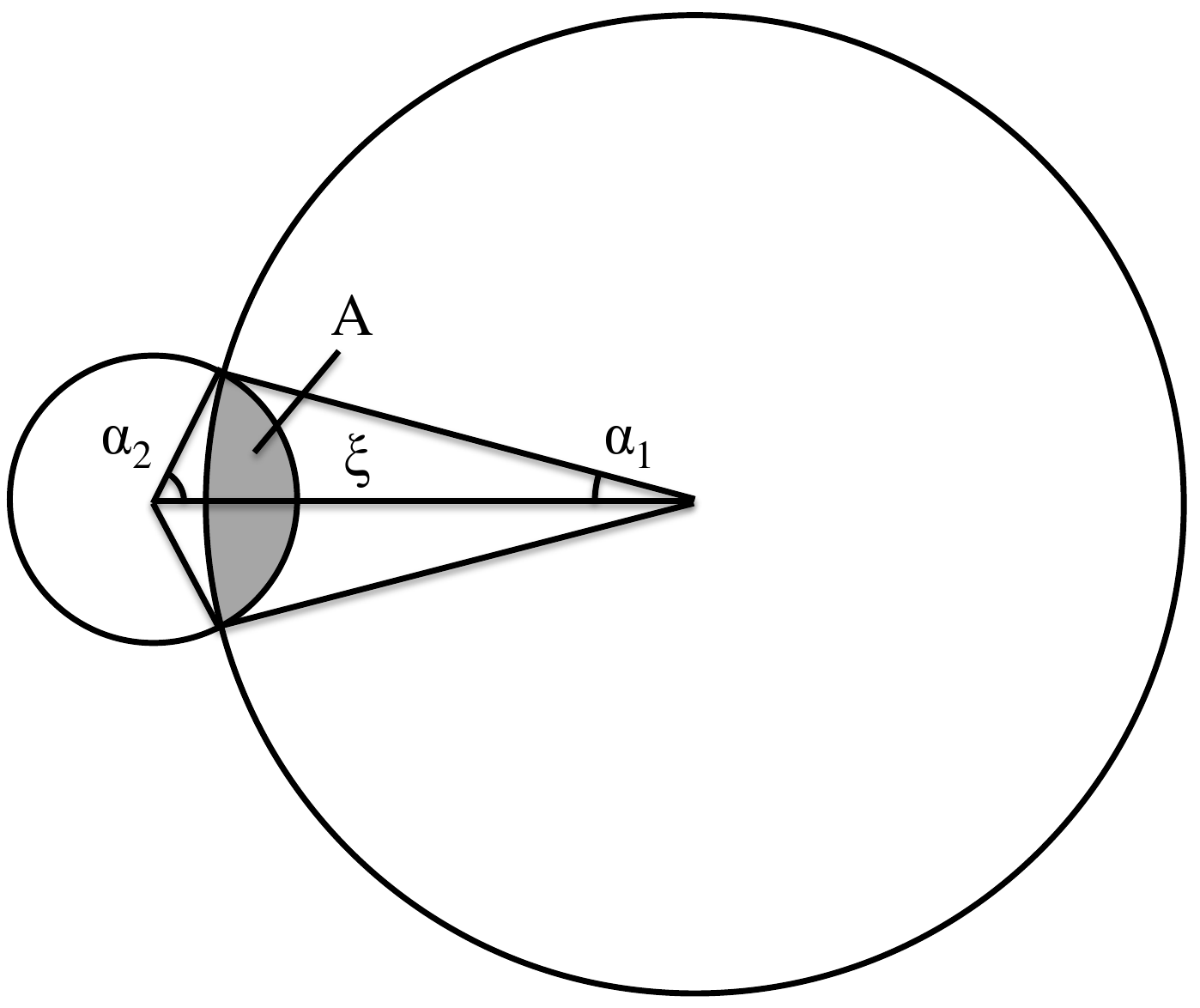}
\caption{ Parameters used to calculate the overlapping surface $A$ (shaded) between the star (large circle) and the blob. Here $\xi$ is the projected distance between the centers of the blob and the star. The angles $\alpha_1$ and $\alpha_2$ are indicated (see Eq.~\ref{eq:surface}).}
\label{fig:geom}
\end{figure}

The observed line profile can be represented by an assumed constant part (atmospheric plus wind) and a variable part, related to the blob(s), as elaborated in Sect.\,\ref{quotient}. To isolate these contributions we make use of subsequent quotient spectra, i.e.,\ the following spectrum divided by the previous one, thereby avoiding -- and ignoring -- the complicated line formation process of the underlying wind of this rapidly rotating star \citep[see][]{hillier:2012}. The justification of this approach is that the equivalent width of the observed variations are usually less than a few percent of the total equivalent width.
We describe a blob as a sphere with radius $r$ centered around the geometric center of a prominence at distance $h$ from the stellar surface. We assume that the optical depth of the  emission (or absorption) profile of the blob itself can be described as a Gaussian function with central optical depth $\tau$ and full width at half maximum $w$ expressed in velocity units. For convenience we express $r$ in stellar radii and put the stellar radius itself equal to unity. The resulting line profile of the blob, normalized by unity, can be described by
\begin{equation}
\label{eq:blob}
F(v)_b=\exp\left[-A\tau e^{-\frac{1}{2}\left((v-v_b)/w\right)^2}\right]\exp\left[(\pi r^2 - A)\tau e^{-\frac{1}{2}\left((v-v_b)/w\right)^2}\right]\,,
\end{equation}
with $v_b$ the location of the center of the blob expressed in velocity units, and $A\leq \pi r^2$ the overlapping surface of the star and the blob. The two factors in Eq.~(\ref{eq:blob}) correspond to the absorption and emission component, respectively. We have
\begin{equation}
\label{eq:vb}
v_b=(1+h)v_{eq}\sin i\sin\phi\cos\theta\,,
\end{equation}
with $h \geq r$ the distance between the surface of the star and the center of the blob expressed in stellar radii (if $h=r$ the blob touches the stellar surface). The parameter $v_{eq}$ is the equatorial rotational velocity, $i$ is the inclination angle, $\phi$ is the longitude of the blob ($\phi=0$ when the blob passes the subsolar central meridian), and $\theta$ is the latitude of the blob ($\theta=\pi/2$ at the north pole and $\theta=0$ at the equator).
The overlapping surface area, $A$, follows from the description of exoplanet transits \citep[Sect.\ 3.2.2]{haswell:2010}:
\begin{equation}
\label{eq:surface}
A = r^2\alpha_1 + \alpha_2 - \frac{1}{2}\sqrt{4\xi^2 -(1+\xi^2-r^2)^2}\,.
\end{equation}
Here $\alpha_1$ and $\alpha_2$ are the half angles subtended by the intersecting arc of the star and the blob, respectively, as seen from their centers, which are separated by their projected distance $\xi$ (see Fig.\,\ref{fig:geom}). Their values follow from
\begin{equation}
\label{eq:alpha1}
\cos\alpha_1=\frac{r^2+\xi^2-1}{2\xi r}\,\, {\rm and}
\end{equation}
\begin{equation}
\label{eq:alpha2}
\cos \alpha_2=\frac{1+\xi^2-r^2}{2\xi}\,, {\rm with}
\end{equation}
\begin{equation}
\label{eq:xi}
\xi=(1+h)\sqrt{(\sin^2\!\phi+\cos^2\!\phi\cos^2\!i)\cos^2\!\theta+\sin^2\!i \sin^2\!\theta}\,.
\end{equation}

\begin{figure}[!ht]
\centering
\includegraphics[width=0.8\columnwidth]{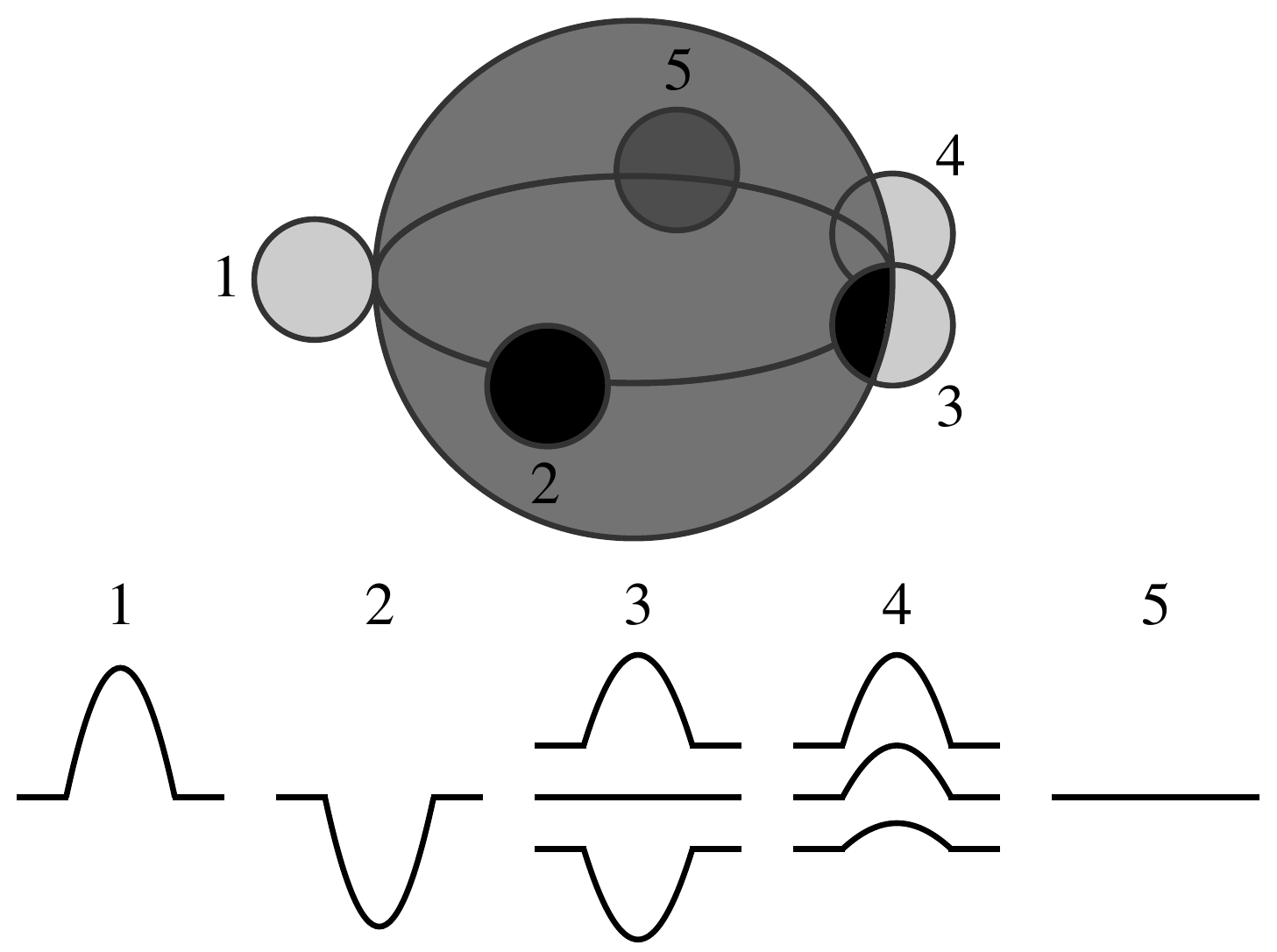}
\caption{ Line profile contributions of a spherical blob, depending on its relative location. {\it Top:} Geometry, here drawn when the blob touches the star ($h=r$). {\it Bottom:} Corresponding sample profiles. (1) The blob is located at the side of the star. (2) The full surface of the blob is in front of the star. (3) The blob partly overlaps the stellar surface. (4) The blob  is partly hidden behind the star. (5) The blob is completely behind the star.}
\label{fig:blobloc}
\end{figure}

\begin{figure}[b!]
\centering
\includegraphics[width=0.8\columnwidth]{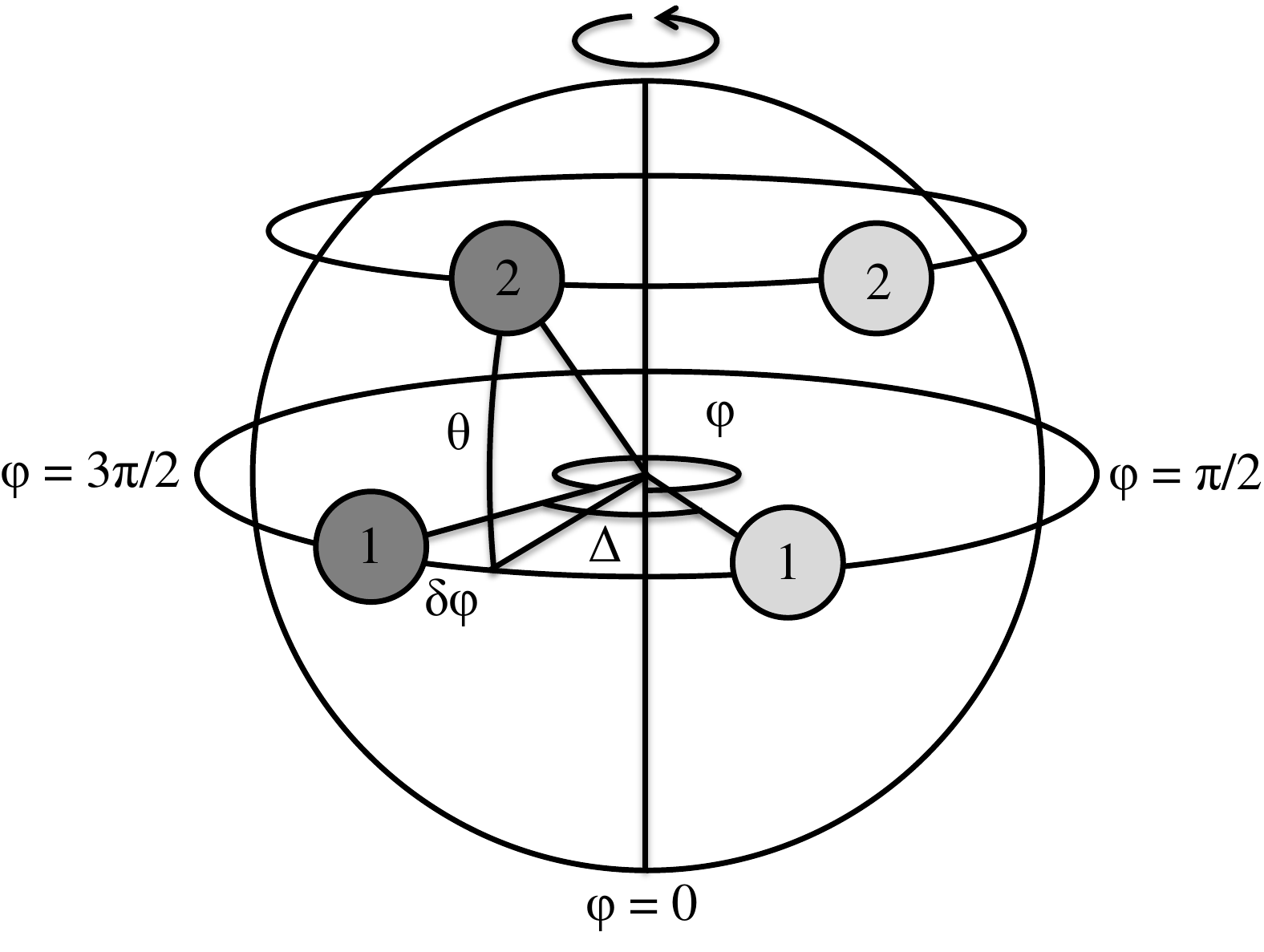}
  \caption{Angles determining the position of multiple blobs. Two blobs, labeled 1 and 2, at the equator and at latitude $\theta$, respectively, are depicted at their initial position (dark gray) at longitude $\phi$ measured from the subsolar meridian, and an instant later (light gray) after being rotated by an angle $\Delta$. A second blob leads the first by an angle $\delta\phi$.}
\label{fig:angles}
\end{figure}

There are five different cases that act differently on the line profile depending on the relative location of a blob (see Fig.\,\ref{fig:blobloc}):
\begin{enumerate}
\item The projected blob is located next to  the star (their surfaces do not overlap). The blob contributes to the line profile as an emission feature with maximum amplitude and with the highest velocity (negative or positive)  of $(1+h)v$sin$i$.
\item The full projected surface of the blob is in front of the star. In this case the blob  gives extra absorption with maximum amplitude at a velocity within the range $\pm(1-r)v$sin$i$.
\item The blob  partly overlaps the stellar surface. The result is a superposition of the emission and absorption components. Depending on the relative size of the overlap area the extra emission or absorption  has a lower amplitude than in the two previous cases. If the overlap area is half of the blob surface, the emission and absorption components cancel each other out.
\item If the blob is partly behind the star it  gives extra emission with a lower amplitude than in the first case.
\item The blob is completely behind the star and does not contribute to the line profile.
\end{enumerate}

As a function of the rotational phase, $\phi$, the contribution to the line profile due to the presence of a single blob  appears as an additional absorption at $\phi=0$, turns into emission at $\phi=\pi/2$, then disappears and reappears as emission at $\phi=3\pi/2$, and finally  returns to absorption after one stellar rotation.

Our description allows for multiple prominences, or blobs, by multiplying Eq.~\ref{eq:blob} for each blob with its respective positional parameters. The position at a given moment in time, $t$, is determined by two angles: its longitude $\phi$ and latitude $\theta$ (Fig.\,\ref{fig:angles}). We assume that the relative position of blobs does not change with time.
The longitude of the first blob is taken as a reference: $\phi_1=\phi_0$. Subsequent blobs have longitude $\phi_1+\delta\phi_j$ with $\delta\phi_j$ the angular distance between the first and subsequent blobs, numbered with subindex $j = 2,3,..., N$.
 The epochs of the first, second, and $n^{th}$ observation are dentoted  $t_1, t_2,..., t_n$; after a time interval $t = t_2 - t_1$ all blobs of a given configuration have turned by an angle $\Delta$, measured from $\phi_0$. The longitude of the first blob at $t = t_n$ is $\phi_1=\phi_0+\Delta(t_n-t_1)/(t_2-t_1)$, whereas the longitude of blob $j$ is  $\phi_1+\delta\phi_j$.

The model can also account for differential rotation. In this case the longitude of blob $j$ related to the longitude of the first blob is $\phi_1+\delta\phi_j+f\phi_1\sin\theta$, where $f$ is the factor determining the velocity of the rotation at a given latitude.
The latitude and the inclination angle dependencies follow from the obvious projection effect, and is taken into account.
With the stellar rotation period denoted by $P$ we have
\begin{equation}
\label{eq.Period}
\Delta=\frac{2\pi}{P}(t_2-t_1)\,.
\end{equation}
Consequently, for an adopted value of $\Delta$ (or rotation period) with a specified configuration of blobs, the position of the blobs is fixed at any given moment of time. This is used for predicting the time evolution of the features in subsequent quotient spectra, to be compared with the observed values.

In summary, the subsequent quotient profiles can be calculated by providing the values of the parameters  $i$, $v_{eq}$, $h_j$, $r_j$, $\tau_j$, $w_j$, $f$, $\theta_j$, $\delta\phi_j$, $\phi_0$, and $\Delta$.

\section{Fit results}
\label{section:fits}

\subsection{Fit strategy}
\label{sub:fits}

The many degrees of freedom in our formalism obviously complicate the search in parameter space for a fit.  For this reason we simplified our calculation by restricting the blobs so that they  all have  the same radius,  are        attached to the stellar surface ($h_i=r_i$), and are also in the equatorial plane ($\theta_i=0$), and therefore without differential rotation ($f=0$). As we are most interested in the lifetimes of the prominences (or blobs in our description) the  value used for the rotation period is essential. After initial trials, we kept $\Delta$ corresponding to $P=4.1$d (see Sect.\,\ref{star}), because in general with this value the fits appeared acceptable for a longer time span of subsequent quotient spectra than for other nearby periods. It is not certain, of course, that this assumption yields the real rotation period, and our fits  therefore cannot be considered unambiguous, but rather as illustrative. The justification of this approach is twofold. First we note that
deviations of the adopted value of rotation period would affect all derived lifetimes, but only to a very minor extent because the best time-resolved datasets cover at most 6 days. Second, searching for the longest stretch of data that can be described by the same set of parameters will give an upper limit  to the derived lifetimes.
After some trials we fixed the radius of all blobs at 0.2~$R_*$ as this would give reasonable values for the competing parameters $\tau_i$ and $w_i$ for a given equivalent width in the quotient domain. In practice, initial parameter values for $\tau_i$, $w_i$, $\phi_i$, and $\delta\phi_i$ are estimated by a trained eye, and when the fit seems reasonable, these values are used as starting values for a final least-squares solution to obtain the best-fit values of the specified parameters.

Our fit strategy is to try to fit as many subsequent quotients as possible with the same set of parameters. We consider a fit to be completed when the fit of a preceding and a following quotient of a series fails.  To emphasize this strategy we added the failed fits as the first and the last plots in all figures with fits, and overplotted the subsequent quotients  as described below and presented in Appendix C.  As a practical example, we consider the fit displayed in Fig.\ \ref{fig:qfit2013b}, third series. The geometry is drawn for spectrum number 16, the first one included in the fit of this series. To reproduce quotient 19/18 we note that the star has rotated only by a minimal amount (about 6$^{\circ}$), as spectrum number 18 was taken 0.7 + 1.1  = 1.8 h later. To reproduce the quotient spectrum (with time interval of 1.9 h) it appears that three blobs are needed.  The rightmost blob produces an absorption in both single spectra, but will appear in the quotient spectrum as an emission and an absorption feature at positive velocities because of the location of the blob on the receding hemisphere and the redwards shifted second  spectral profile of the blob.   Similarly, the middle blob is largely responsible for the absorption in the quotient spectrum at negative velocity, whereas the leftmost blob gives the emission at negative velocities. There are two locations for the leftmost blob that would have the same effect on the quotient spectrum. One of these locations appeared to be in better agreement with the other quotients in the series, which we therefore adopted. The subsequent quotient spectrum could also be fitted with this configuration, but quotient 21/20 could not.  Adding more blobs could make one fit somewhat better, but will destroy other fits. To minimize arbitrariness when fitting the next series, in all cases we tried to keep as many blobs as possible from the previous configuration while aiming at the minimum number of blobs to obtain an acceptable fit over as many subsequent fits as possible.

We now describe the results for our datasets.


\begin{figure*}[!t]
\begin{minipage}{\columnwidth}
\centering
\includegraphics[width=0.53\linewidth]{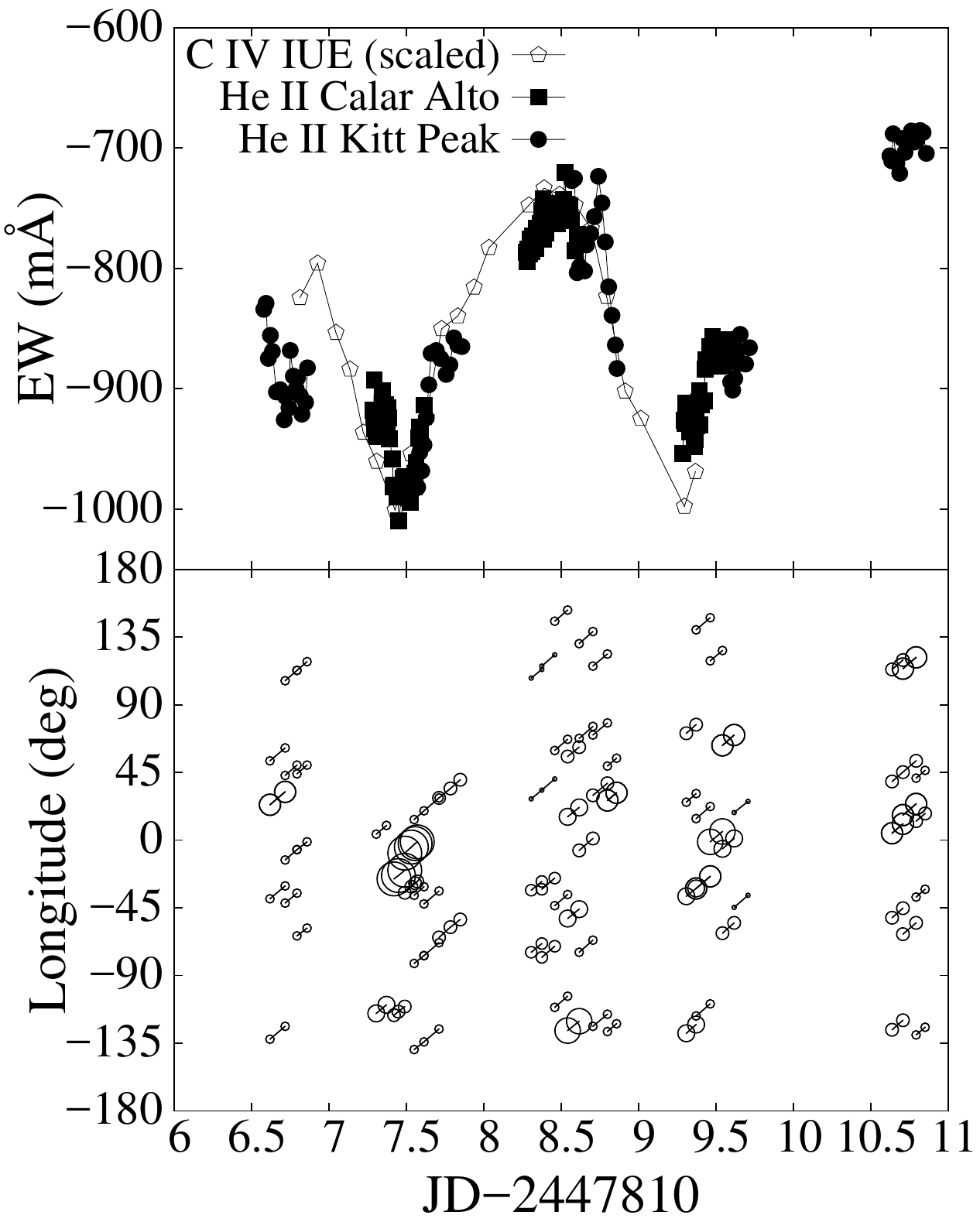}\par
\caption{Equivalent width of the \ion{He}{II} $\lambda$4686 line in 1989 and simultaneous \ion{C}{IV} blue edge wind variability (top, taken from \citealt{henrichs:1991}) plotted along with the derived properties of the blobs with the symbol size proportional to the optical thickness (bottom).  The vertical axis denotes the longitude of each blob as derived from the fits. The length of the connecting solid lines indicates the lifetime of the blobs. The linear trend is caused by the rotation.}
\label{fig:EWblobph1989}
\end{minipage}\hfill
\begin{minipage}{\columnwidth}
\centering
\includegraphics[width=0.6\linewidth]{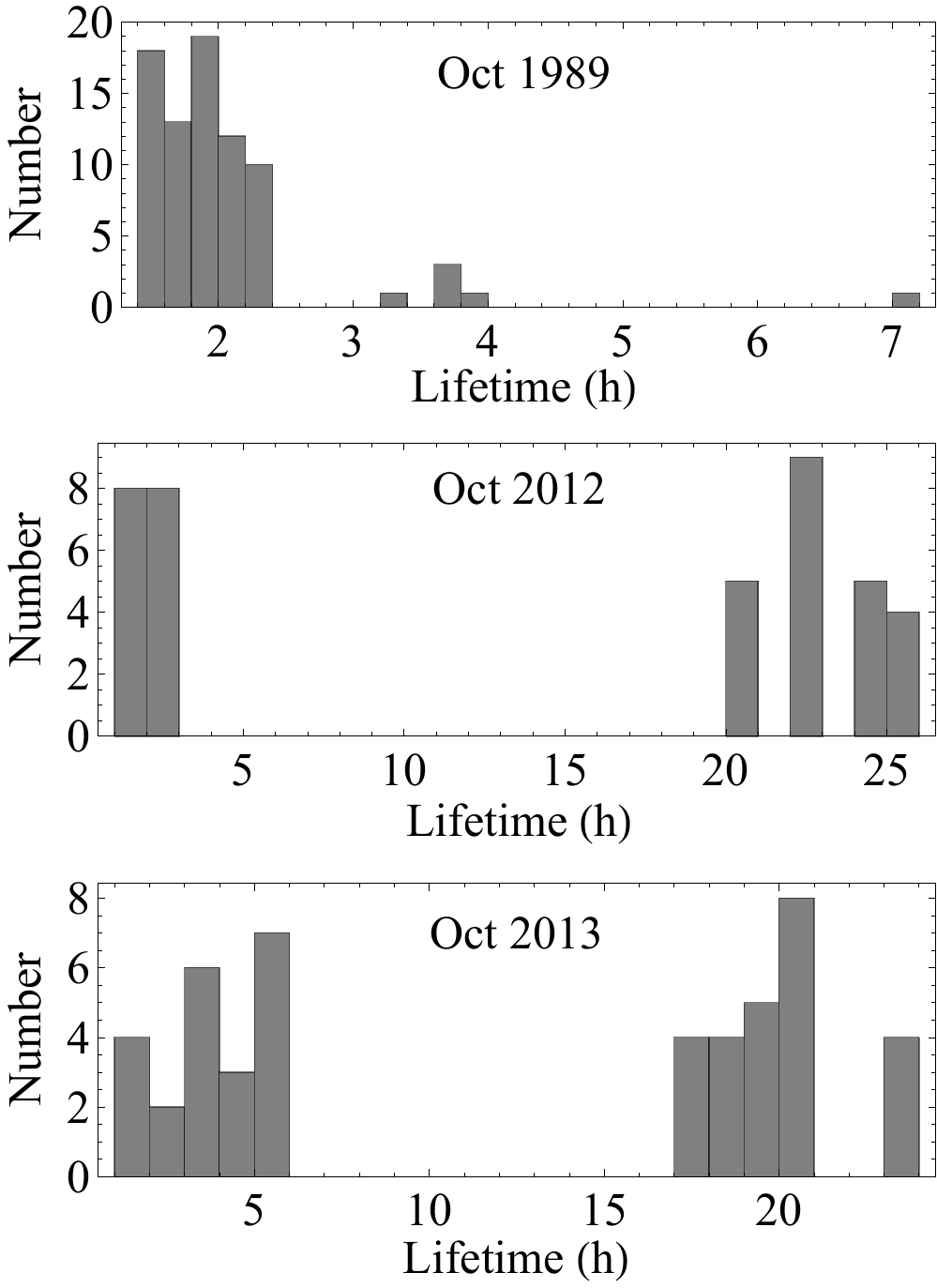}
\caption{Histograms showing the number of fitted blobs as a function of lifetime in  1989, 2012, and 2013.}
\label{fig:histo_all}\
\end{minipage}
\end{figure*}

\begin{figure*}[!b]
\includegraphics[width=0.87\linewidth]{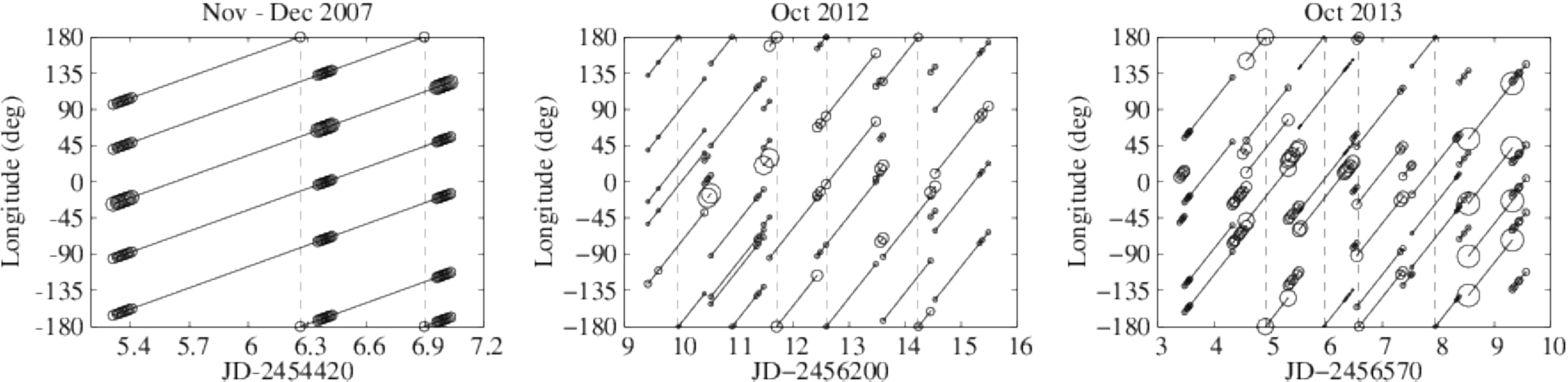}
  \caption{ Same as the bottom panel of Fig.\,\ref{fig:EWblobph1989}, but for the   2007, 2012, and 2013 datasets. The vertical dashed lines denote the continuation of the solid line where the longitude changes from 180$^\circ$ to $-180^\circ$.}\label{fig:blobph_all}
\end{figure*}

\subsection{The 1989 dataset}
\label{sub:1989analysis}

In total 180 spectra were available, often with a time resolution of 15 min that already showed significant changes between subsequent spectra. To limit the number of the time-consuming fits to the most significant variable epochs we rebinned the spectra in the time domain onto a grid of an average of 2 hours, except  during the second night when the spectra changed so quickly that we used averages of 0.5 hours. This resulted in a new set of 33 spectra, with 32 corresponding subsequent quotients.
The fits are presented in Appendix~\ref{A1989}. We emphasize that in all sequences the first and last panels display  a failed fit on purpose, signifying the beginning and ending of what we consider  one episode, and hence constraining the
lifetime of a given configuration.
The main result is that in 1989 there are always multiple blobs present, from 2 to 5, but mostly 4, with lifetimes from 1.4 to 7.2\,h. The typical optical depth is $\tau = 0.2$, but values vary between 0.1 and 0.8 with a typical FWHM of $w = 70$\,km\,s$^{-1}$, varying between 50 and 90\,km\,s$^{-1}$. A graphical summary of the configurations, lifetimes, and properties of the blobs as a function of longitude is presented in the bottom panel of Fig.\,\ref{fig:EWblobph1989}. Solid lines connect the longitudes for each identified blob, with the length representing the lifetime. The linear trend is caused by the rotation. The size of the circles is proportional to the optical thickness $\tau$ of the blob.


As mentioned above, the simultaneous optical and UV observations for this dataset show a puzzling phase coherence of the equivalent width of the optical emission of the \ion{He}{II} $\lambda$4686 line measured within $v = [-400,400]$\,km\,s$^{-1}$ and blue-edge variability of the \ion{C}{IV} line at $-2300$\,km\,s$^{-1}$ \citep[Fig.\,\ref{fig:EWblobph1989}, top panel, taken from][]{henrichs:1991}, which has  never been explained. In this figure, the highest edge velocity occurs around day 8.5. A direct relation between the optical thickness and the edge velocity (here represented by the equivalent width) is not apparent in this figure.  However, as mentioned in the introduction, since the origin of DACs must be close to the surface of the star, a delay is expected between the onset of a DAC and the effect on the blue edge or, in terms of the quantities in Fig.\,\ref{fig:EWblobph1989}, this same delay is expected between the occurrence of the blobs with the strongest optical depth and the maximum EW of the \ion{C}{iv} line. In the diagram this delay is about 1 day, but the onset of a DAC for this dataset cannot be determined with sufficient confidence from the UV spectra. From \citet[][Fig.\,26, lower panel]{kaper:1999} they derive $\sim$1.5 days for the 1991 dataset when the DAC development looked different. From the analysis of \cite{massa:2015} of a number of stars with sufficiently high time resolution of UV spectra, we estimate that this delay varies between 0.5 and 1.5 days in $\xi$ Per, and around 1 day with a very large uncertainty for $\lambda$ Cep. Although it seems suggestive, we cannot confirm a causal relationship between the strongest blobs and the highest edge velocity, which means that the phase coherence in the upper panel of the plot could still be a coincidence in the sense that the 1-day delay coincides with  half of the 2-day quasi-cycle period.
The left-hand panel of Fig.\,\ref{fig:histo_all} shows the histogram of the number of blobs in our model fitting as a function of the lifetime. Most of the blobs survive only a few hours, which is not a consequence of the limited coverage (see the fits in the figures in Appendix~\ref{A1989}).

\subsection{The 2007 dataset}
Spectra were obtained during 2 hours on 20--22 November and  16 December 2007, with an average of 6--8 spectra per night.  All acceptable fits to the subsequent quotient spectra are listed in Appendix~\ref{A2007}. Three or five blobs were needed with optical depths $\tau$ = 0.07 to 0.77, and FWHM of $w$ = 72 to 95\,km\,s$^{-1}$. The derived lifetime of the blobs is mostly 1.7\,d, but is not well constrained because of the limited coverage. The time lapse between the last two nights is 23.9\,d, and reaching a fit over this time span is likely a coincidence; therefore,  the left-hand panel of Fig.\,\ref{fig:blobph_all} contains  the lifetimes only for 20--22 November. We note that we were able to  make several acceptable fits for different series of subsequent quotients that contained a given quotient with a large amplitude. This obviously illustrates the ambiguity in the interpretation, but this is of less concern (see Sect.\,\ref{section:unique}).

\subsection{The 2012 dataset}

Subsequent quotient spectra were calculated over 7 sequential nights in 2012 (see Appendix~\ref{A2012}). From 2 to 5 blobs were needed for a fit with optical depths $\tau$ = 0.14 to 0.7 (but mostly 0.2) and FWHM of $w$\,$=$\,$40$ to 93\,km\,s$^{-1}$.  A histogram with the number of blobs as a function of lifetime is depicted in Fig.\,\ref{fig:histo_all}, middle panel. The derived lifetimes of the blobs are from 1.2 h to 25.1 h (i.e., 1.05 d). The lifetimes of the blobs cluster around 23 h (0.96 d), implying that many survive until the next night. The evolution of the blobs is displayed in the middle panel of Fig.\,\ref{fig:blobph_all}.

\subsection{The 2013 dataset}
The subsequent quotient spectra with the corresponding time resolution for the 7 sequential nights in 2013 with the best fits are given in Appendix~\ref{A2013}.
The quotients show less variability than in 1989 for spectra separated by less than 1~h apart.
Like before, we were able to  make more acceptable fits for different series of subsequent quotients that contained a given quotient with a large amplitude, again illustrating the ambiguity in the interpretation. From 2 to 5 blobs with optical depth $\tau$ = 0.08 to 0.81 (but mostly 0.2) were needed with a FWHM of $w$ = 55 to 192\,km\,s$^{-1}$ (mostly 60--80\,km\,s$^{-1}$). The evolution of the blobs is displayed in Fig.\,\ref{fig:blobph_all}, right-most panel. A histogram with the number of blobs as a function of their lifetimes is given in the right-hand panel of Fig.\,\ref{fig:histo_all}. The derived lifetimes are between 1.6 and 24.7\,h with many surviving until the following night, as in October 2012.

\subsection{Summary of fit results}

Our results suggest that, within the framework of our model, there are multiple blobs (2 to 5)  with typical lifetimes of a few hours present around the star, but also quiet phases without much variation over several hours, especially in 2013. There is a wide variety among the blob properties regarding optical depth and velocity dispersion. The configurations clearly differ in the four datasets.  The most likely picture that emerges is that the stellar activity is variable and that the lifetimes differ from year to year. This could be concluded, for example, from the notion that the amplitudes of the subsequent quotient spectra taken within 1 h in 1989 are larger than in 2013, and that more quotients in a row over a longer time stretch could be fitted in the 2013 dataset.


\section{Discussion}
\label{section:discussion}

\subsection{Uniqueness of fits}
\label{section:unique}

Since the fits of a series of subsequent quotient spectra depend on an adopted rotation period, it is obvious that these fits cannot be unique. The many restrictions we imposed on the parameter range, like spherical geometry, latitude, radius, and distance to the surface, also limit the validity of our application. For instance, using a slightly different rotation period will give results with slightly different parameters and lengths of longest stretch of data that can be fitted.
We also cannot be sure that blobs survive until the next night in the last two datasets, although consistent fits could be obtained. In fact,  these longer lifetimes were not found in the 1989 dataset with much better time coverage during 24 hours, which hints that the longer lifetimes may not be real, but a consequence of lack of data coverage.
From our point of view, however, the important outcome is the number of blobs and lifetimes, which will not change much. We therefore emphasize that the fit results should not be taken at face value, but rather as an indication that our model can be applied, which will hopefully serve as a starting point for further research with a more physical basis. The various (quasi-)periodicities in the different spectral lines for a given dataset may provide insight into the stratification of the line forming regions, assisted with calculations of the type  presented by RH15 (their Fig.\,7).

\begin{figure}[!b]
\includegraphics[width=0.98\columnwidth]{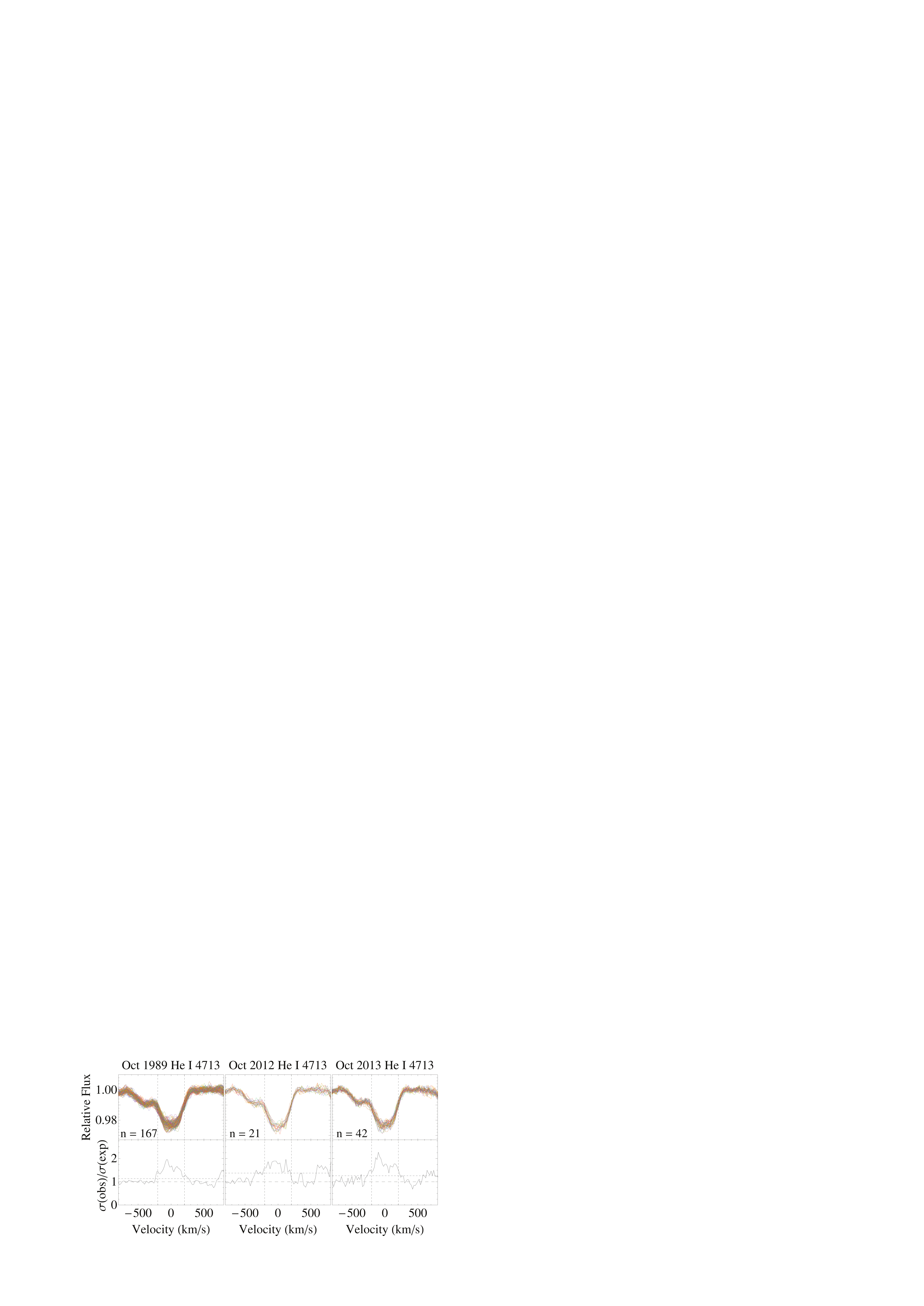}
\caption{Same as Fig.\,\ref{fig:TVS4686}, but for \ion{He}{I} $\lambda$4713 with S/N = 900 for the 1989, 2012, and 2013 datasets, with $n$ the number of spectra. We note that the variability above the 1$\%$ significance level (indicated by the upper short-dashed line in the lower panels) is restricted to a velocity range approximately  within the $v$sin$i$ limits. }
\label{fig:TVS4713}
\end{figure}

\begin{figure*}[!t]
\centering
\includegraphics[width=0.98\linewidth]{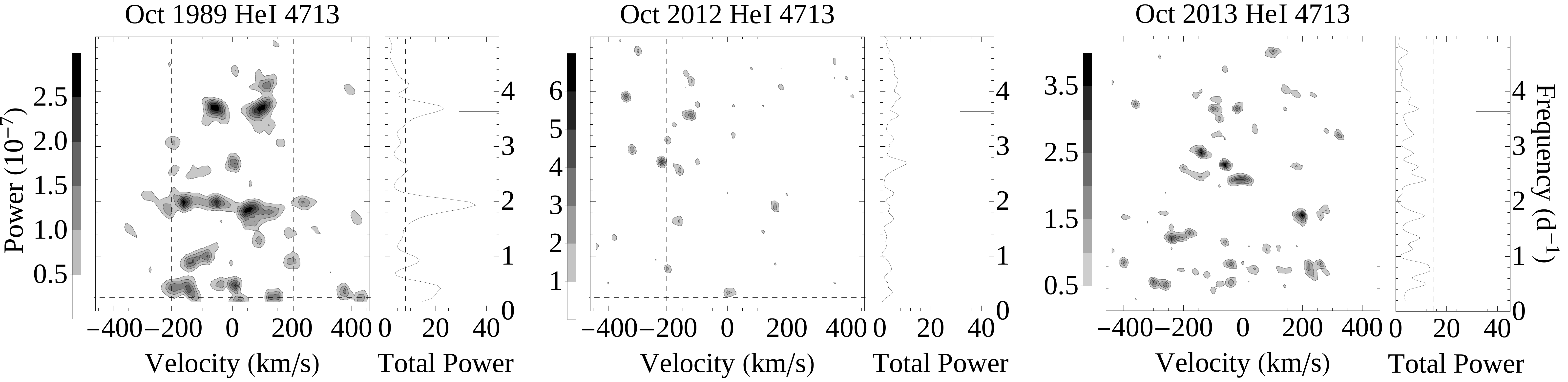}
\caption{Same as Fig.\,\ref{fig:powerHe2_4686}, but for \ion{He}{i} $\lambda$4713. The full profiles are displayed in Fig.\,\ref{fig:TVS4713}. We note that in 2012 and 2013 the total power has no peaks above the noise level. The two NRP frequencies reported by \cite{dejong:1999} are marked by horizontal tickmarks.}
\label{fig:powerHe1_4713}
\end{figure*}

\subsection{Effect of non-radial pulsations}
\label{nrpdiscussion}

\afterpage{
\begin{figure}[!t]
\centering
\includegraphics[width=0.98\columnwidth]{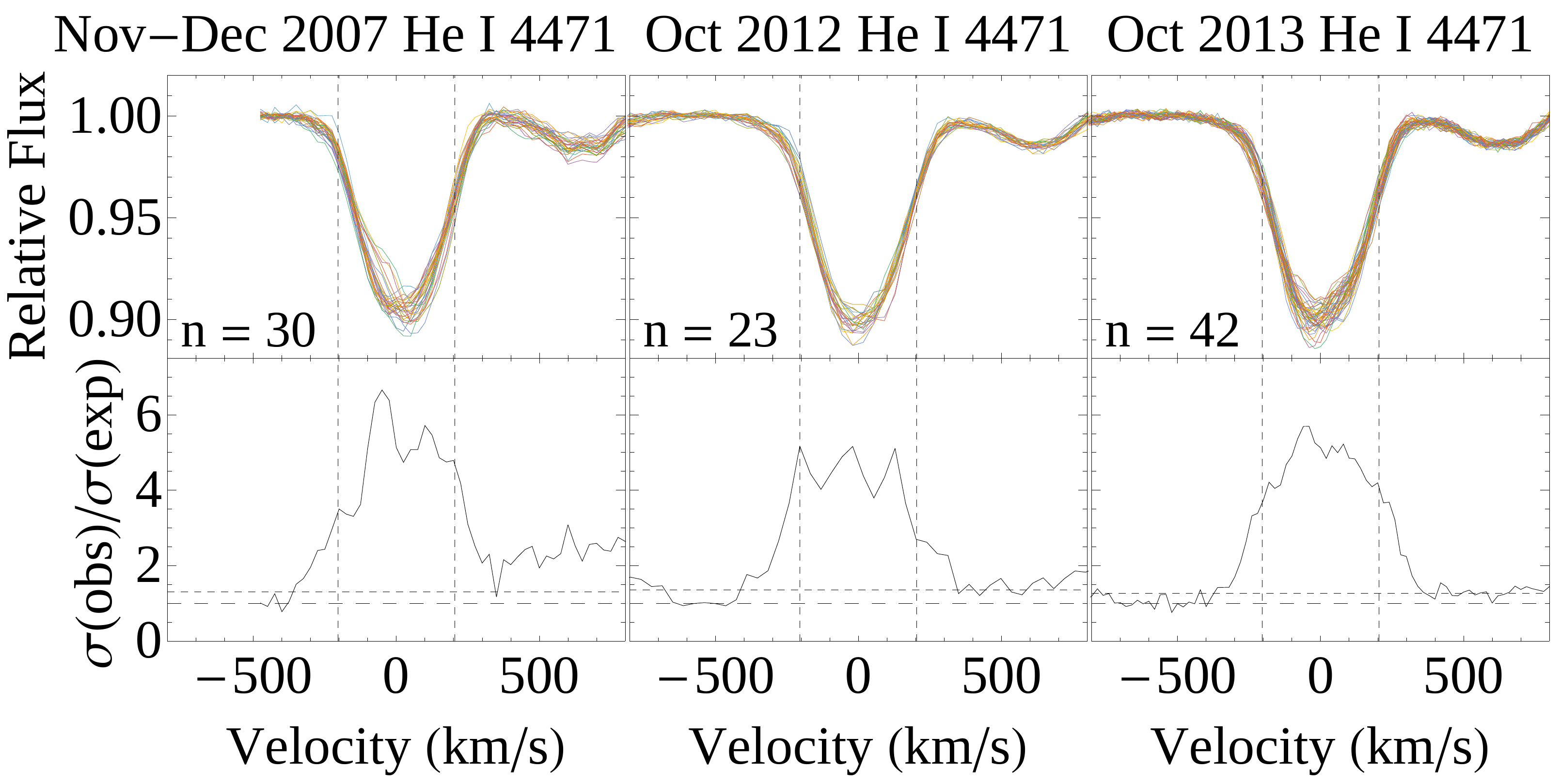}
\caption{Same as Fig.\,\ref{fig:TVS4713}, but for \ion{He}{I} $\lambda$4471 with S/N = 870 for the   2007, 2012, and 2013 datasets. We note that in all three panels the variability above the 1$\%$ level extends to nearly 2 times the $v$sin$i$ limits.}
\label{fig:TVS4471a}
\end{figure}

\begin{figure}[!t]
\centering
\includegraphics[width=0.98\columnwidth]{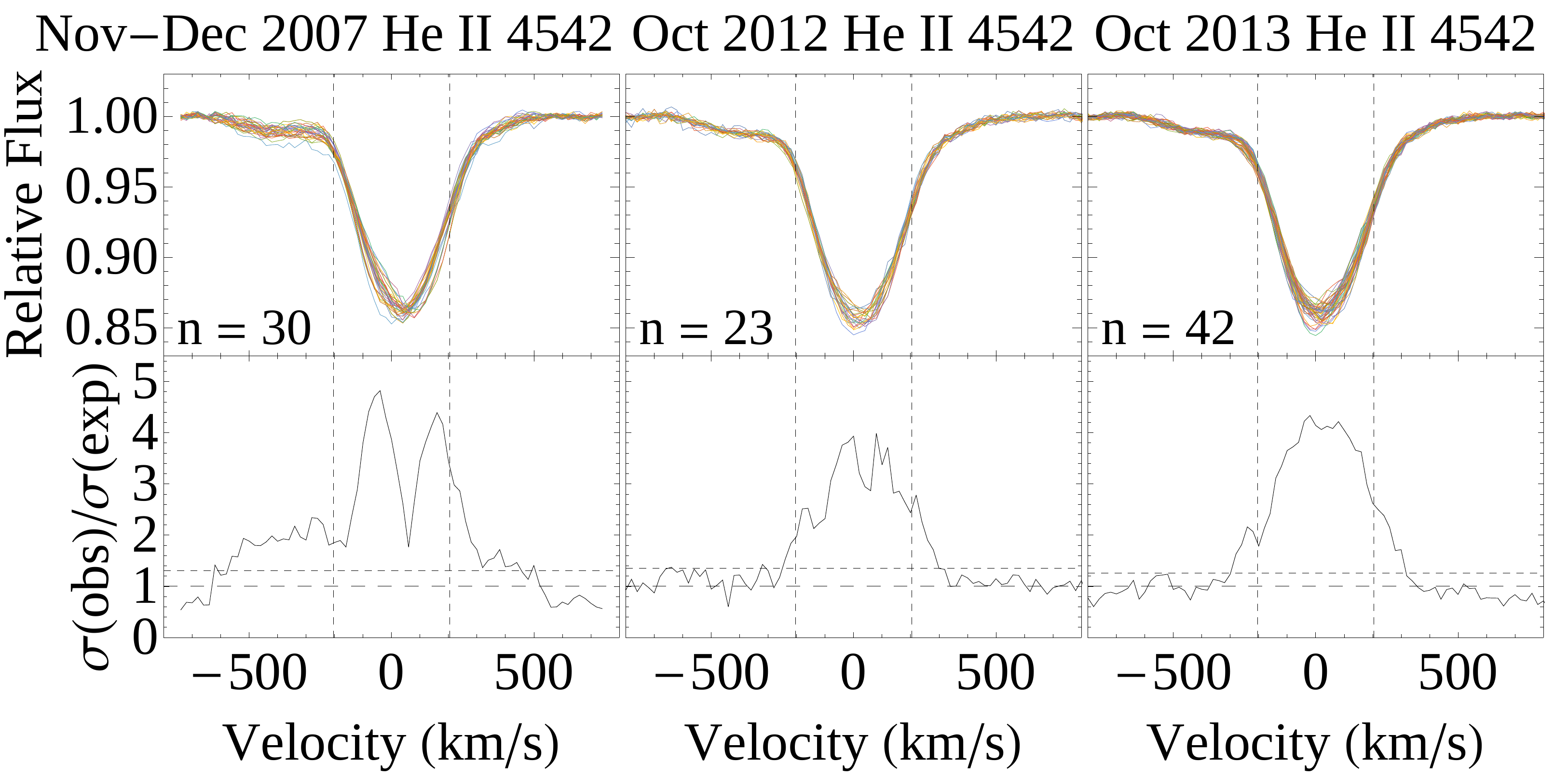}
\caption{Same as Fig.\,\ref{fig:TVS4713}, but for \ion{He}{II} $\lambda$4542 with S/N = 590 for the   2007, 2012, and 2013 datasets.}
\label{fig:TVS4542a}
\end{figure}
}

\begin{figure*}[!t]
\includegraphics[width=1.90\columnwidth]{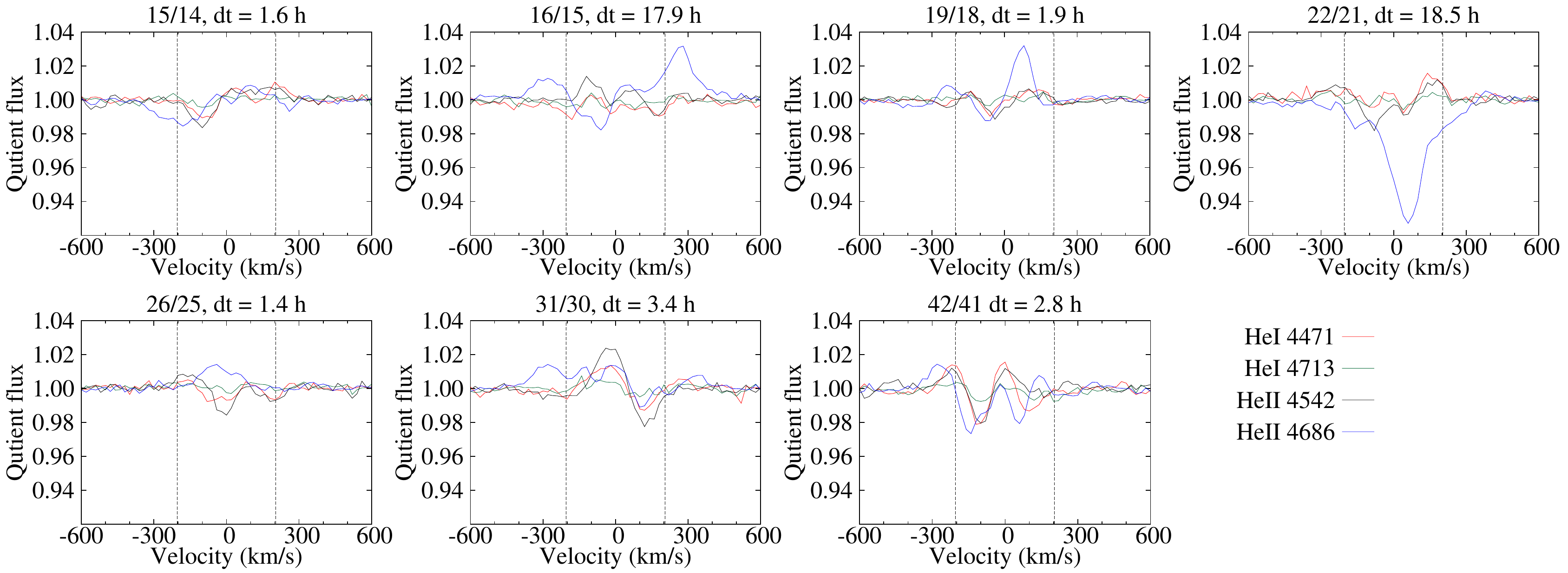}
  \caption{ Same as Fig.\,\ref{fig:quotients}, but only for the lines $\lambda$4471 and $\lambda$4542, studied by \cite{uuh-sonda:2014}, and the subsequent quotients of $\lambda$4713 and $\lambda$4686  for comparison.}
\label{fig:US14quotients}
\end{figure*}

In the paper by \citet[hereafter DJ99]{dejong:1999} two non-radial pulsation (NRP) modes with periods of 12.3 and 6.6~h were reported. In first order (for spherical stars) an NRP signal should be present in all lines, and in the light of the results presented above the question arises whether the spectral line variability that we attribute to phenomena above the stellar surface would  affect our analysis. Conversely, a contamination by the lower wind, as found above in most spectral lines, could have prevented a proper NRP analysis as presented by DJ99, which was solely based on the weak atmospheric \ion{He}{I} $\lambda4713$ line.
 \citet[hereafter   US14]{uuh-sonda:2014} have reported results of a multi-epoch spectroscopic monitoring campaign of
$\lambda$~Cep that took place between December 2009 and September 2011 and have searched for the presence of NRP in several spectral lines, which, unfortunately, did not include \ion{He}{I} $\lambda4713$. The authors could neither confirm neither the periodicities nor pulsation modes reported by DJ99, although $10.9 \pm 0.4$ h was considered to be a candidate NRP period. Importantly, these and other derived periodicities appeared to be different at different epochs, which made US14 conclude that any periodicity might reflect transient phenomena, and that there is no stable period that can be  attributed with certainty to NRP. This questions the validity of the NRP modes reported by DJ99. Since DJ99 used the same dataset as analyzed above and we have two more datasets that contain most of the same lines as in US14, we address here the issues raised in this paper.
A  point of concern is the S/N of the data. \cite{dejong:1999} based their analysis on data which were rebinned to a S/N~$\approx 900$, which allowed them to detect variations in this \ion{He}{I} line below $0.5\%$ of the continuum, i.e.,\ the approximate amplitude of the reported NRP.   Figs.\,\ref{fig:TVS4713} and \ref{fig:powerHe1_4713} show a comparison between our 1989 data and our new datasets (2012 and 2013), which were sampled at the same S/N to allow a proper comparison. We conclude that in 1989 all significant variations above the 1$\%$ significance level occurred essentially within $\pm v$sin$i$. This justified the implicit assumption by DJ99 that the variability is likely dominated by an atmospheric phenomenon, which they interpreted as NRP. This is supported by \cite{martins:2015} who showed that of all optical hydrogen and helium lines, this line is formed deepest in the photosphere, although modeled for a somewhat cooler star than $\lambda$ Cep. Our 2012 dataset is undersampled to confirm any conclusion regarding NRP. Our 2013 dataset is also undersampled for an NRP analysis in a period range of 0.2--0.5 d, but the overall TVS looks similar to the TVS of 1989, in the sense that hardly any significant variability occurs outside the $v$sin$i$ limits, and that the amplitude of variations in 1989 and 2013 are comparable.

We have two lines in common with US14.  The first is the  \ion{He}{i} $\lambda4471$ line, which is four times stronger (see Fig.\,\ref{fig:TVS4471a}), where the variability extends to nearly $\pm 2v$sin$i$ in all three of our  datasets and which points to a significant wind contamination. Large-amplitude NRP could cause variations outside the $v$sin$i$ limits, but if the small signal in \ion{He}{I} $\lambda4713$ represents NRP, it would be buried under the much larger amplitude caused by variations in the lower wind region as is observed in nearly all the other lines we studied.
The variability we find at $\lambda4471$ is in accordance with US14 (their Fig.\,1), which is also within $\approx$ $2v$sin$i$ in their dataset (sampled with a S/N $\leq405$).  Given the episodic or stochastic nature of the variability that we found in the lower wind, we conclude that the absence of the reported NRP periodicities in the $\lambda$4471 line is most likely due to this contamination, which dominates in this line over the NRP signal. Additional support for this conclusion comes from inspection of subsequent quotient spectra, illustrated in Fig.\,\ref{fig:US14quotients} for our 2013 data of the \ion{He}{I} $\lambda$4471, $\lambda$4713 and \ion{He}{II} $\lambda$4542, $\lambda$4686 lines. In all cases the amplitudes of the \ion{He}{I} $\lambda$4713 subsequent quotients are tiny and are the smallest; in particular, they are smaller than the \ion{He}{I} $\lambda$4471 quotients.

The other line in common with US14 is the much stronger \ion{He}{II} $\lambda$4542 line, with a central depth of 15$\%$ (see Fig.\,\ref{fig:TVS4542a}). The amplitudes of the subsequent quotients of this line are in all cases larger than those of the \ion{He}{I} $\lambda$4471 quotients, which signifies an even larger impact of the lower wind on this line, and hence an even more inhibiting  detection of NRP.  In this line the velocity range of variability  both in our datasets and in US14 is somewhat smaller than in the $\lambda$4471 line, which may be due to the higher excitation level, possibly associated with a latitude dependency in this rapidly rotating star, for which the polar regions are about 3$\%$ hotter than near the equator (RH15).

Interestingly in this aspect, US14 also performed a red-noise analysis, following  \cite{blomme:2011}, to measure the extent of stochastic processes that contribute to power in a Fourier analysis. They found that the mean lifetime of dominant structures in the line-profile variability is on average $0.028\pm 0.013$ d, or $< 1$\,h, and that about one half of the power contributes to the observed random variations. This is  shorter than but not necessarily inconsistent with the lifetime of the blobs in our analysis.

\subsection{Subsurface convection zone as the possible source of stellar prominences}
\citet[hereafter   CL09]{cantiello:2009} showed the existence of subsurface convection layers in luminous massive stars, caused by the iron opacity peak in the so called iron convection zone (FeCZ).  They found a correlation between the properties of such a convective region and microturbulence at the surface.  Preliminary results of 3D simulations were presented by \cite{cantiello:2011a} who showed that these convection zones could indeed produce observed microturbulent velocity fields.  They also argued that magnetic fields at the equipartition level are generated in the FeCZ, which could reach the surface thanks to buoyancy. The field strength could be enhanced up to kG fields by dynamo action in the presence of shear and rotation, which may result in localized magnetic spots.  They argue that such spots may be linked to several observable phenomena such as the occurrence of DACs (Section\,\ref{section:wind}), line profile variability (as studied in the current paper), non-thermal X-ray emission (as found for $\lambda$\,Cep by RH15, variable on the rotational timescale and  formed within 2$R_*$), excitation of non radial pulsations, and wind clumping since they are all related to surface inhomogeneities. \cite{blomme:2011} suggested that the red noise detected in CoRoT light curves of early-type stars is also a consequence of this convection zone, perhaps in interaction with pulsation. According to CL09, the lifetime
of these spots depends on the turnover time of the convection in these subsurface layers.
A first estimate by \cite{cantiello:2011b} for this turnover time, which could  in principle be derived from the equipartition field strength,
is indeed relatively short (hours-days), but a reliable quantitative estimate is difficult to derive, and has to await more detailed modeling. In addition there is the much longer timescale determined by the convective properties of the FeCZ and the rate of mass removal at the surface, which could be related to the monthly/yearly long-term changes we observe.
According to \cite{cantiello:2011b}, an upper limit to the lifetime of magnetic spots is given by the time taken by the mass loss to remove the magnetic material, provided no magnetic annihilation occurs in the convective layer. For a 60 M$_{\odot}$ star ($\lambda$ Cep is $\sim$50 M$_{\odot}$) this limit is about 3 months, but 4 years in the case of dissipative processes like magnetic reconnection.
We tentatively identify the occurrence of the short-lived magnetic spots with the footpoints of the stellar prominences modeled above. Better estimates of the lifetimes of these spots will tell whether these lifetimes are indeed in the range of what we derived, but of course the lifetime of prominences could well be shorter (as in the Sun).

\subsection{X-ray emission and stellar prominences}

Possible support for our picture of stellar prominences in massive, luminous stars can be derived from RH15, who found that most of the X-rays from $\lambda$ Cep are emitted from a region starting at 0.1\,$R_*$ to  less than 2\,$R_*$ above the photosphere. From our perspective this would be the average extent of the stellar prominences,  which we have independently chosen to be 0.4\,$R_*$ with geometric center at 0.2\,$R_*$ from the surface, as argued in Sect.\,\ref{sub:fits}. \cite{rauw:2015} find the X-ray spectra of $\lambda$ Cep consistent with emission from wind-embedded shocks, which could be generated by the action of localized magnetic fields by magnetically heating the plasma, giving rise to the  prominences  proposed by RH15.

\subsection{Relation to clumping}

Stellar winds of early-type stars are highly clumped (see, e.g.,\ the extensive review by \citealt{puls:2008} and references therein). The exact cause of this clumping is not clear, but all researchers agree that the onset must happen very close to the stellar surface (see, e.g.,  \citealt{puls:2006}, \citealt{fullerton:2006}, \citealt{najarro:2011}, and \citealt {sundqvist:2011}). In the last paper the authors suggest that strong clumping might even trigger NRP. Interestingly, the multi-diagnostic study of $\lambda$ Cep by \cite{sundqvist:2011} shows that the clumping factor close to the star turns out to be $\sim$28 (their Table 3) with an average interclump distance  \citep[or porosity length;][]{owocki:2004} of $\sim$$0.5~R_*$, which is roughly similar to the configuration of blobs that we figured in our analysis above. The question arises whether this reflects the same phenomenon, in other words, Does the azimuthal inhomogeneity associated with our proposed stellar prominences give the same observational effect as the clumps in these papers? and/or, In the light of the discussion in the previous subsections, is the surface density of magnetic spots due to dynamo effects in the subsurface FeCZ similar to the projected surface density of clumps? A positive answer to these intriguing
questions, which we cannot provide,  may bring together a number of surface related phenomena on a quantitative basis by further research.

\subsection{Future work and other stars}
Preliminary results obtained with the HERMES spectrograph at the Mercator telescope of other O-type stars like $\xi$ Per, 19 Cep, $\alpha$ Cam, $\lambda$ Ori, and $\zeta$ Ori
showed behavior very similar to what we found in $\lambda$ Cep. It is clear that the approach to solving the problems addressed in this paper is currently data limited.
Space-based photometry of $\xi$ Per by \cite{ramiaramanantsoa:2014},
showed a frequency range in the lightcurve similar to what we find for the frequencies of prominences. In their interpretation, bright spots that generate frequencies below 2 d$^{-1}$ last only several days.
  The question here is whether these frequencies can be identified with the occurrence of spots and/or prominences (which may have a shorter lifetime than spots), the presence of which will likely  have a small but measurable effect on the light curve.
A ground-based spectroscopic campaign with 24 h coverage lasting several days taking place simultaneously with space-based photometry is likely to provide many of the answers to the questions raised in this study.

Progress from an analysis point of view can be expected by considering the stratification of the formation regions of the different spectral lines, such as was done by \cite{martins:2015} for somewhat cooler stars than $\lambda$ Cep, which have a strong diagnostic value.


\section{Conclusions}
In our analysis of four datasets, which were taken years apart, containing time-resolved, high-resolution optical spectroscopy of the O6.5I(n)fp star $\lambda$ Cep, we have identified many periodicities in the range of 0.4 to 4.1 d in hydrogen and helium lines, which vary not only from dataset to dataset, but also among different lines in each dataset. We showed that relative changes (expressed in the form of subsequent quotient spectra) in various optical absorption and emission lines are often very similar. The timescale of the cyclic behavior in the wind lines of this star is in the same range as the (quasi-)periodicities in the optical lines. This confirms similar conclusions from other previous work, and points strongly to a phenomenon that is driven by both short (hours-days) and longer (months-years) timescales. We have developed a simplified phenomenological model to explain cyclical variability where multiple spherical blobs attached to the surface represent magnetic-loop structures, which we call stellar prominences by analogy with solar prominences. We applied our model to subsequent quotient spectra and fitted the free physical parameters (optical depth and velocity dispersion) and plotted them as a function of phase of the adopted rotation period of 4.1\,d.
Our proposed model applied to the \ion{He}{ii} $\lambda$4686 line can be fitted with 2\,to\,5 equatorial blobs with lifetimes between $\sim$1 and 24 h. These numbers are representative in spite of the many simplifying assumptions.

Given the irregular timescales involved, our results support a model in which the azimuthal distribution of DACs corresponds to the locations of stellar prominences (or magnetic-loop structures) attached to the surface. This configuration could explain the observed variability in optical and UV lines. Application of this model to other stars can put constraints on the strength and lifetime of these magnetic structures, to be compared with theoretical predictions. The action of the subsurface convection zone \citep{cantiello:2009} would be the most likely driving mechanism that generates short-lived magnetic spots (i.e., bright spots)  as the source of prominences, in particular because of the two different timescales involved. Simultaneous space-based photometry, X-ray, and ground-based spectroscopy is the most promising approach to make progress in this field, which we believe is applicable to more than 90$\%$ of luminous, massive stars.


\begin{acknowledgements}
The authors are very grateful to the many observers who contributed to this project: Douglas Gies at Kitt Peak and Regina Zwarthoed at Calar Alto during the 1989 campaign; Kang Dong-il at the Bohyunsan Optical Astronomy Observatory, South Korea; Samayra Straal at the Mercator telescope in 2012; and Manos Zapartos, Marieke van Doesburgh, and Mieke Paalvast for observing and for a preliminary analysis of the 2013 dataset, also acquired at the Mercator telescope. The generous time allocation and support for the latter two projects from the IvS, Leuven University, in particular Hans van Winckel, is highly appreciated and acknowledged. The authors would like to thank the participants of the $6^{\rm th}$ MiMeS workshop in Saclay in May 2012 and in particular the attendees of the Conference on Magnetism and Variability in O stars, September 2014, in Amsterdam at the retirement of HFH for enlightening discussions, in particular Lex Kaper and Alex de Koter.
NS fondly acknowledges the warm hospitality and financial support of the Anton Pannekoek Institute for Astronomy in Amsterdam where most of this work was done. She is also grateful to NOVA, the Leids Kerkhoven Bosscha Fonds, and Lex Kaper, who provided additional financial support. She acknowledges Saint-Petersburg State University for research grants 6.38.73.2011, 6.50.1555.2013, and 6.38.18.2014. This research has made use of the SIMBAD database, operated at CDS, Strasbourg, France

\end{acknowledgements}

\bibliographystyle{aa}
\bibliography{references}

\begin{appendix}
\section{Variability in spectral lines}

\begin{figure}[!h]
\includegraphics[width=0.93\columnwidth]{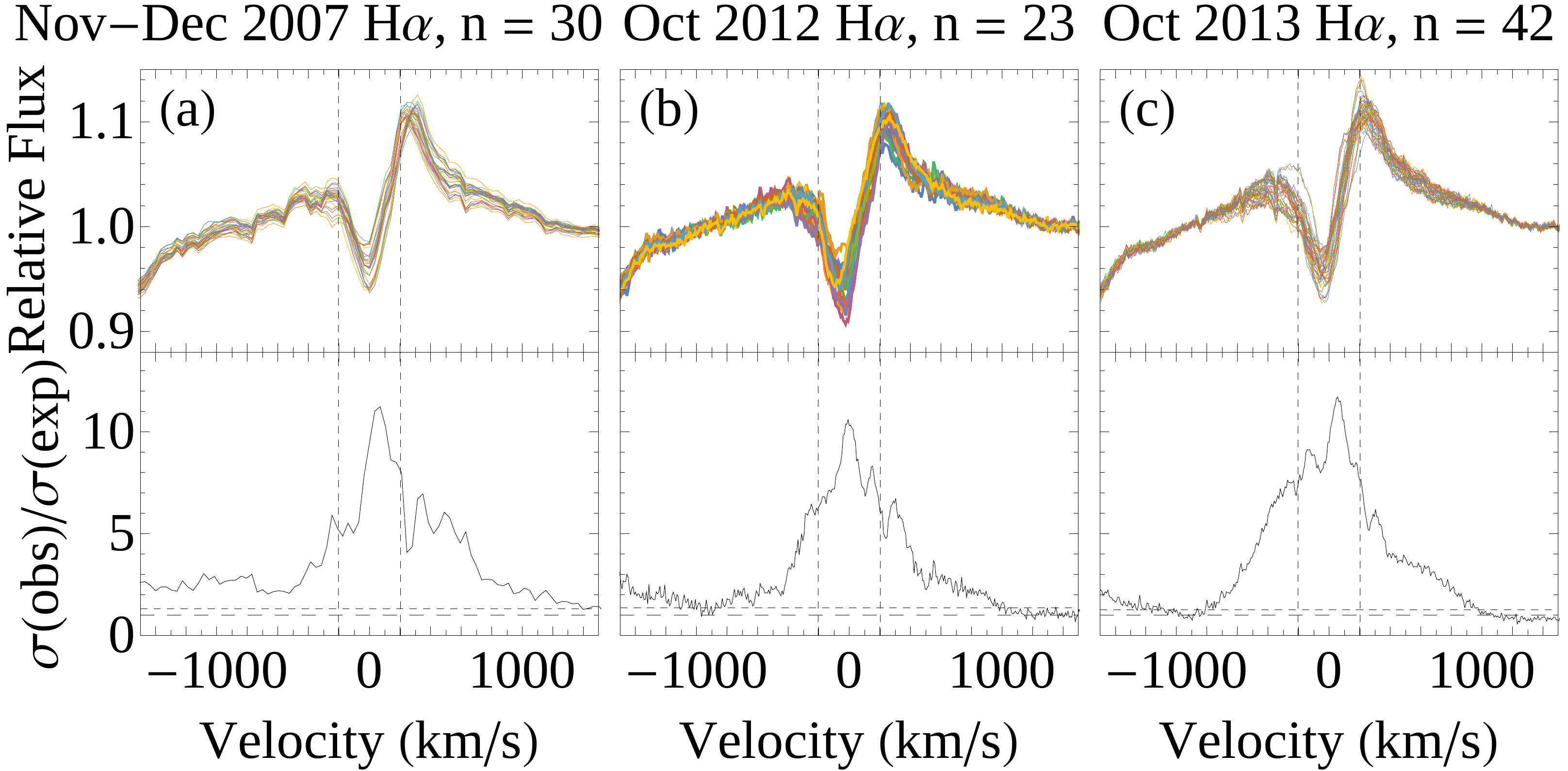}
\caption{{\it Top panels:} Overplotted H$\alpha$ $\lambda$6563 spectra with $n$ the number of spectra, which is the same for each panel in Figs.\,\ref{fig:TVShb}--\ref{fig:TVS4200}.  Vertical dashed lines are drawn at $\pm v$sin$i$. {\it Lower panels:} Temporal variance spectra expressed as the ratio of the observed to the expected standard deviation, $\sigma_{\rm obs}/\sigma_{\rm exp}$. The lower horizontal long-dashed line is drawn at unity. The upper horizontal short-dashed line is the 1$\%$ significance level. From left to right: {\it (a)} acquired at BOAO in 2007, spanning 27 days. {\it (b)} acquired at the Mercator telescope in 2012, spanning 6 days. {\it (c)} acquired at the Mercator telescope in 2013 spanning 6 days. The spectra in the panels have been rebinned to a velocity grid giving the same S/N of 560 per bin. }
\label{fig:TVSha}
\end{figure}

\begin{figure}[!h]
\includegraphics[width=0.87\columnwidth]{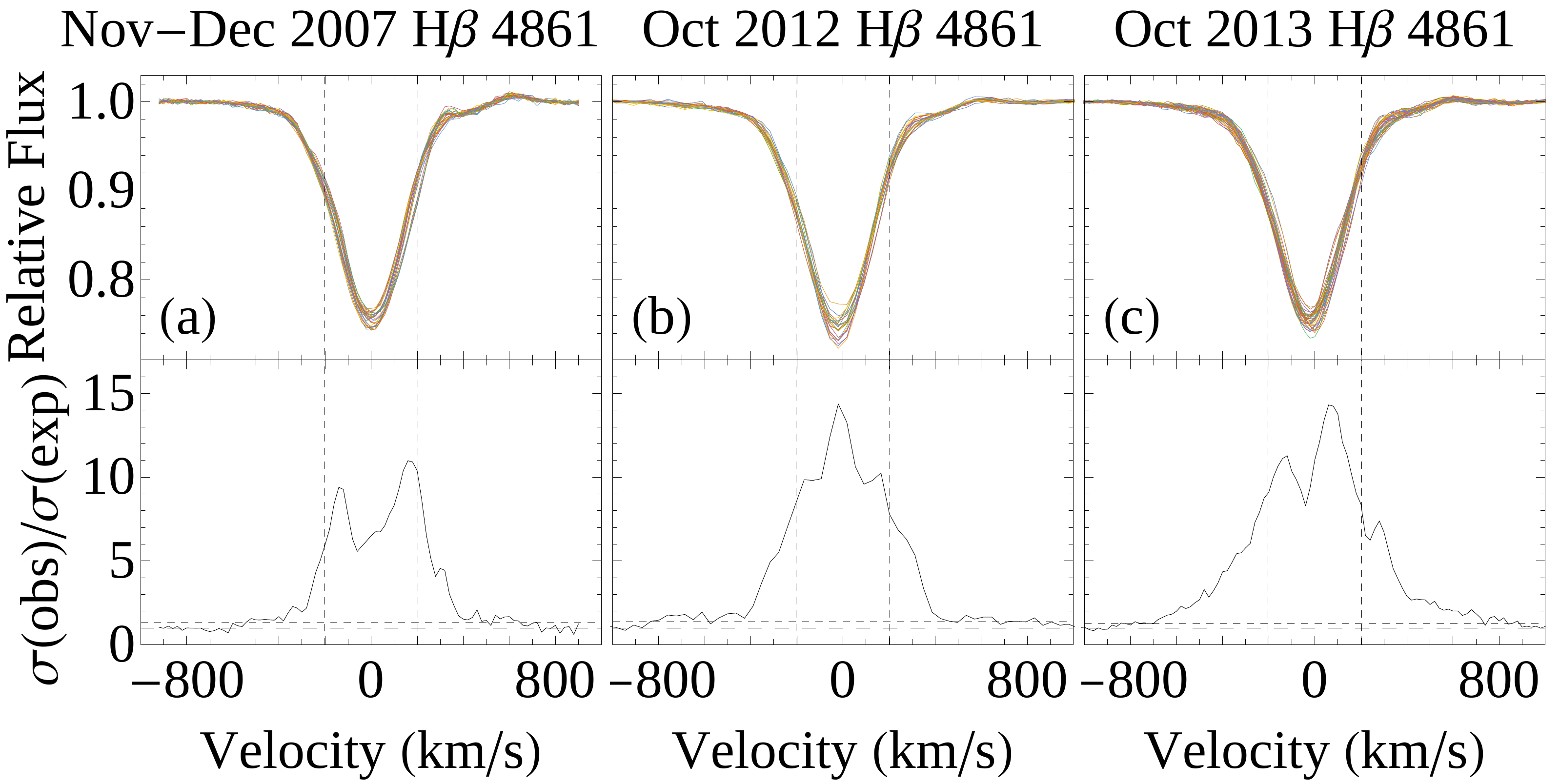}
\caption{Same as Fig.\,\ref{fig:TVSha}, but for H$\beta$ $\lambda$4861 with S/N = 1040.}
\label{fig:TVShb}
\end{figure}

\begin{figure}[!h]
\includegraphics[width=0.87\columnwidth]{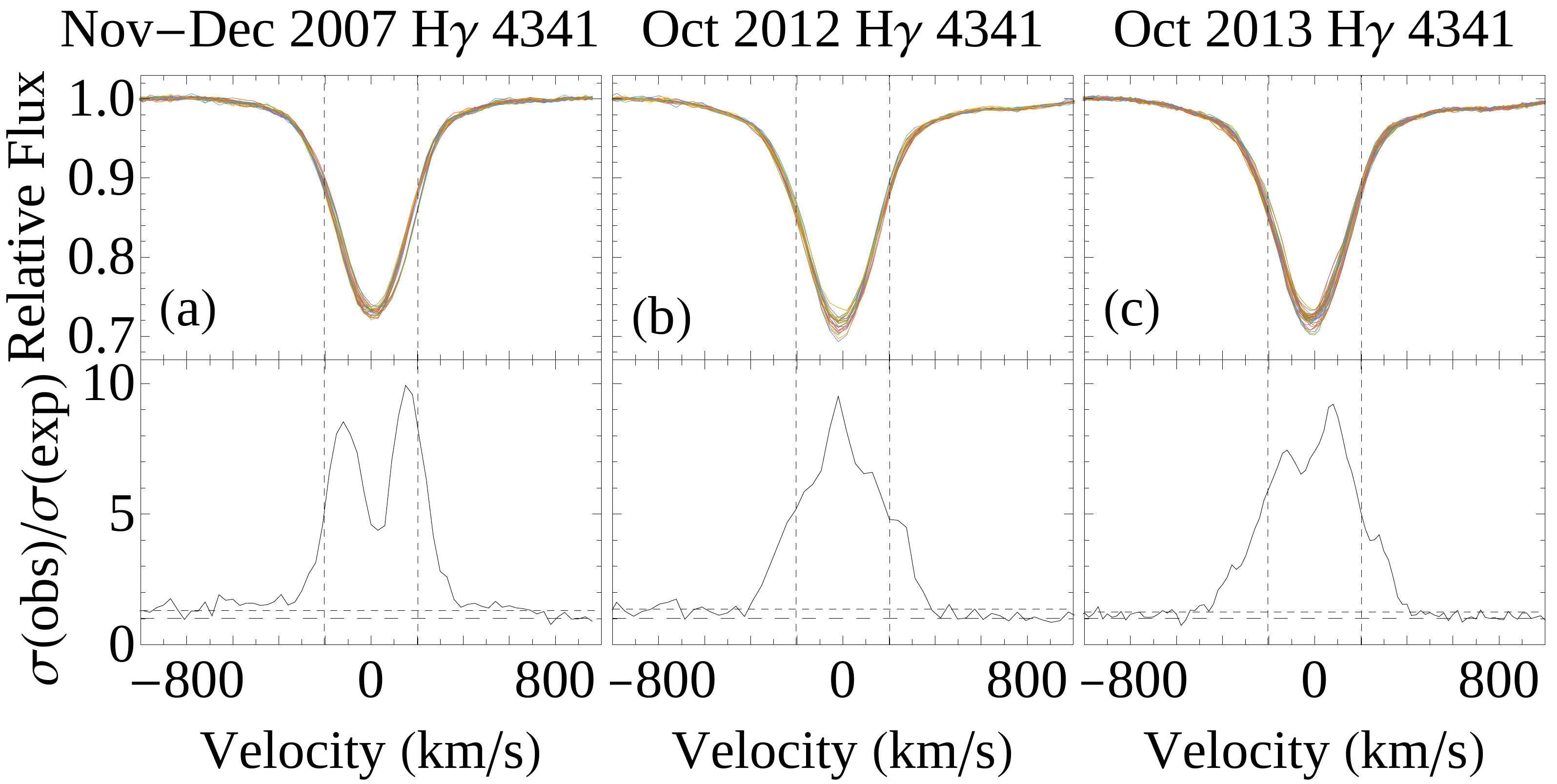}
\caption{Same as Fig.\,\ref{fig:TVSha}, but for H$\gamma$ $\lambda$4341 with S/N = 820.}
\label{fig:TVShg}
\end{figure}

\begin{figure}[!ht]
\includegraphics[width=0.87\columnwidth]{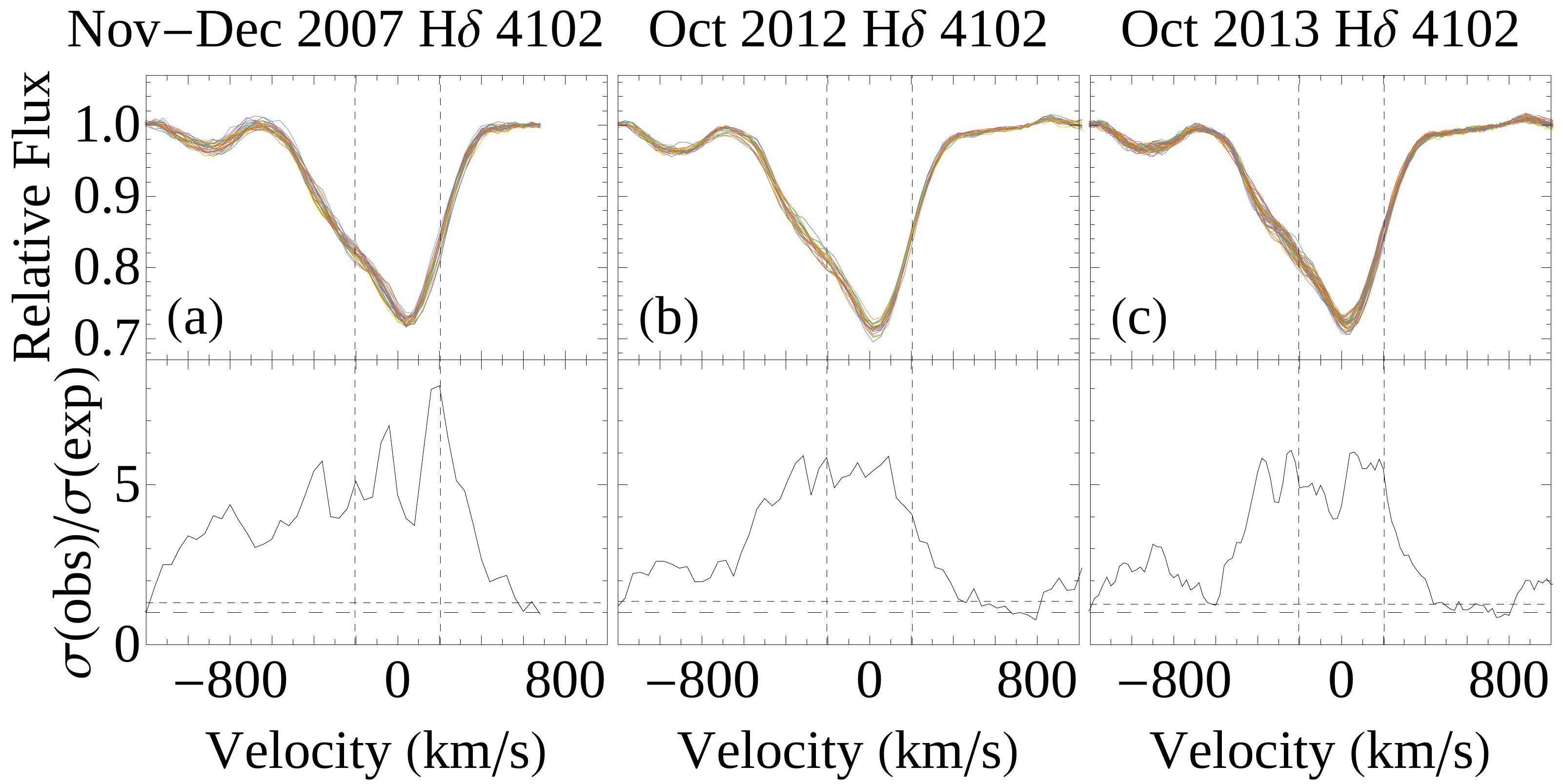}
\caption{Same as Fig.\,\ref{fig:TVSha}, but for H$\delta$ $\lambda$4102 with S/N = 620.}
\label{fig:TVShd}
\end{figure}

\begin{figure}[!ht]
\includegraphics[width=0.87\columnwidth]{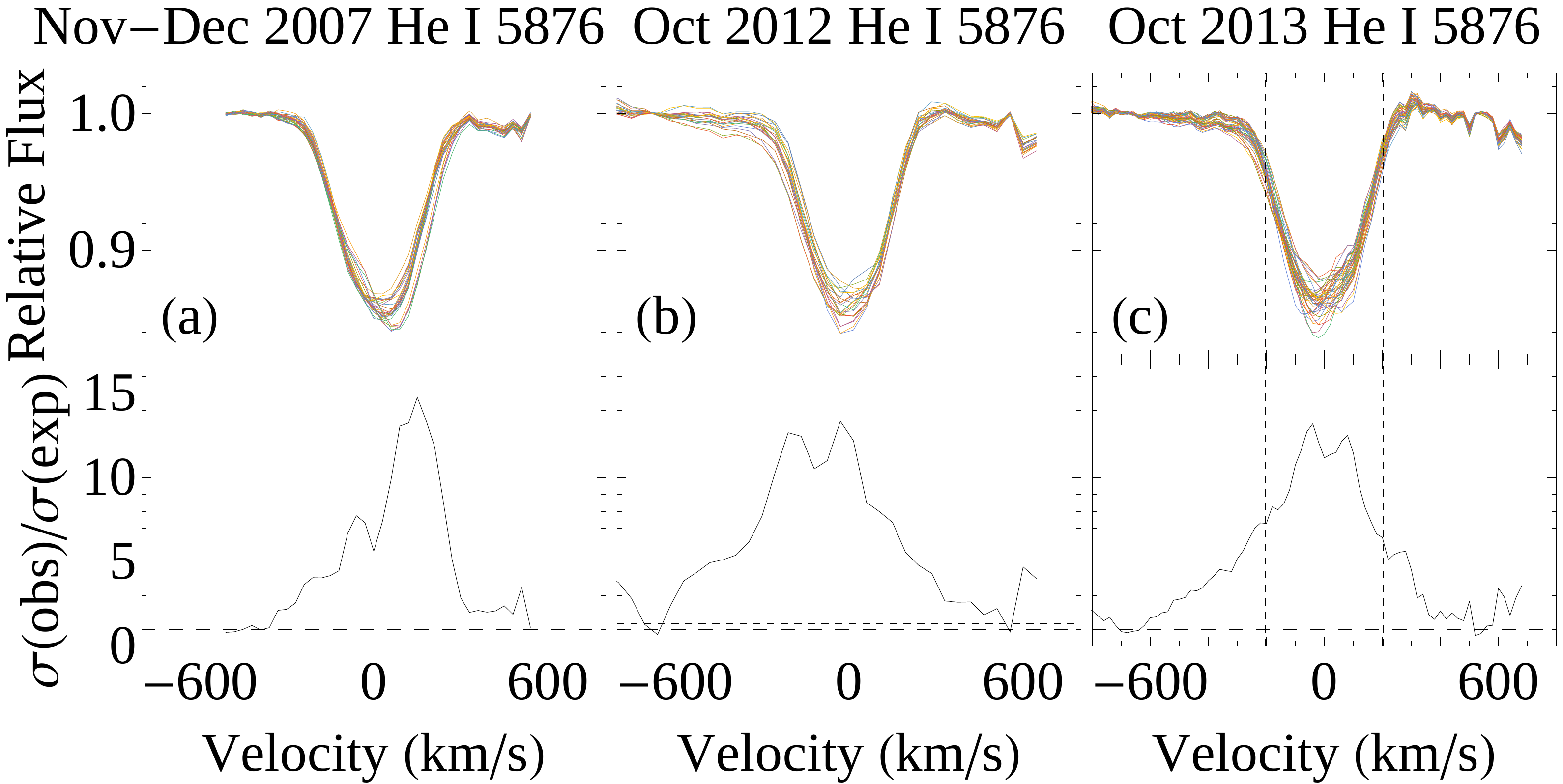}
\caption{Same as Fig.\,\ref{fig:TVSha}, but for \ion{He}{I} $\lambda$5876 with S/N = 1120.}
\label{fig:TVS5876}
\end{figure}

\begin{figure}[!ht]
\includegraphics[width=0.87\columnwidth]{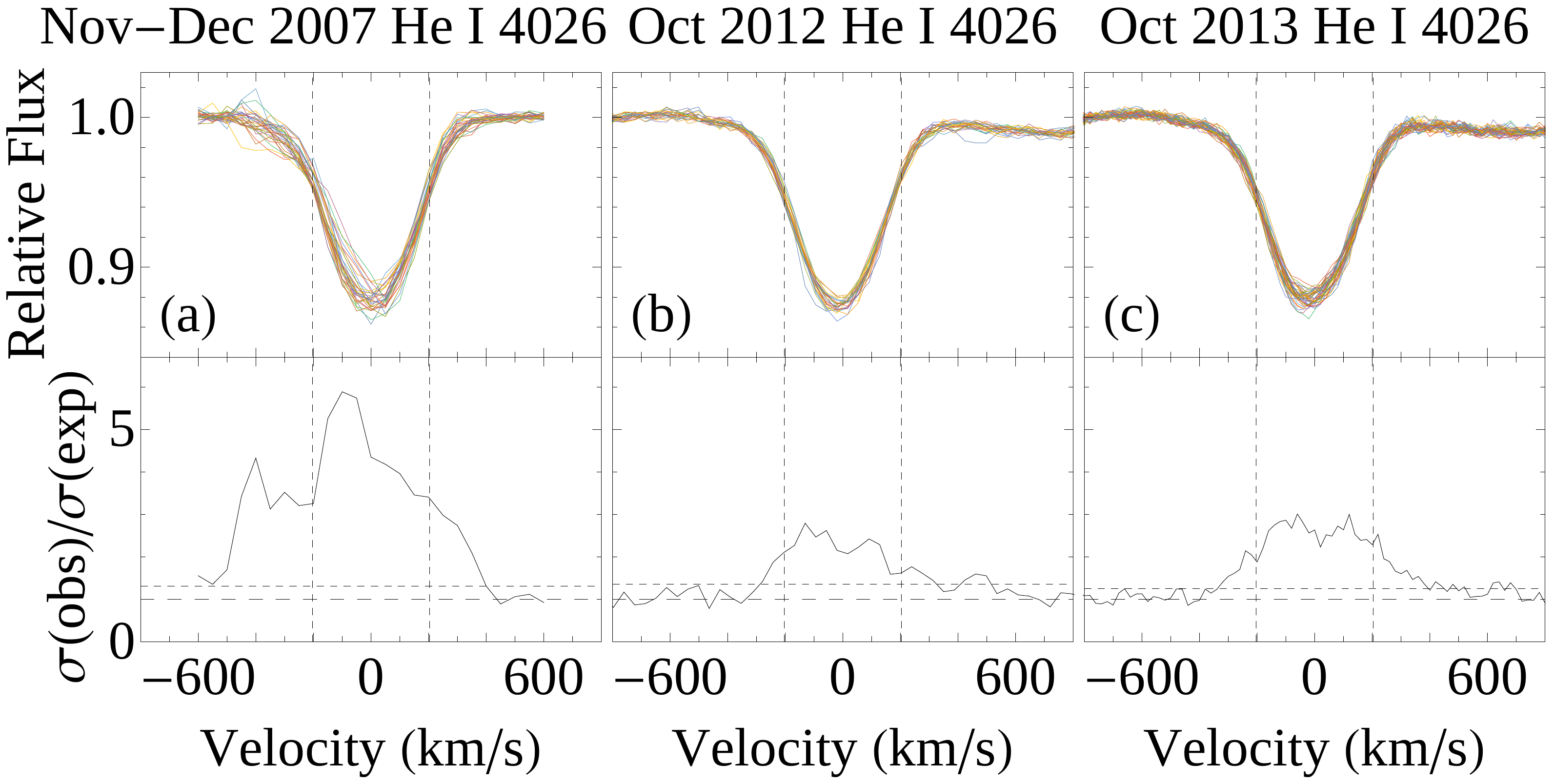}
\caption{Same as Fig.\,\ref{fig:TVSha}, but for \ion{He}{I} $\lambda$4026 with S/N = 560.}
\label{fig:TVS4026}

\end{figure}

\begin{figure}[!ht]
\includegraphics[width=0.87\columnwidth]{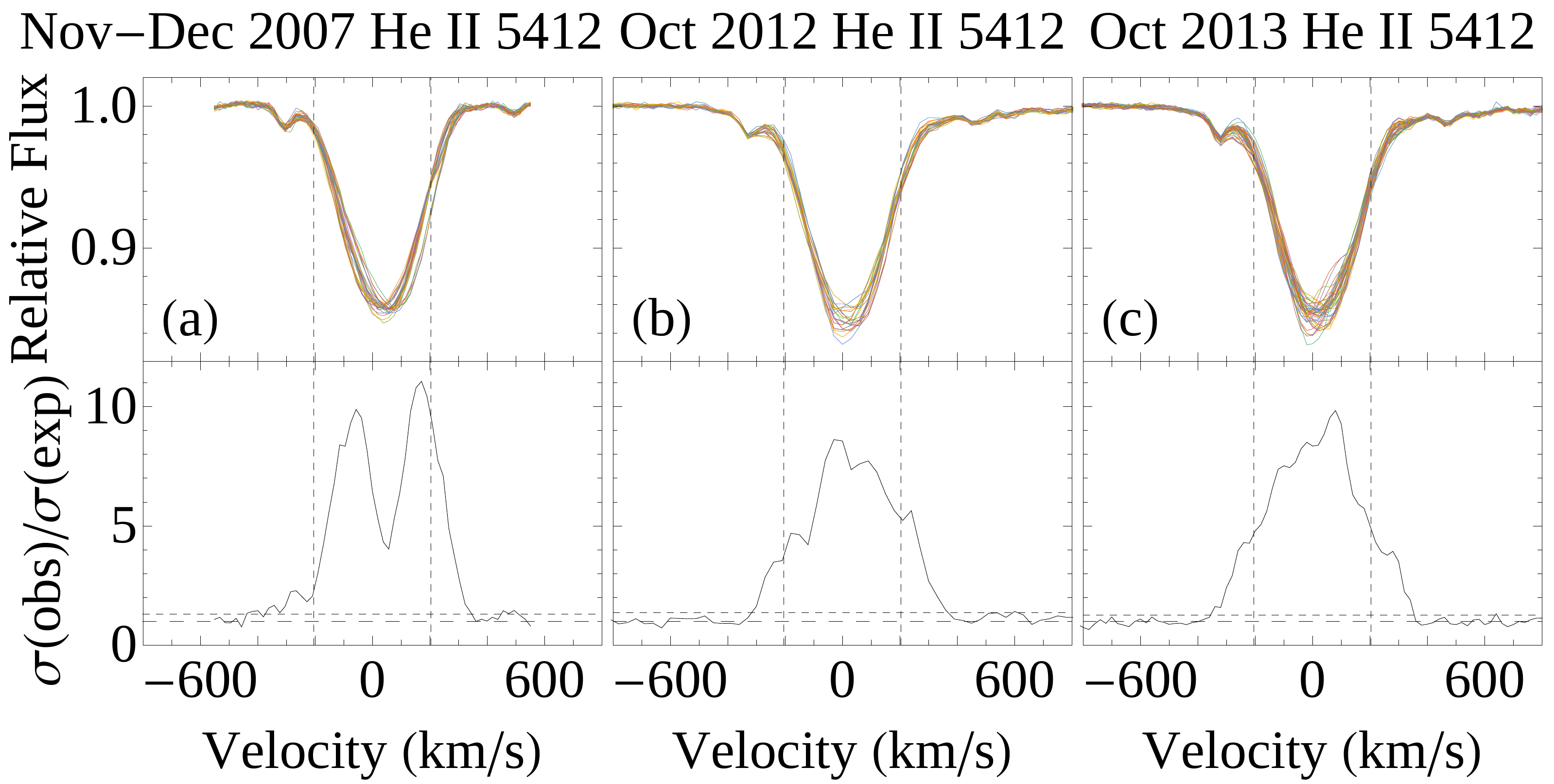}
\caption{Same as Fig.\,\ref{fig:TVSha}, but for \ion{He}{II} $\lambda$5412 with S/N = 1020.}
\label{fig:TVS5412}
\end{figure}

\begin{figure}[!ht]
\includegraphics[width=0.87\columnwidth]{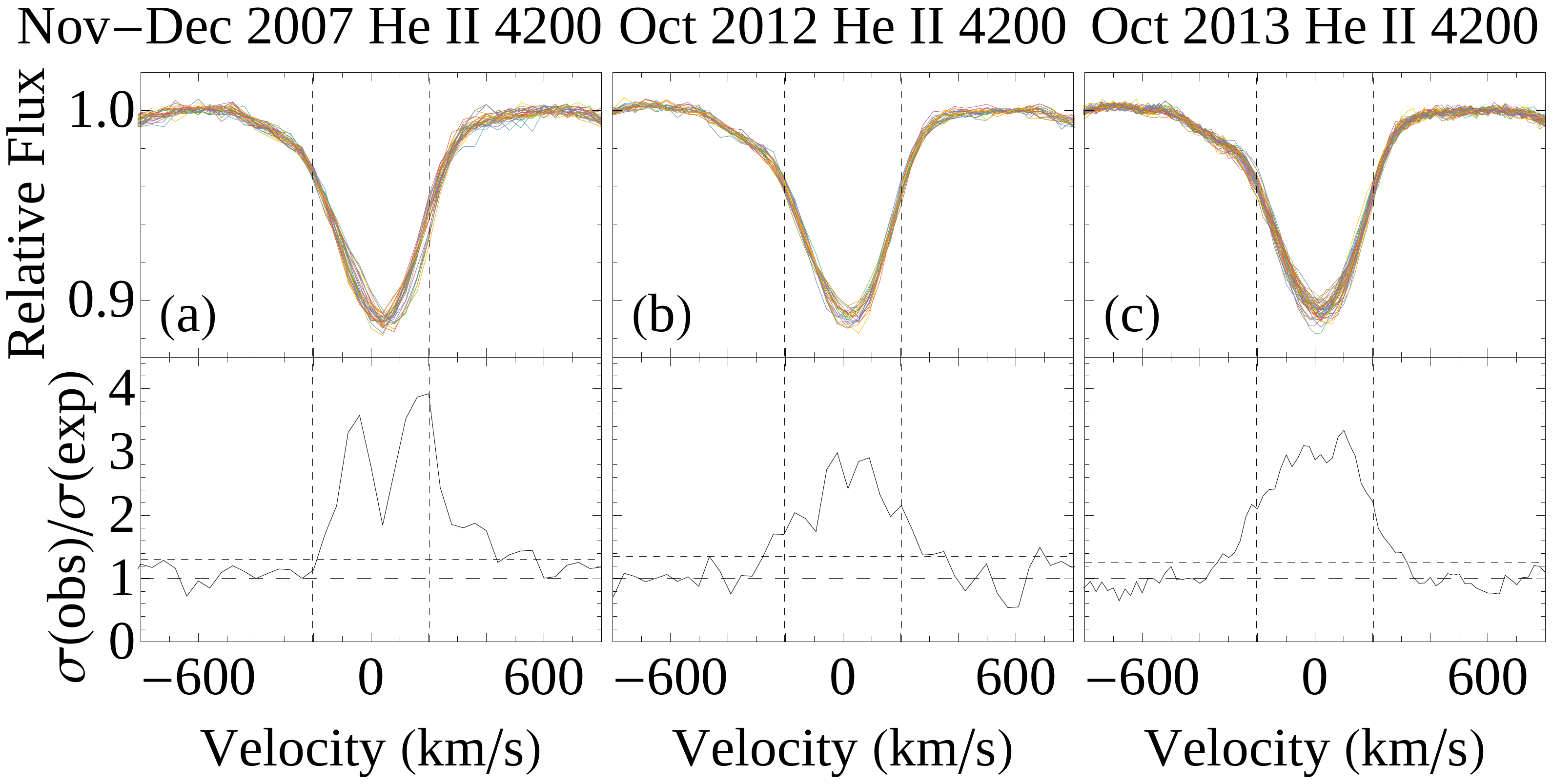}
\caption{Same as Fig.\,\ref{fig:TVSha}, but for \ion{He}{II} $\lambda$4200 with S/N = 610.}
\label{fig:TVS4200}
\end{figure}

\clearpage
\onecolumn

\section{Periodograms}

\begin{figure*}[!ht]
\vspace*{-0cm}
\includegraphics[width=0.65\columnwidth]{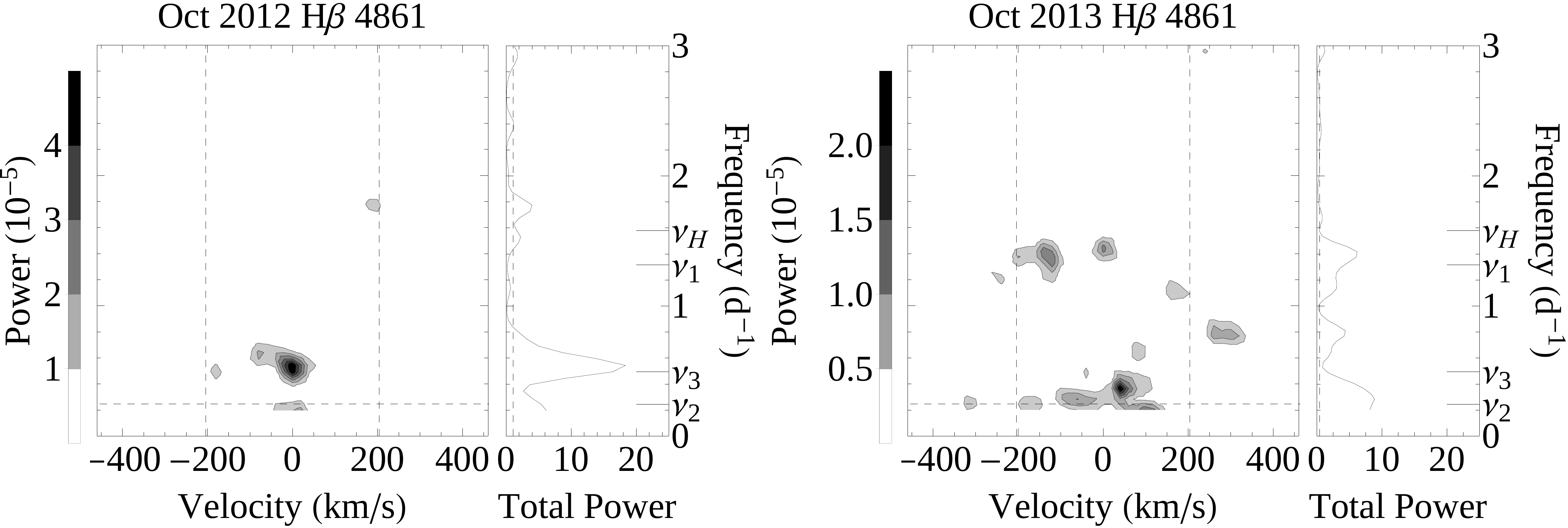}
\caption{Same as Fig.\,\ref{fig:powerHa}, but for H$\beta$ $\lambda$4861 spectra for the   2012 and 2013 datasets. The full profiles are displayed in Fig.\,\ref{fig:TVShb}. }
\label{fig:powerHb}
\end{figure*}

\begin{figure*}[!h]
\includegraphics[width=0.65\columnwidth]{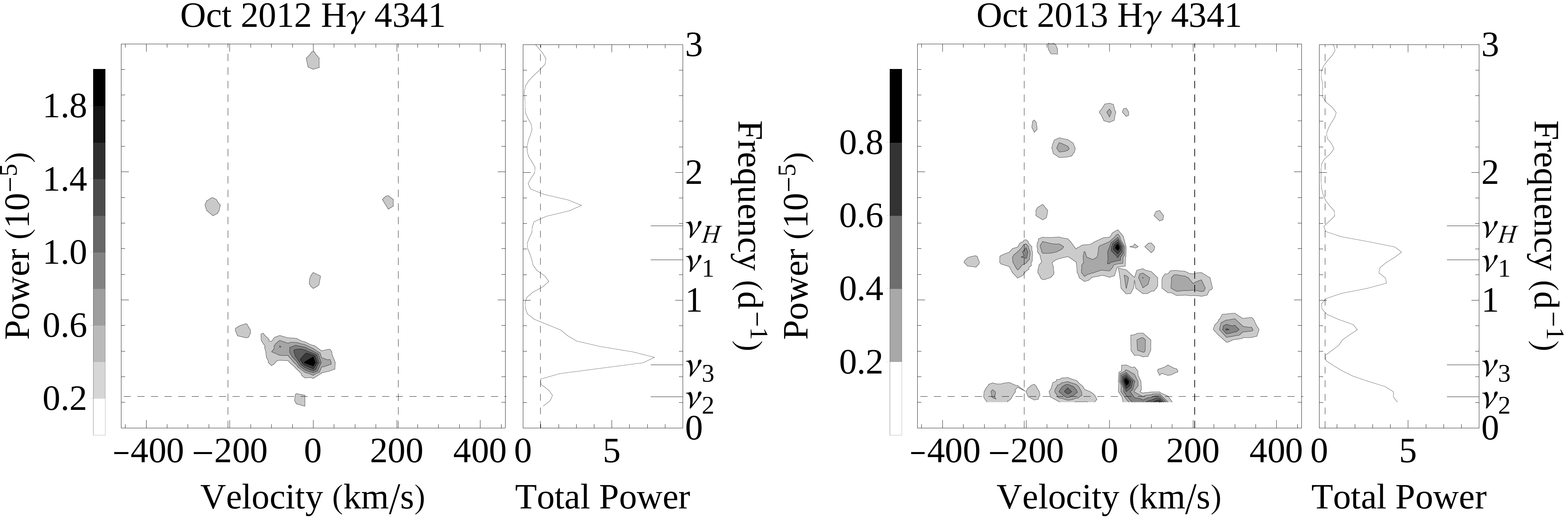}
\caption{Same as Fig.\,\ref{fig:powerHb}, but for H$\gamma$ $\lambda$4341  spectra for the 2012 and 2013 datasets. The full profiles are displayed in Fig.\,\ref{fig:TVShg}. }
\label{fig:powerHg}
\end{figure*}

\begin{figure*}[!h]
\includegraphics[width=0.65\columnwidth]{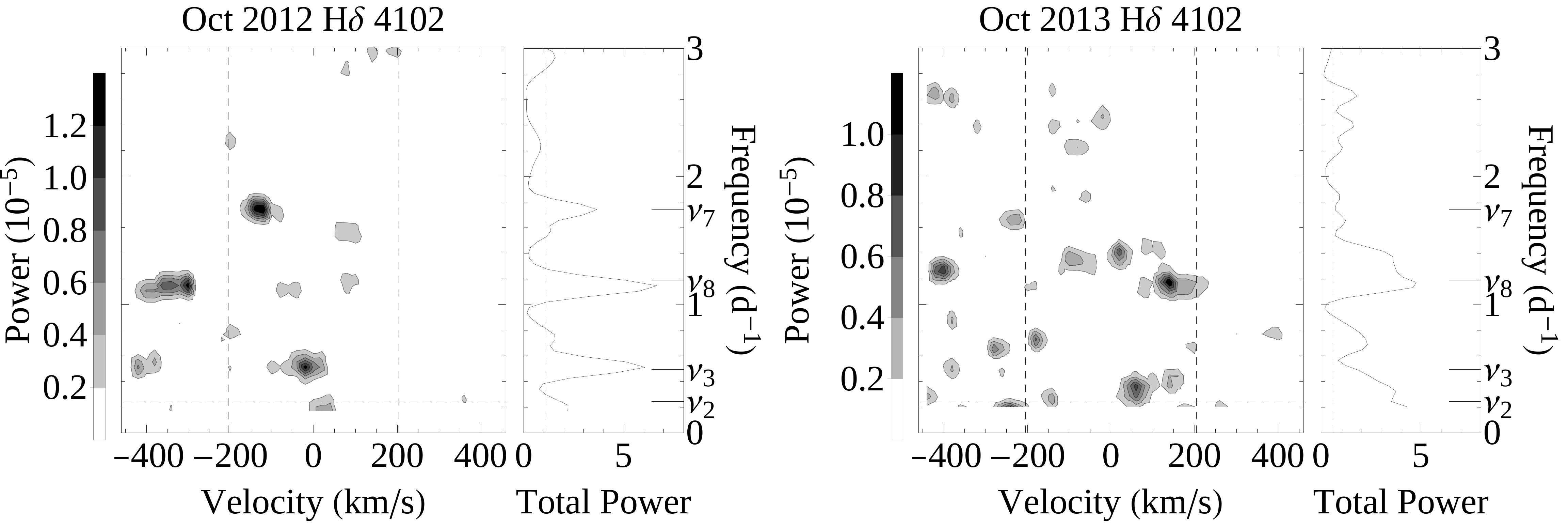}
\caption{Same as Fig.\,\ref{fig:powerHb}, but for H$\delta$ $\lambda$4102  spectra for the 2012 and 2013 datasets. The full profiles are displayed in Fig.\,\ref{fig:TVShd}. }
\label{fig:powerHd}
\end{figure*}

\begin{figure*}[!h]
\includegraphics[width=0.65\columnwidth]{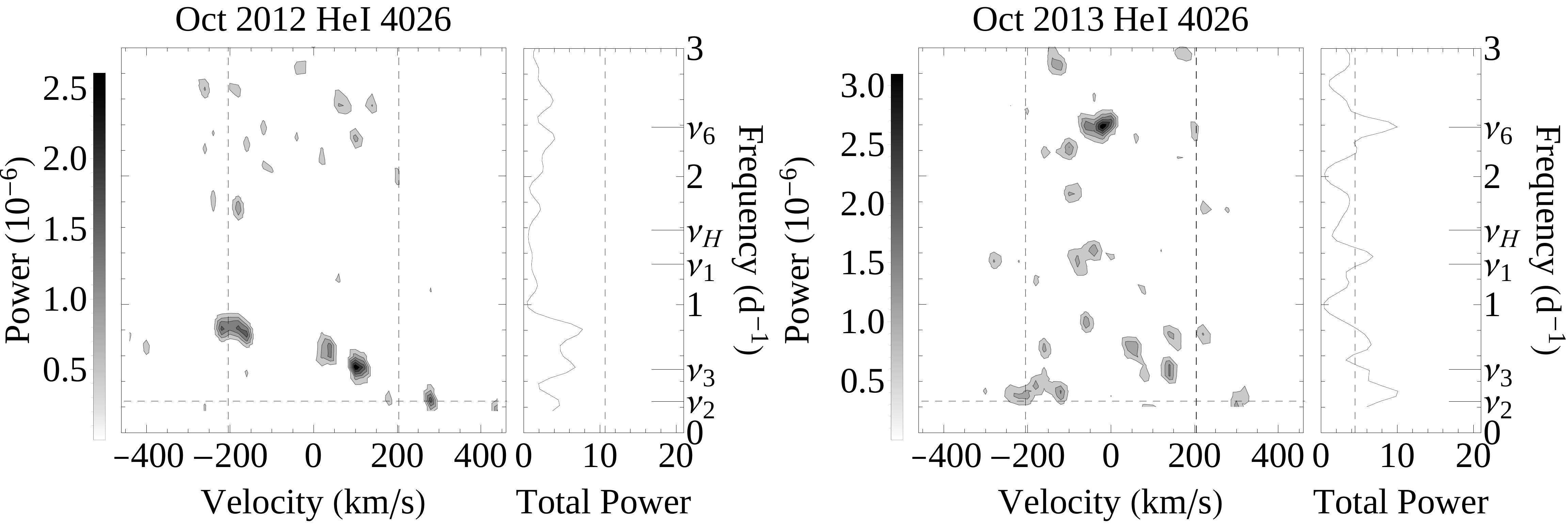}
\caption{Same as Fig.\,\ref{fig:powerHb}, but for \ion{He}{i} $\lambda$4026  spectra for the 2012 and 2013 datasets. The full profiles are displayed in Fig.\,\ref{fig:TVS4026}. }
\label{fig:power4026}
\end{figure*}

\begin{figure*}[!ht]
\includegraphics[width=0.65\columnwidth]{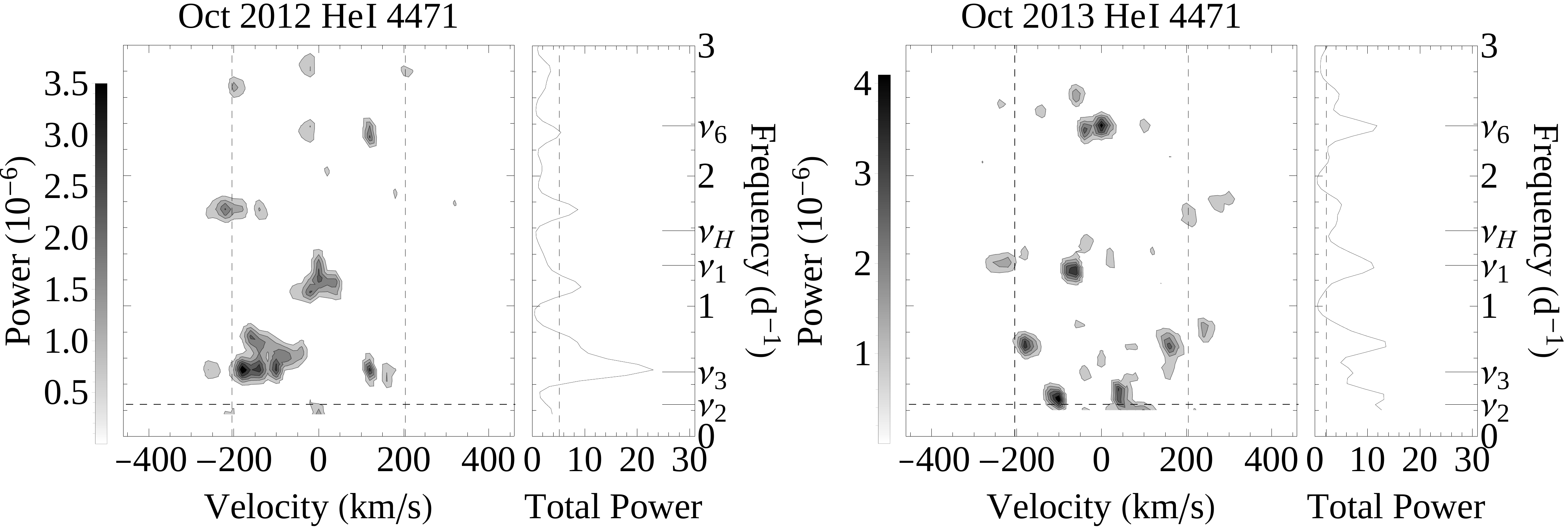}
\caption{Same as Fig.\,\ref{fig:powerHb}, but for \ion{He}{i} $\lambda$4471  spectra for the 2012 and 2013 datasets. The full profiles are displayed in Fig.\,\ref{fig:TVS4471a}. }
\label{fig:power4471}
\end{figure*}

\begin{figure*}[!ht]
\includegraphics[width=0.65\columnwidth]{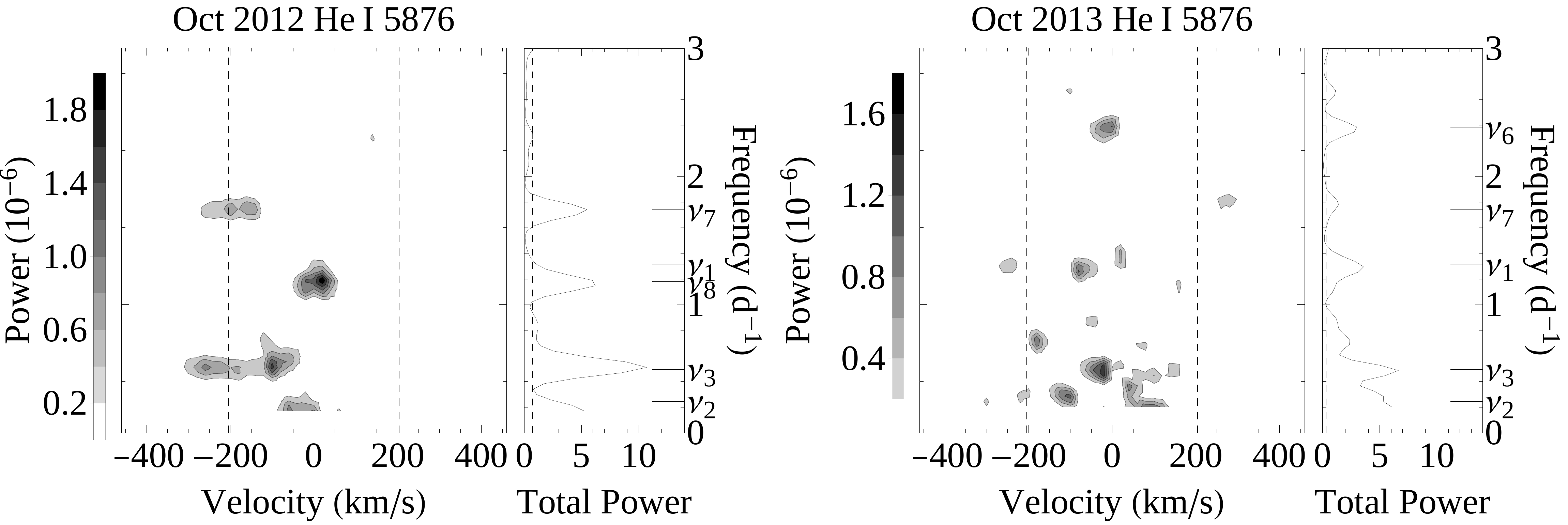}
\caption{Same as Fig.\,\ref{fig:powerHb}, but for \ion{He}{i} $\lambda$5876 spectra for the 2012 and 2013 datasets. The full profiles are displayed in Fig.\,\ref{fig:TVS5876}. }
\label{fig:power5876}
\end{figure*}

\begin{figure*}[!ht]
\includegraphics[width=0.65\columnwidth]{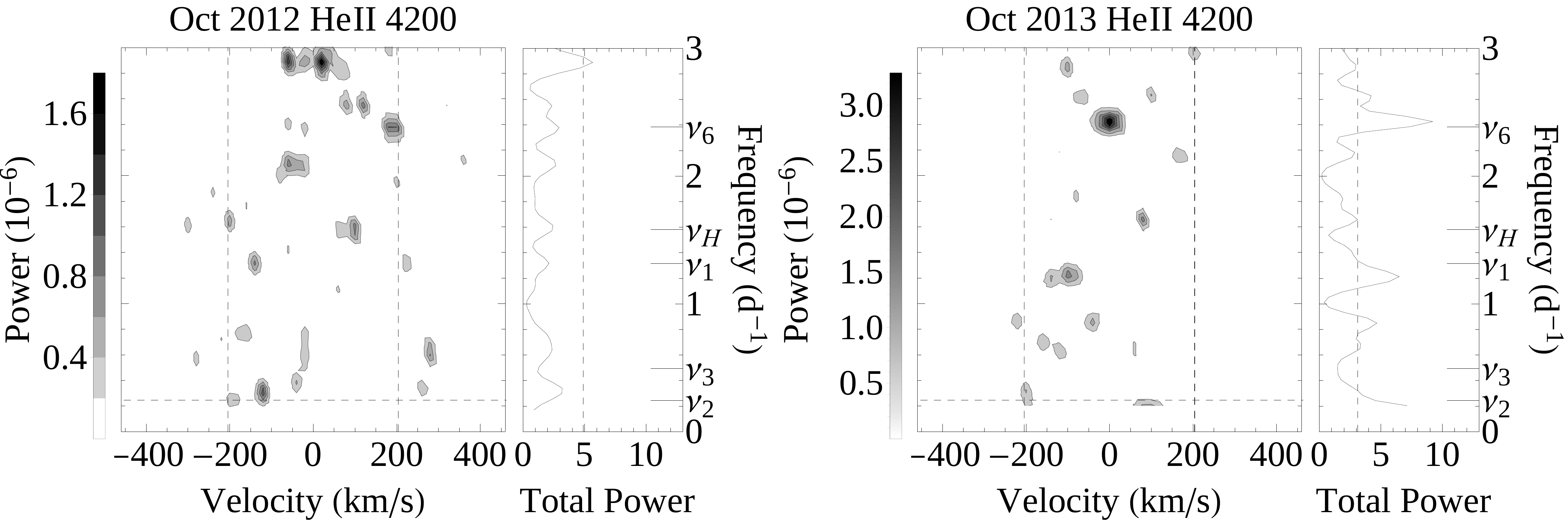}
\caption{Same as Fig.\,\ref{fig:powerHb}, but for \ion{He}{ii} $\lambda$4200 spectra for the 2012 and 2013 datasets. The full profiles are displayed in Fig.\,\ref{fig:TVS4200}. }
\label{fig:power4200}
\end{figure*}

\begin{figure*}[!ht]
\includegraphics[width=0.65\columnwidth]{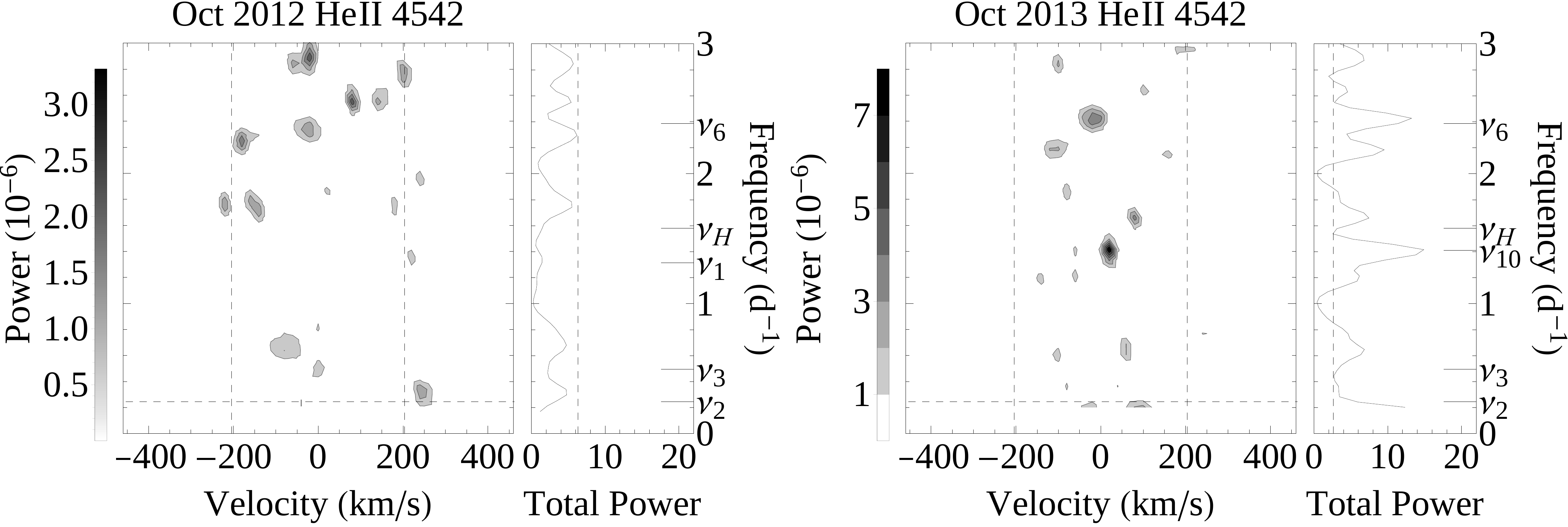}
\caption{Same as Fig.\,\ref{fig:powerHb}, but for \ion{He}{ii} $\lambda$4542 spectra for the 2012 and 2013 datasets. The full profiles are displayed in Fig.\,\ref{fig:TVS4542a}. }
\label{fig:power4542}
\end{figure*}

\begin{figure*}[!ht]
\includegraphics[width=0.65\columnwidth]{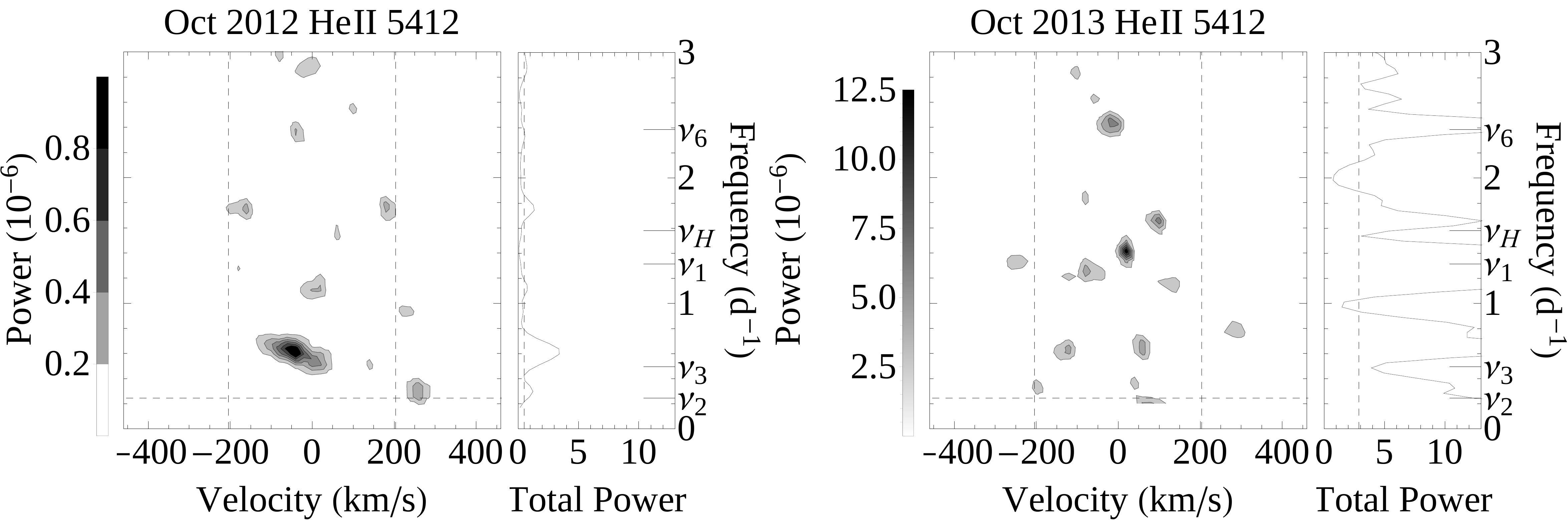}
\caption{Same as Fig.\,\ref{fig:powerHb}, but for \ion{He}{ii} $\lambda$5412 spectra for the 2012 and 2013 datasets. The full profiles are displayed in Fig.\,\ref{fig:TVS5412}. }
\label{fig:power5412}
\end{figure*}


\onecolumn
\section{Model fits of subsequent quotient spectra}
\subsection{The 1989 dataset}
\label{A1989}

\begin{figure*}[h!]
\includegraphics[width=0.74\columnwidth]{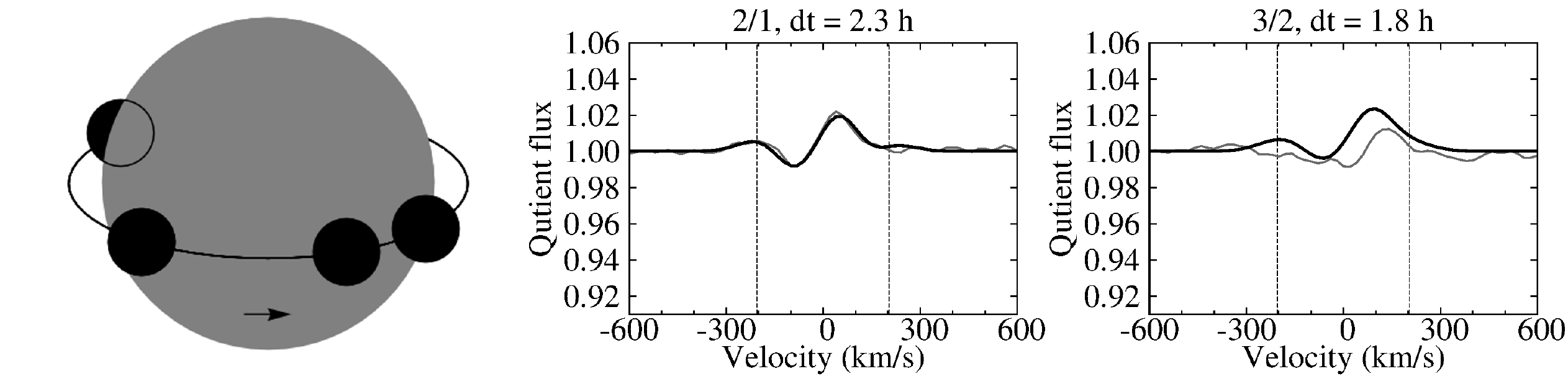}

\includegraphics[width=\columnwidth]{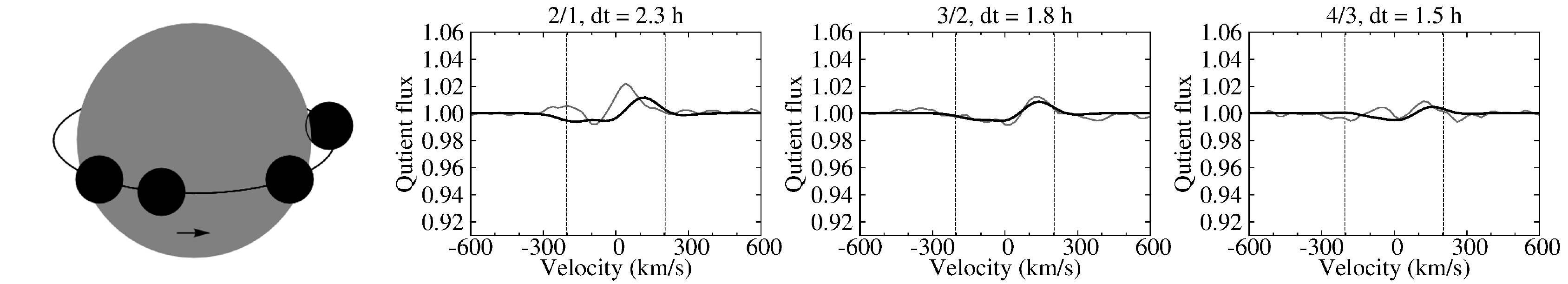}

\includegraphics[width=\columnwidth]{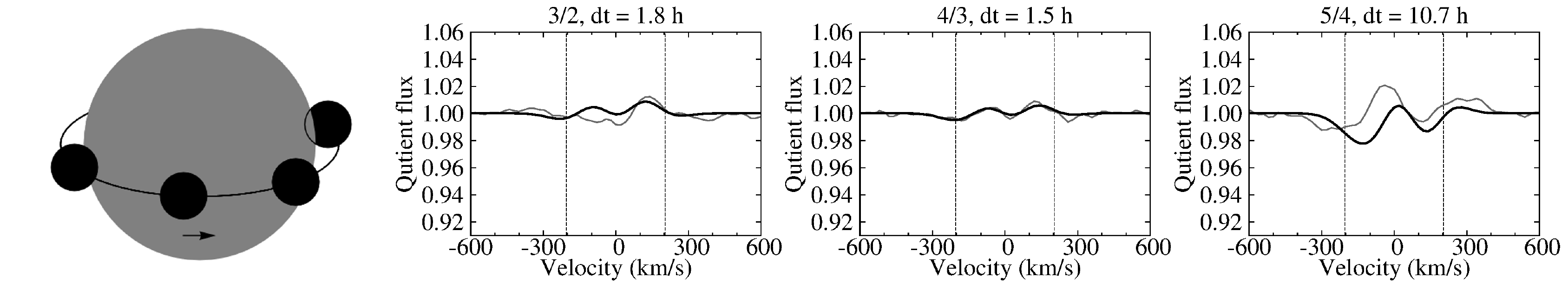}

\includegraphics[width=\columnwidth]{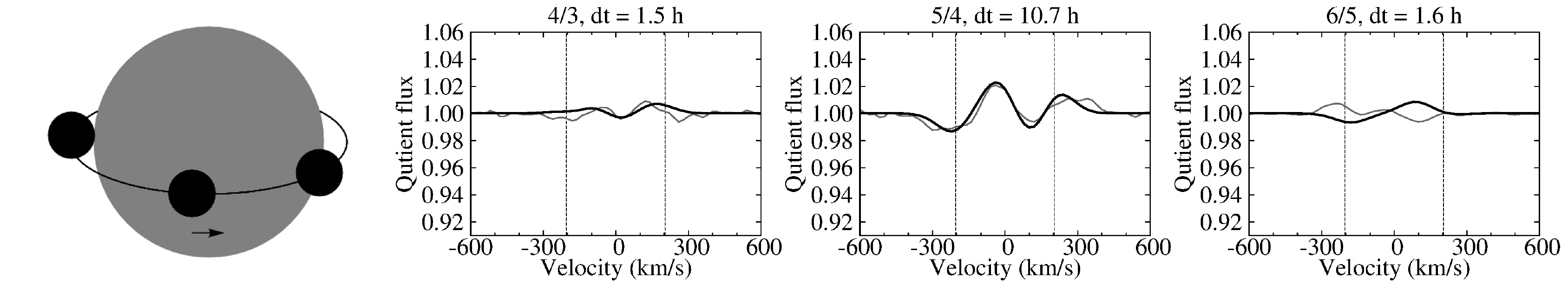}

\includegraphics[width=\columnwidth]{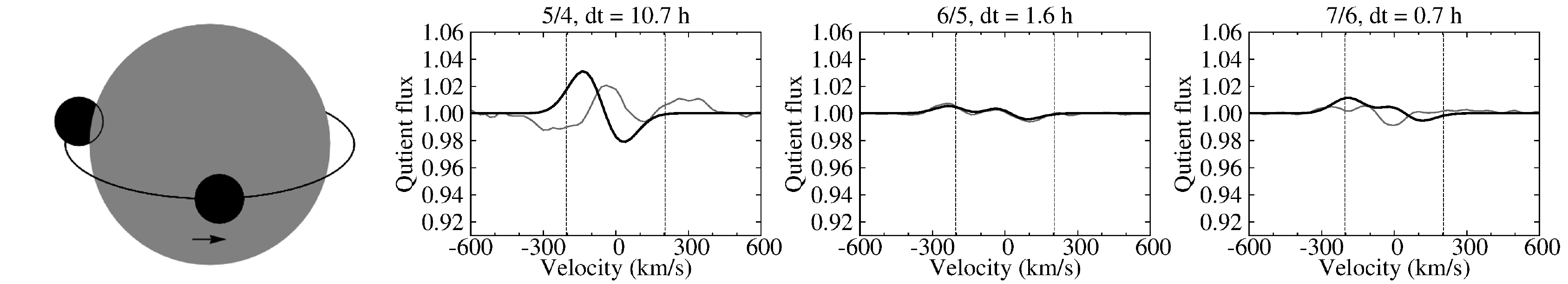}
\caption{Model fits (black thick lines) for subsequent quotient spectra (gray thin lines) of the 1989 dataset, spectra 1--7. The top label gives the sequence numbers of the two spectra of the quotient, followed by their time interval in hours. The geometry, which is depicted at the beginning of each series, is used for all figures in the sequence and is shown for the epoch of the first spectrum of the second panel in the series. The star rotates with the blobs staying in their same relative position.  The first and last figures of each series (except the very first and very last for obvious reasons) contain intentionally failed fits to signify the extent over which the fitted configuration, carried around by the rotation, survives. In this dataset most configurations have short lifetimes, unlike the other datasets. The fits continue in the next figure.}
\label{fig:qfit1989a}
\end{figure*}

\begin{figure*}[ht!]
\includegraphics[width=\columnwidth]{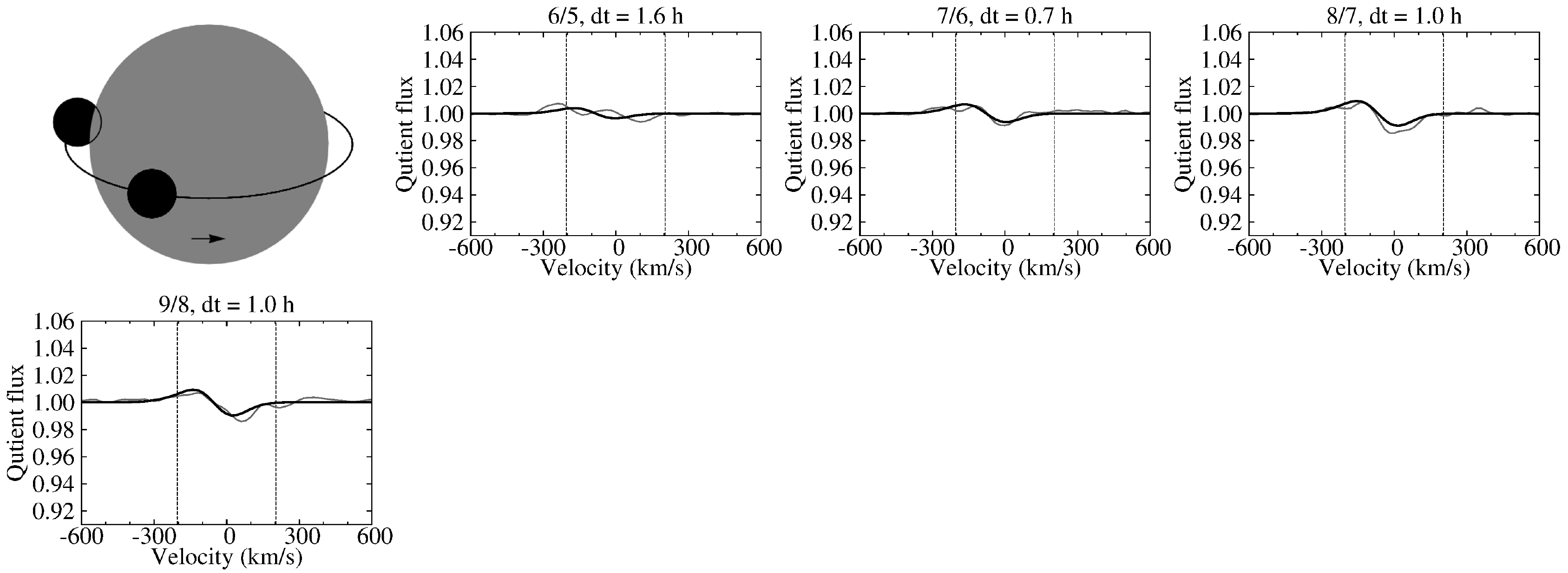}

\includegraphics[width=\columnwidth]{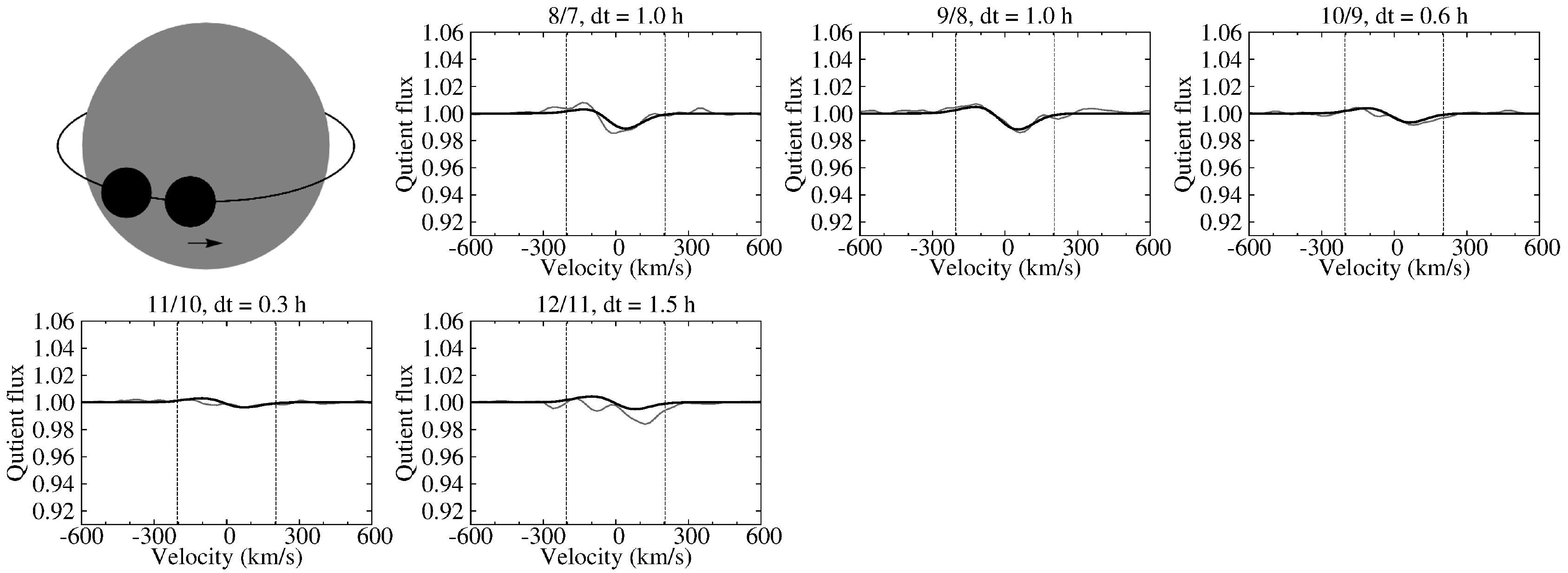}
\includegraphics[width=\columnwidth]{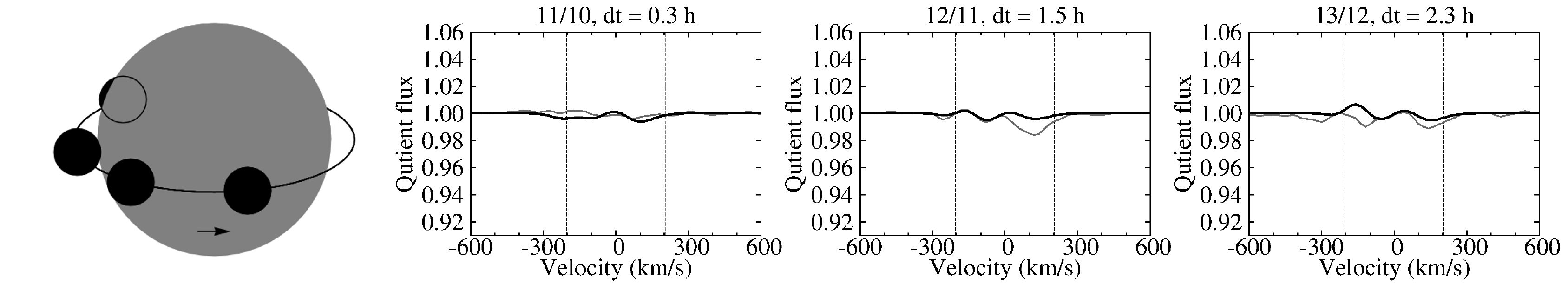}

\includegraphics[width=\columnwidth]{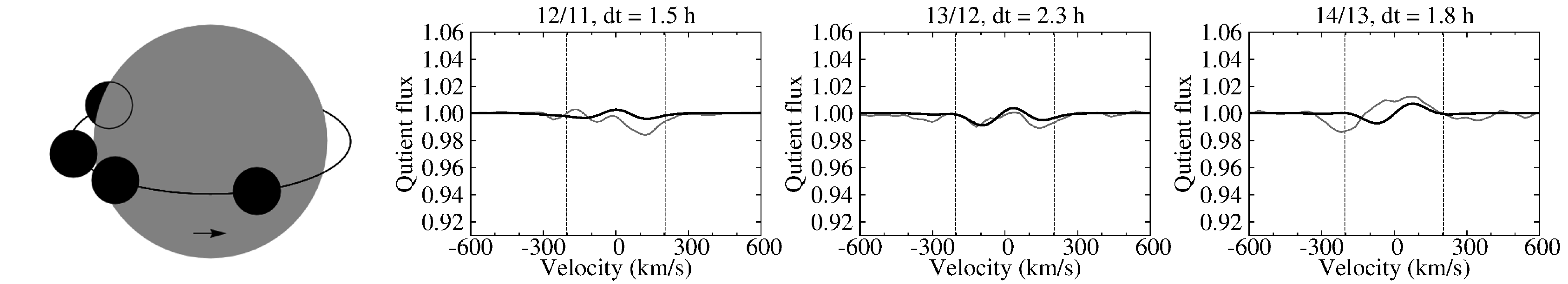}
\caption{Same as Fig.\,\ref{fig:qfit1989a}, but for spectra 5--14.}
\label{fig:qfit1989b}
\end{figure*}

\begin{figure*}[ht!]
\includegraphics[width=\columnwidth]{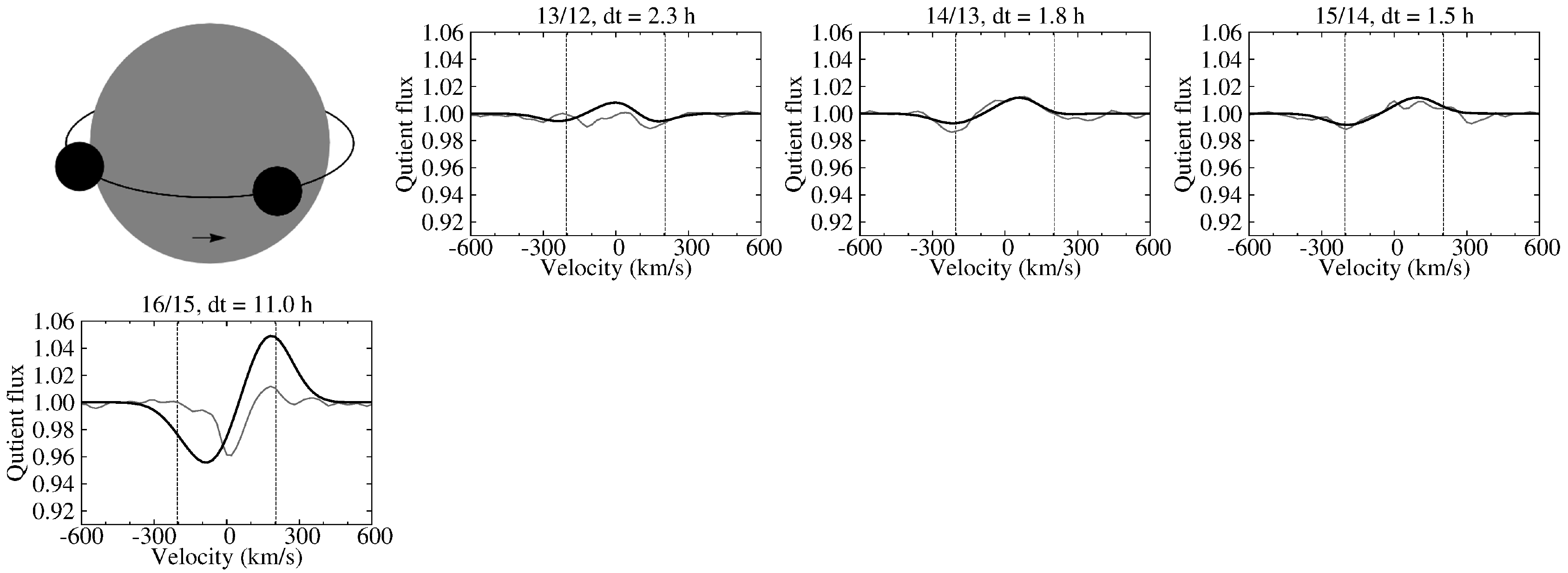}

\includegraphics[width=\columnwidth]{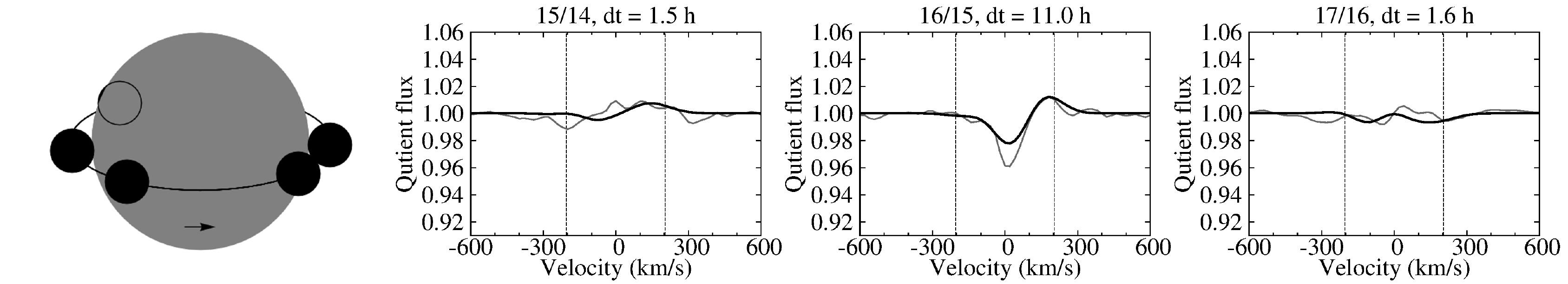}

\includegraphics[width=\columnwidth]{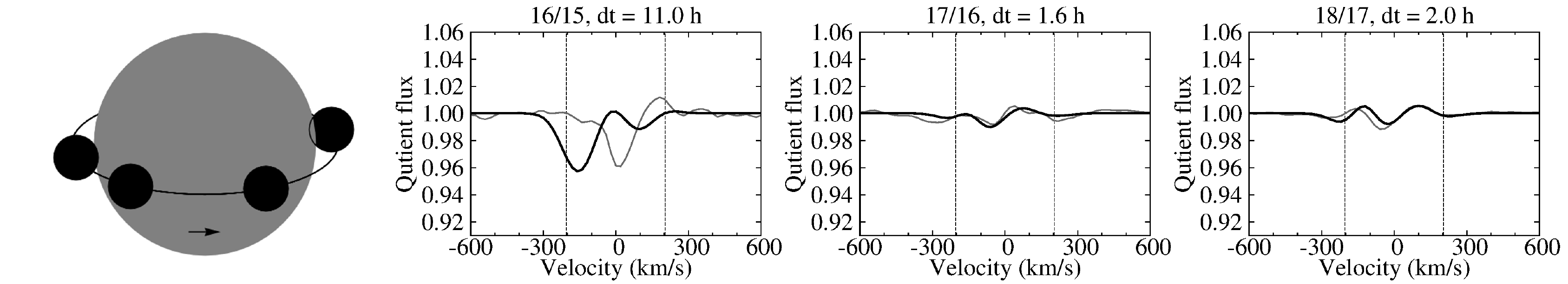}

\includegraphics[width=\columnwidth]{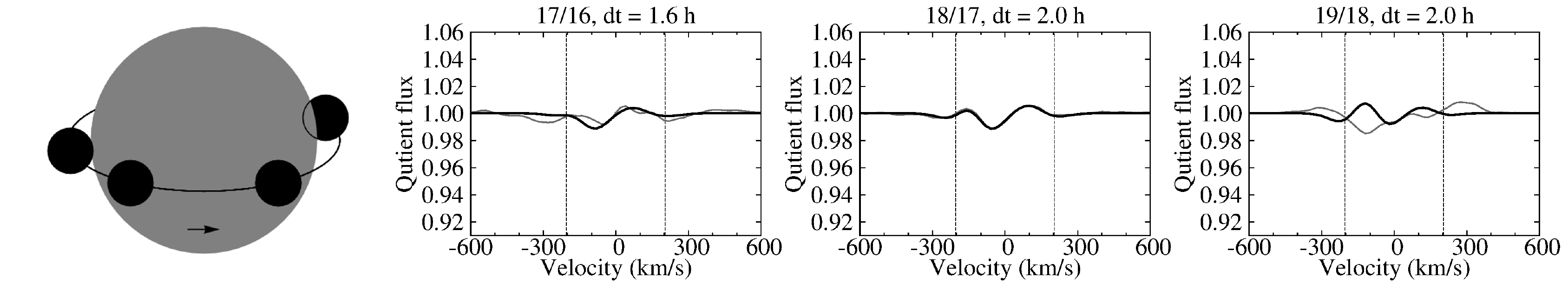}

\includegraphics[width=\columnwidth]{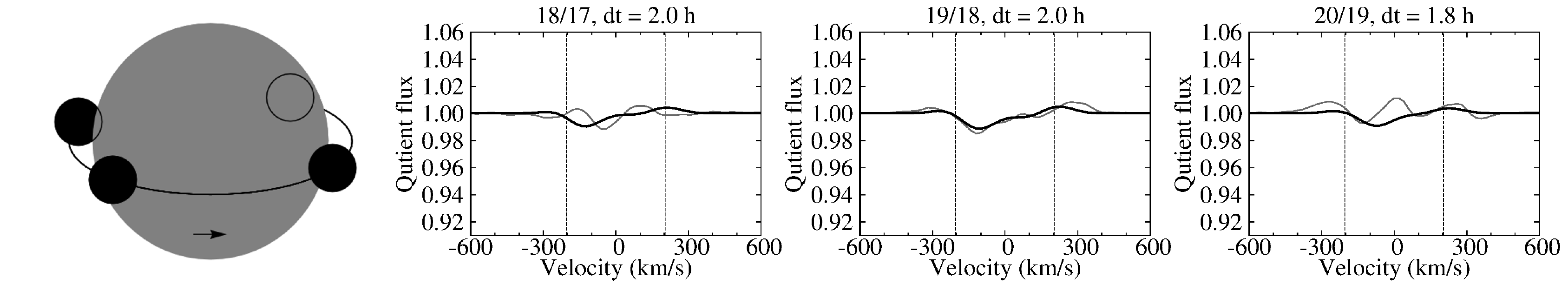}

\includegraphics[width=\columnwidth]{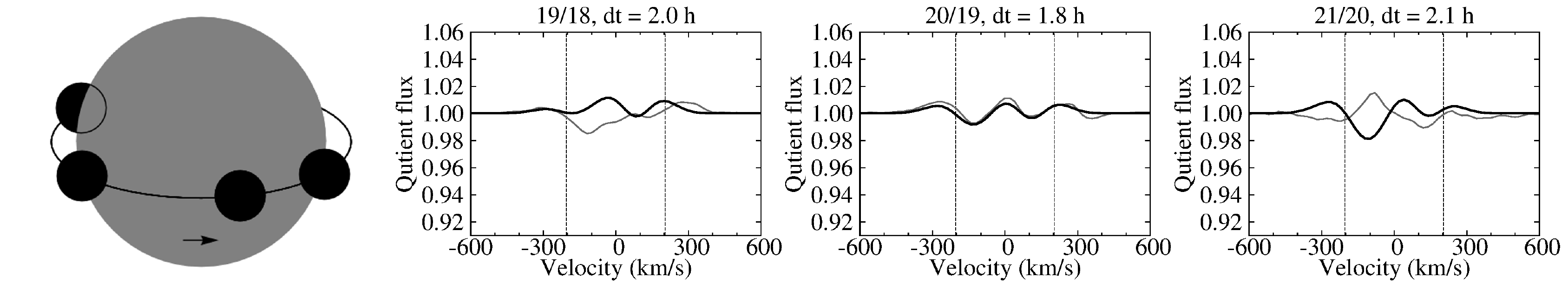}
\caption{Same as Fig.\,\ref{fig:qfit1989a}, but  for spectra 12--21.}
\label{fig:qfit1989c}
\end{figure*}

\begin{figure*}[ht!]
\includegraphics[width=\columnwidth]{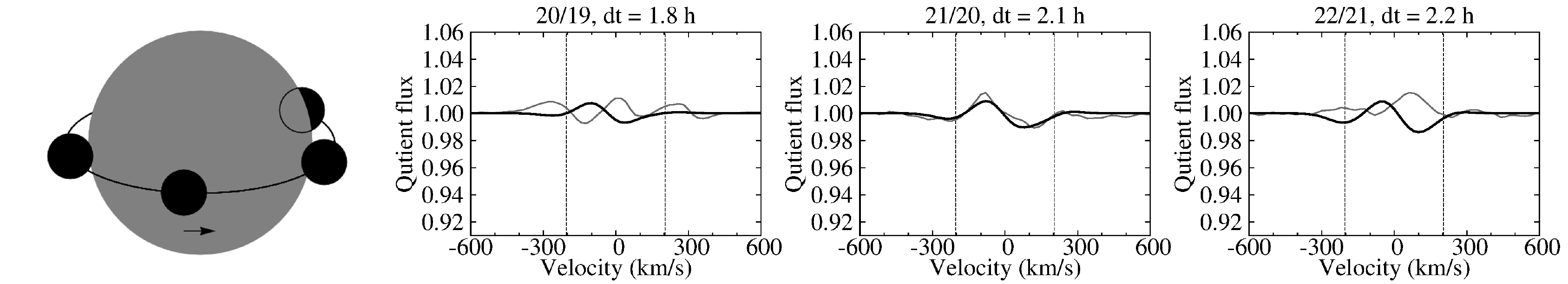}

\includegraphics[width=1\columnwidth]{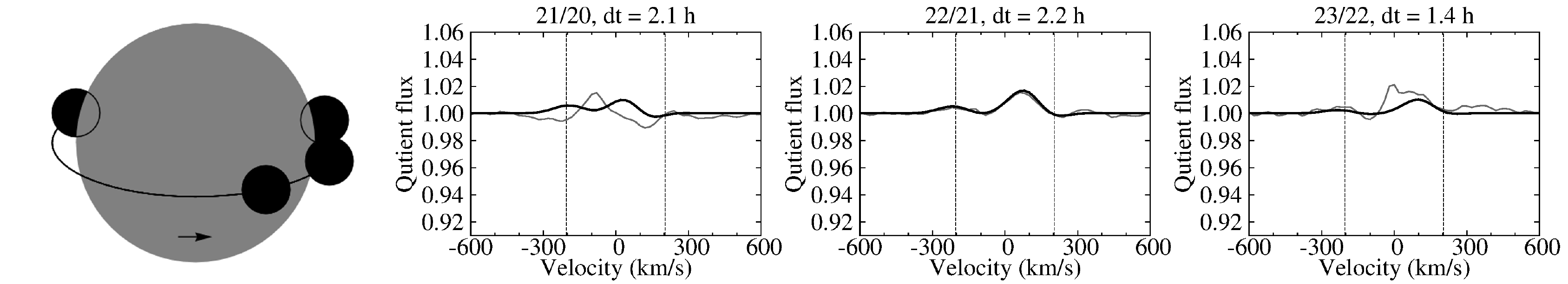}

\includegraphics[width=1\columnwidth]{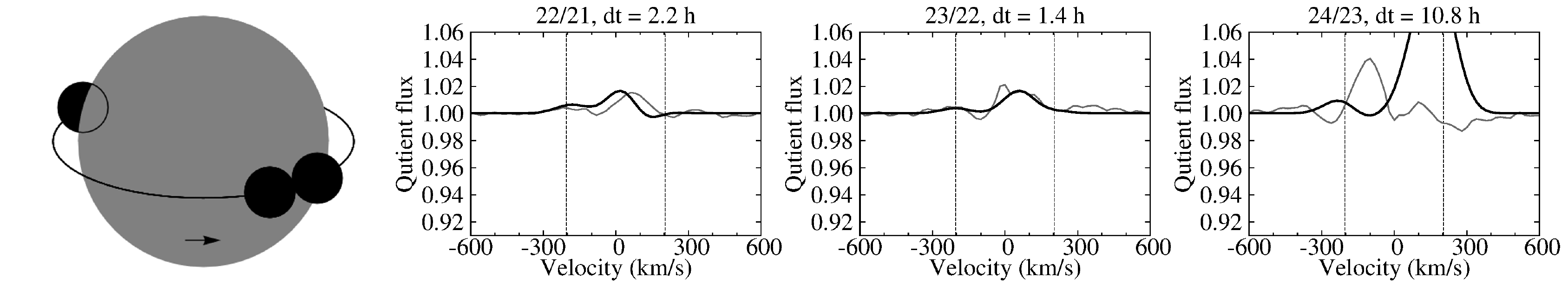}

\includegraphics[width=1\columnwidth]{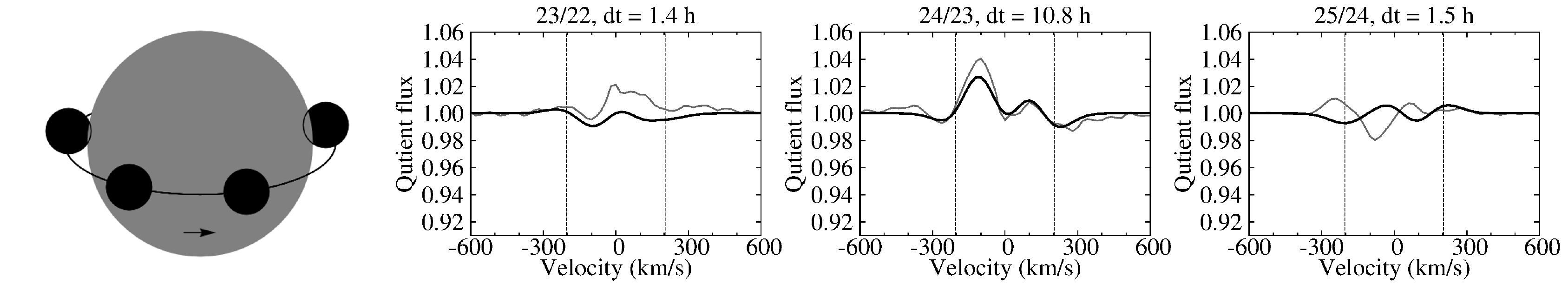}

\includegraphics[width=1\columnwidth]{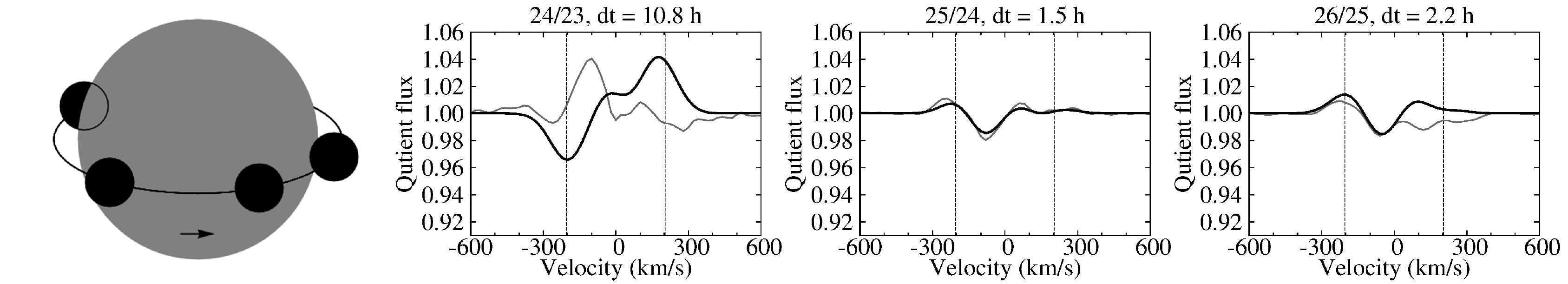}

\includegraphics[width=1\columnwidth]{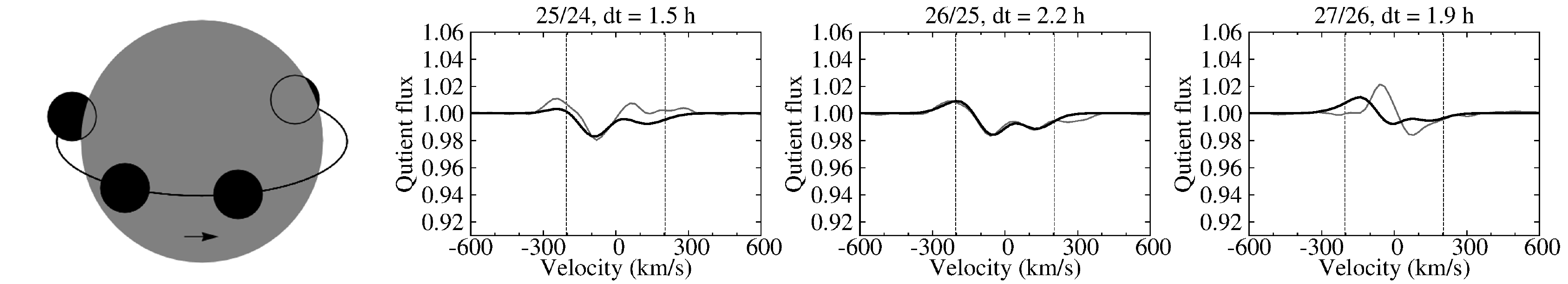}

\includegraphics[width=1\columnwidth]{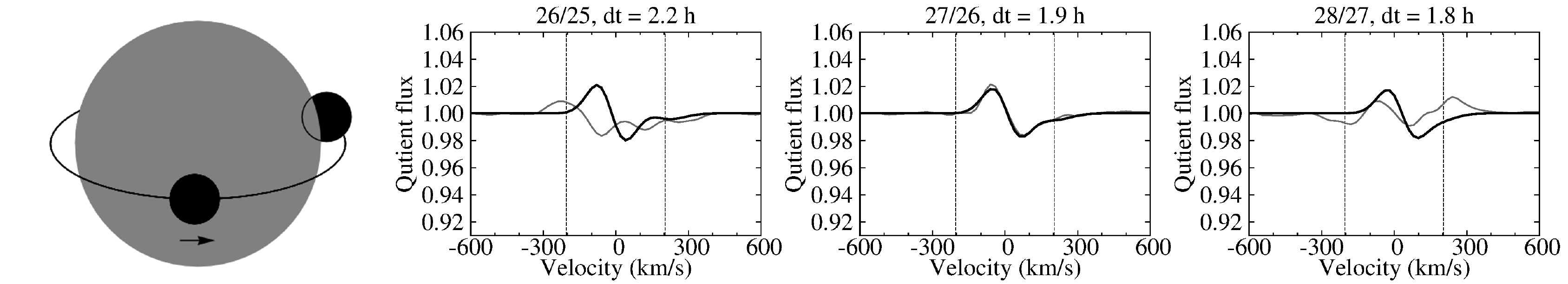}
\caption{Same as Fig.\,\ref{fig:qfit1989a}, but  for spectra 19-28.}
\label{fig:qfit1989d}
\end{figure*}

\begin{figure*}[ht!]
\includegraphics[width=1\columnwidth]{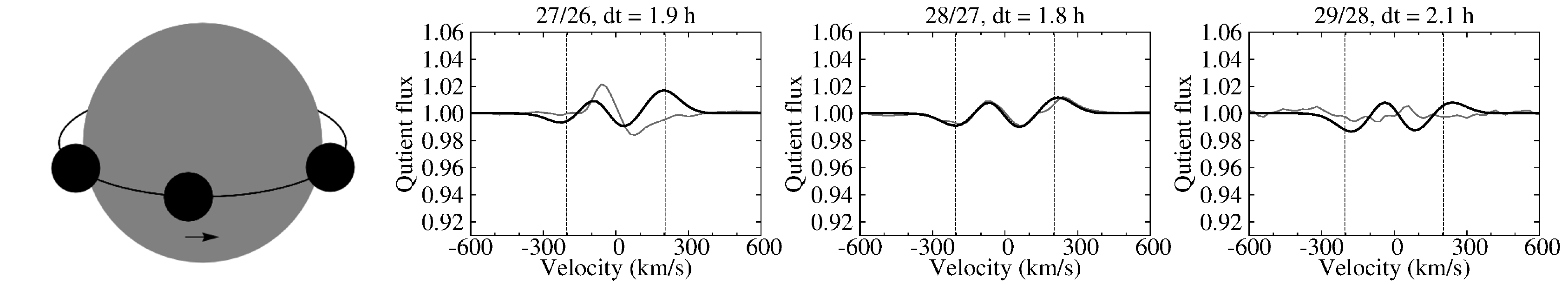}

\includegraphics[width=1\columnwidth]{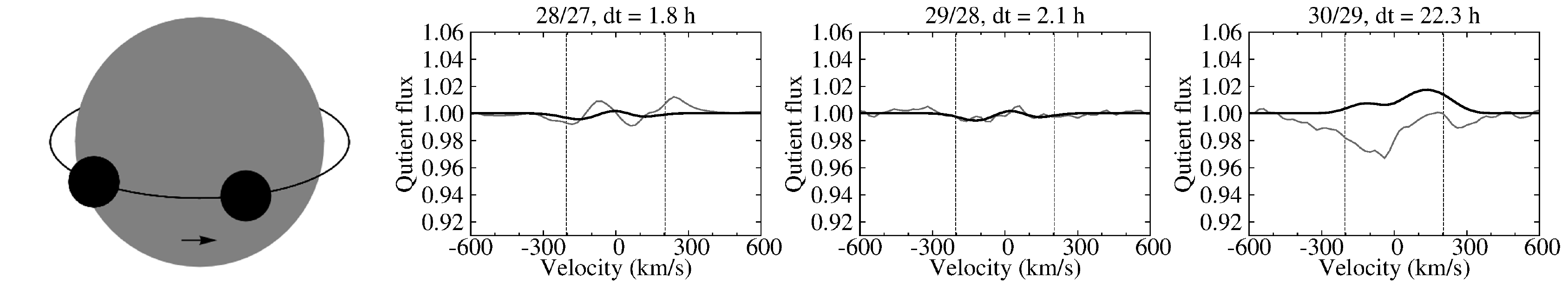}

\includegraphics[width=1\columnwidth]{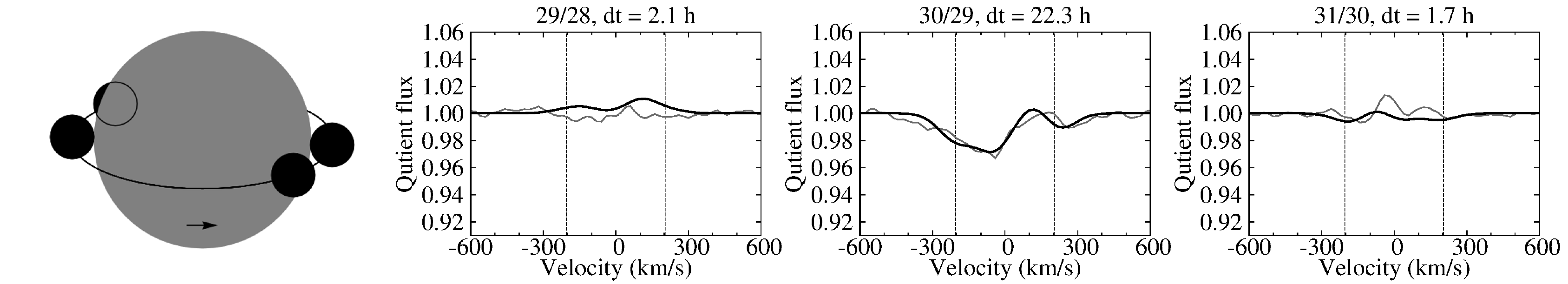}\
\includegraphics[width=1\columnwidth]{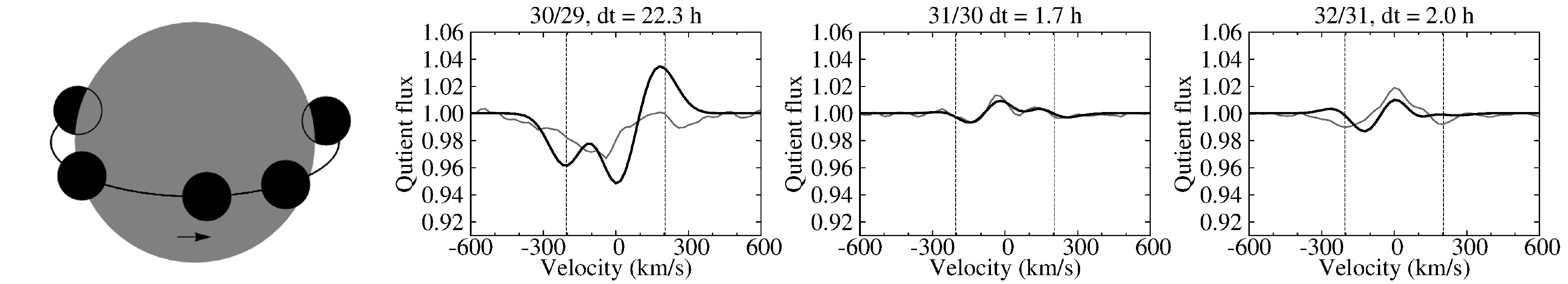}

\includegraphics[width=1\columnwidth]{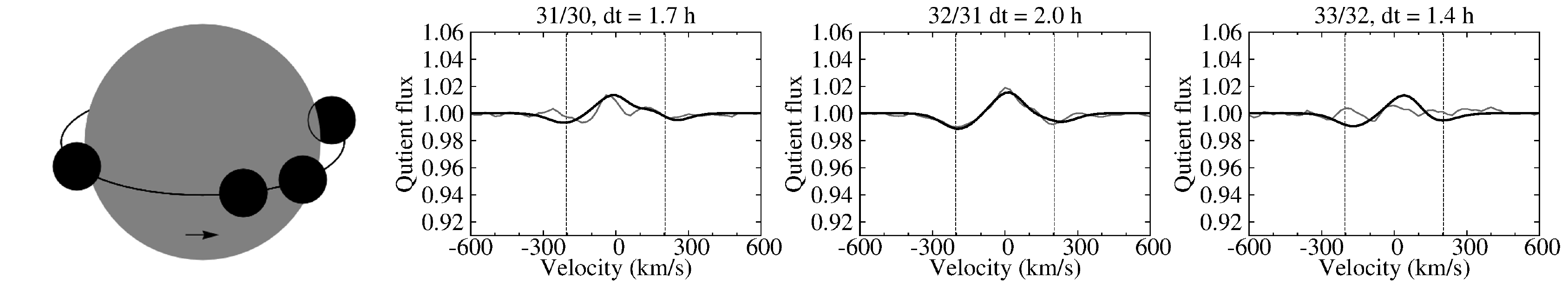}

\includegraphics[width=0.72\columnwidth]{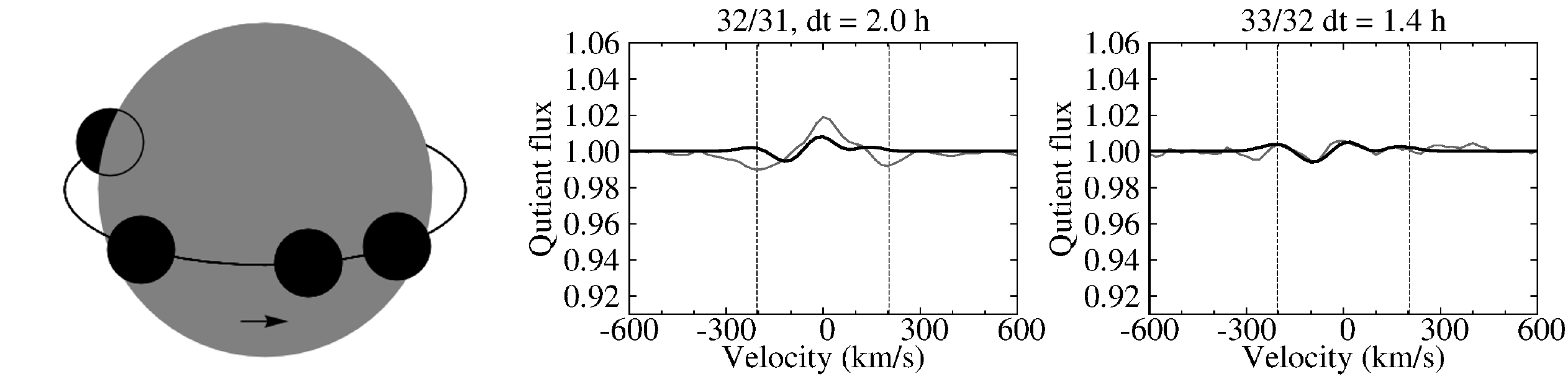}
\caption{Same as Fig.\,\ref{fig:qfit1989a}, but for spectra 26-33.}
\label{fig:qfit1989e}
\end{figure*}

\clearpage

\subsection{The 2007 dataset }
\label{A2007}

\begin{figure*}[ht!]
\includegraphics[width=0.95\columnwidth]{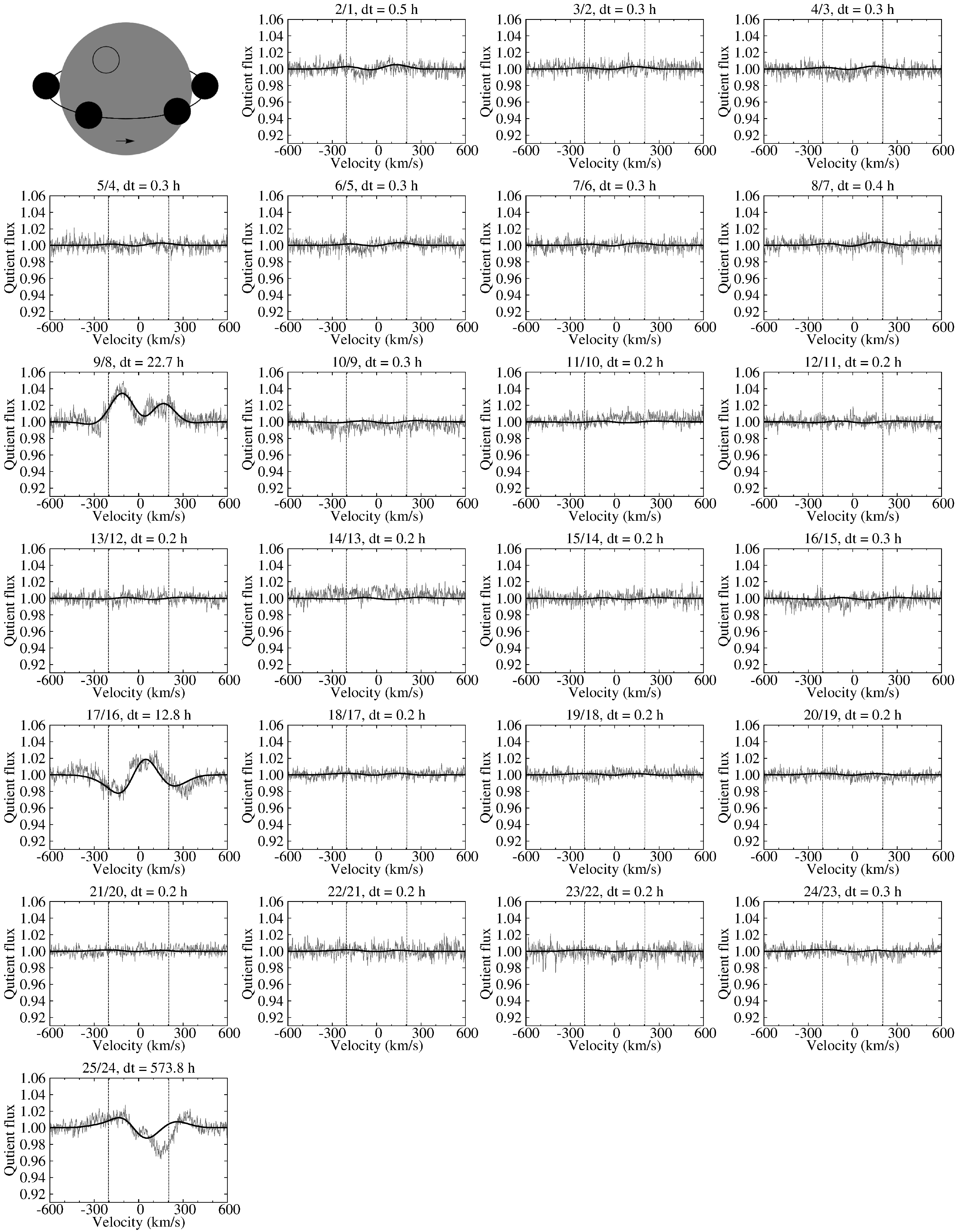}
\caption{Same as Fig.\,\ref{fig:qfit1989a}, but  for the   2007 dataset, spectra 1--25.}
\label{fig:qfit2007a}
\end{figure*}

\begin{figure*}[ht!]
\includegraphics[width=0.95\columnwidth]{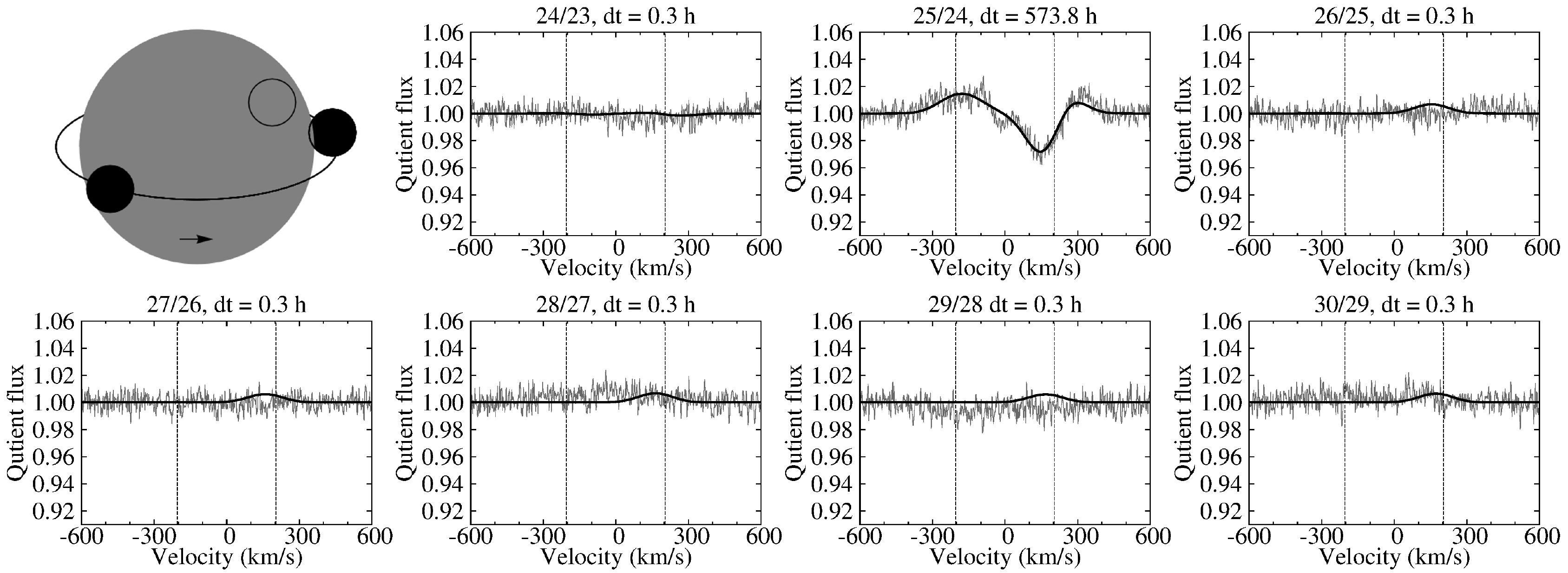}
\caption{Same as Fig.\,\ref{fig:qfit2007a}, but for  spectra 23--30.}
\label{fig:qfit2007b}
\end{figure*}


\subsection{The 2012 dataset }
\label{A2012}

\begin{figure*}[ht!]
\vspace*{-0cm}
\includegraphics[width=0.95\columnwidth]{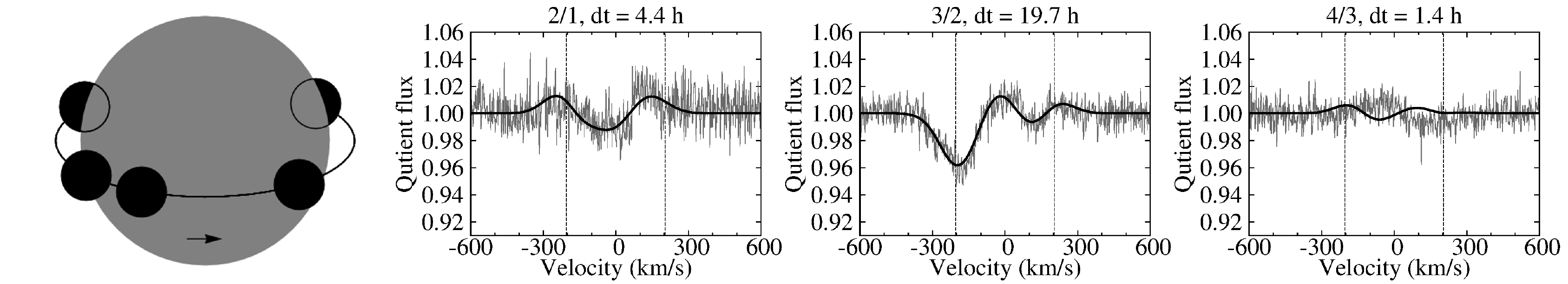}

\includegraphics[width=0.95\columnwidth]{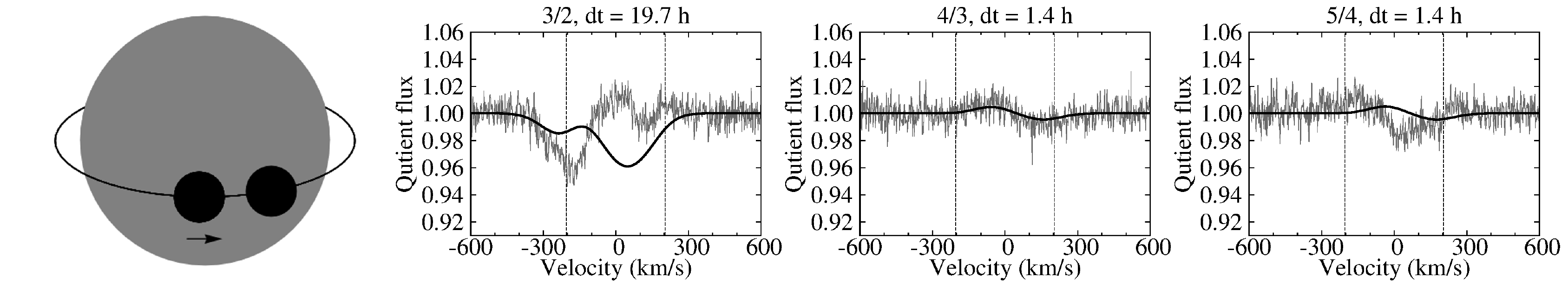}

\includegraphics[width=0.95\columnwidth]{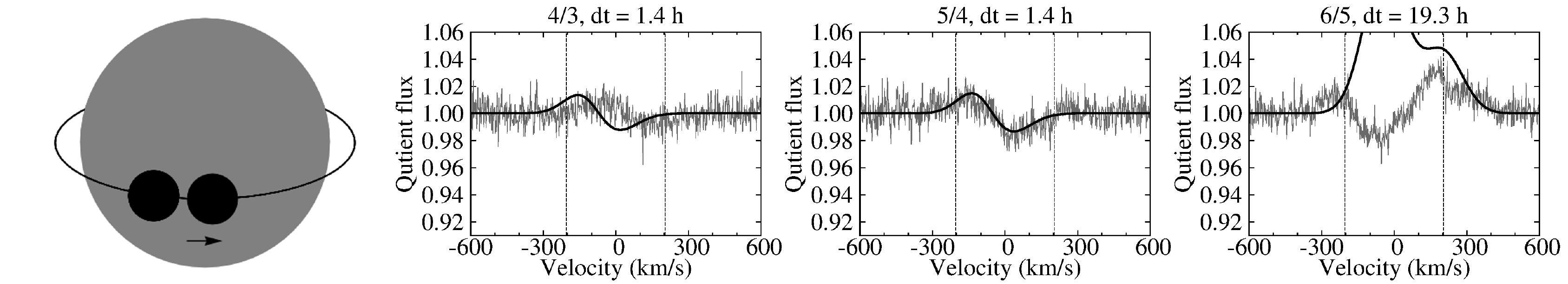}
\caption{Same as Fig.\,\ref{fig:qfit1989a}, but for the 2012 dataset, spectra 1--6.}
\label{fig:qfit2012a}
\end{figure*}

\begin{figure*}[ht!]
\includegraphics[width=0.95\columnwidth]{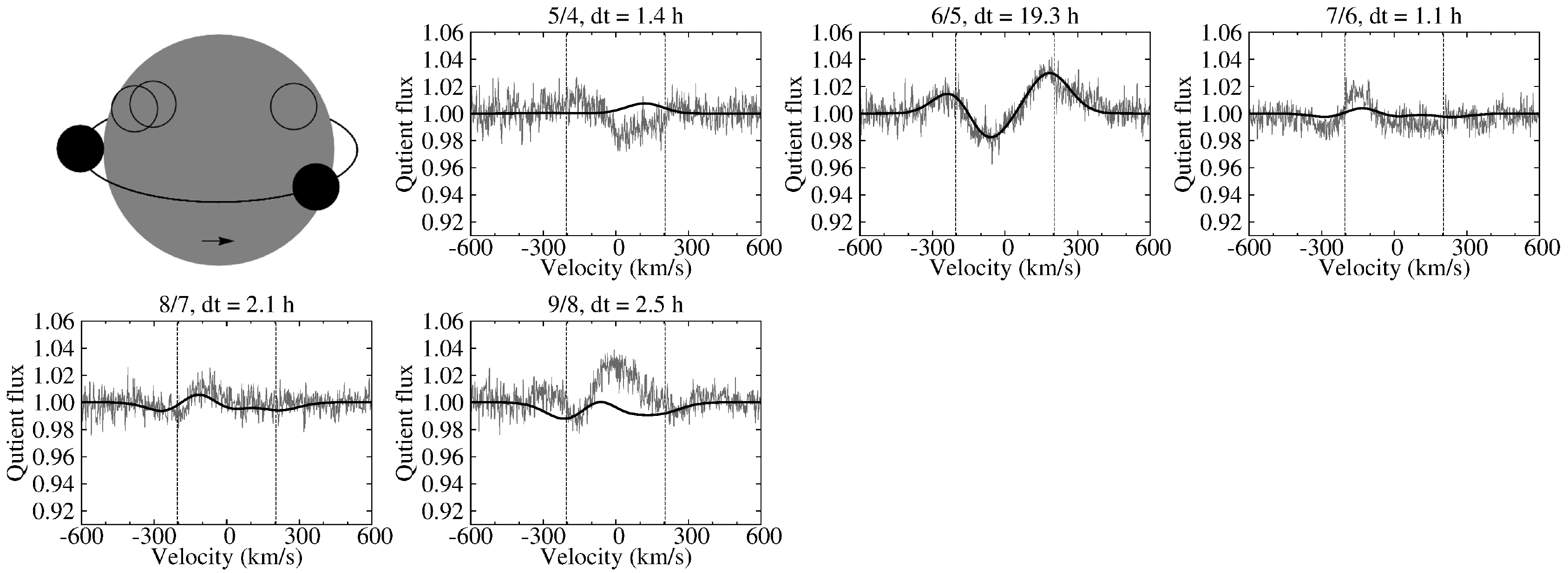}

\includegraphics[width=0.95\columnwidth]{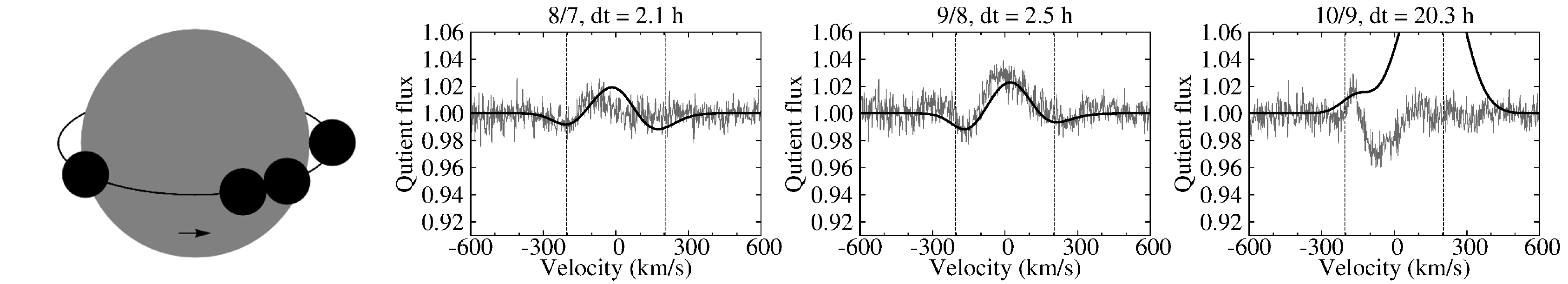}

\includegraphics[width=0.95\columnwidth]{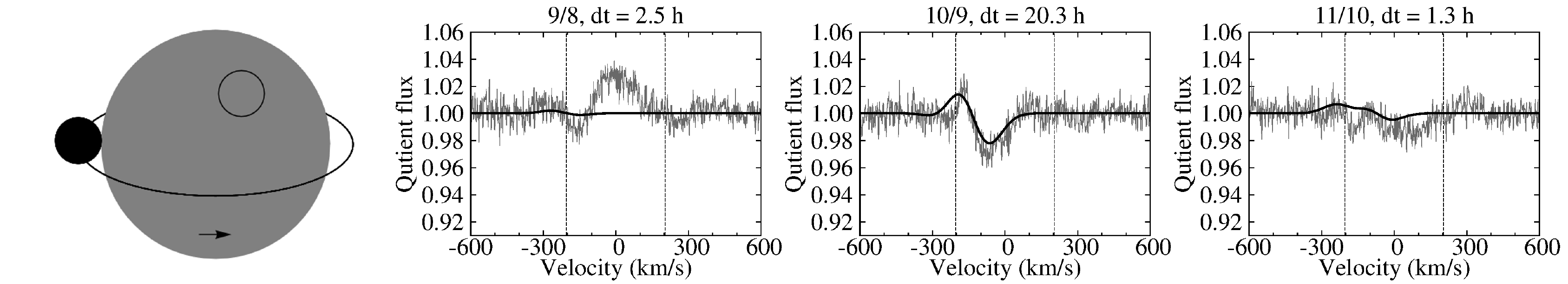}

\includegraphics[width=0.95\columnwidth]{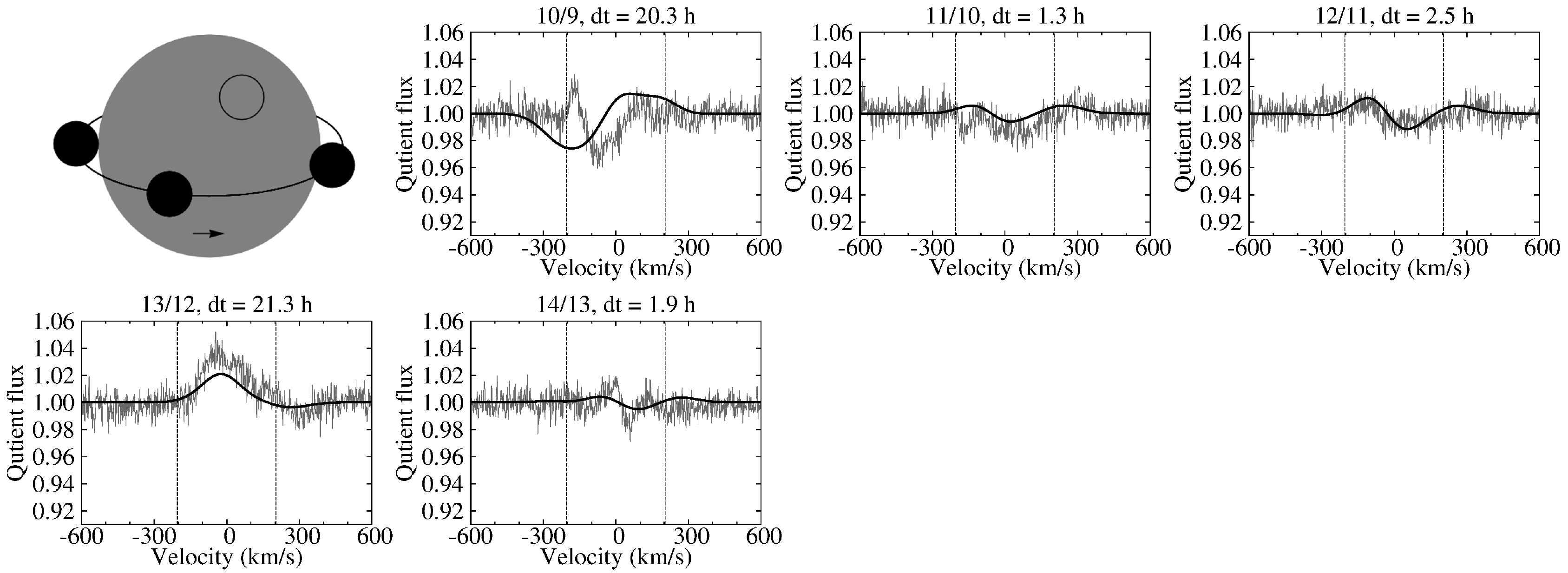}
\includegraphics[width=0.95\columnwidth]{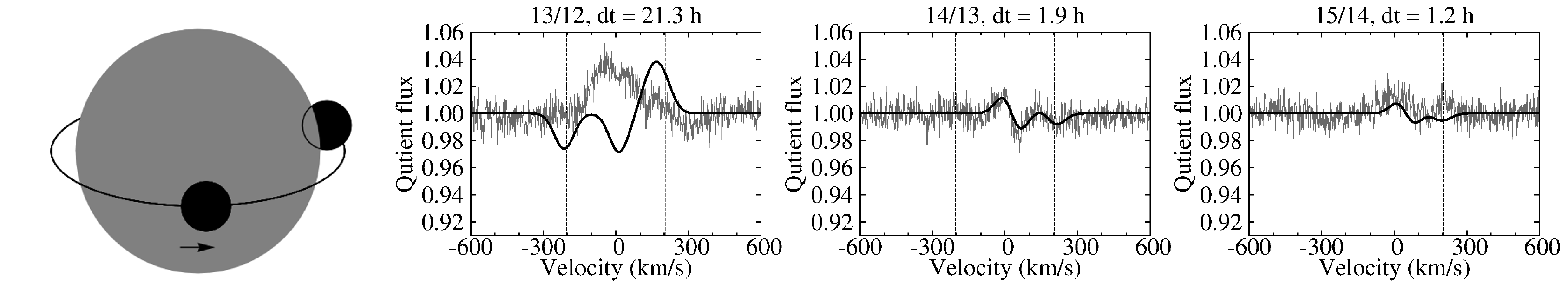}
\caption{Same as Fig.\,\ref{fig:qfit2012a}, but for  spectra 4--15.}
\label{fig:qfit2012b}
\end{figure*}

\begin{figure*}[ht!]

\includegraphics[width=0.95\columnwidth]{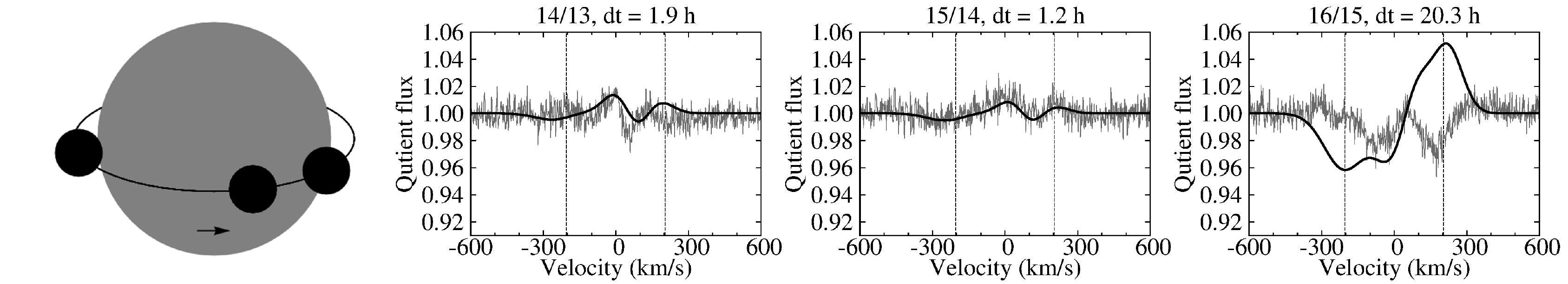}

\includegraphics[width=0.95\columnwidth]{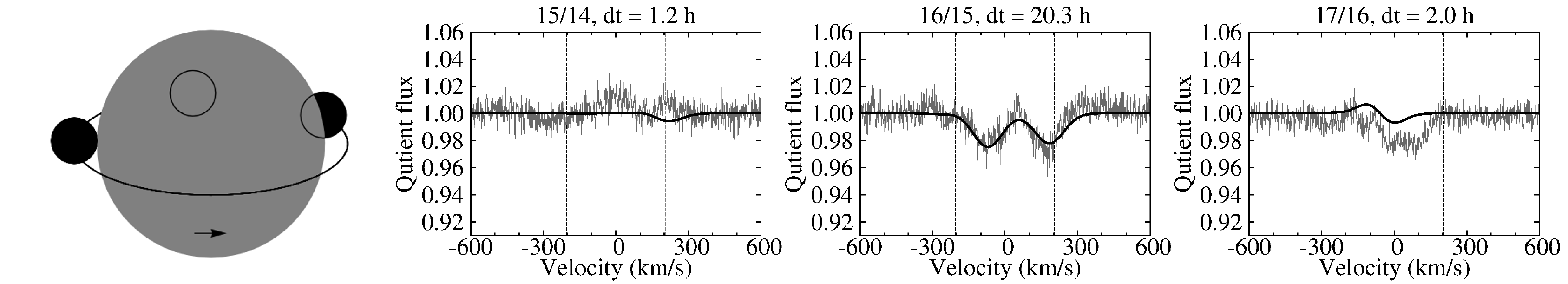}

\includegraphics[width=0.95\columnwidth]{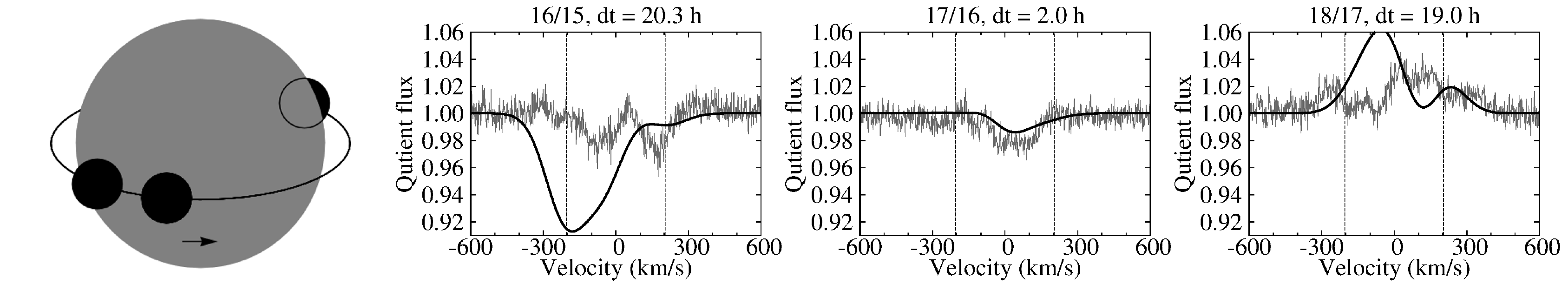}

\includegraphics[width=0.95\columnwidth]{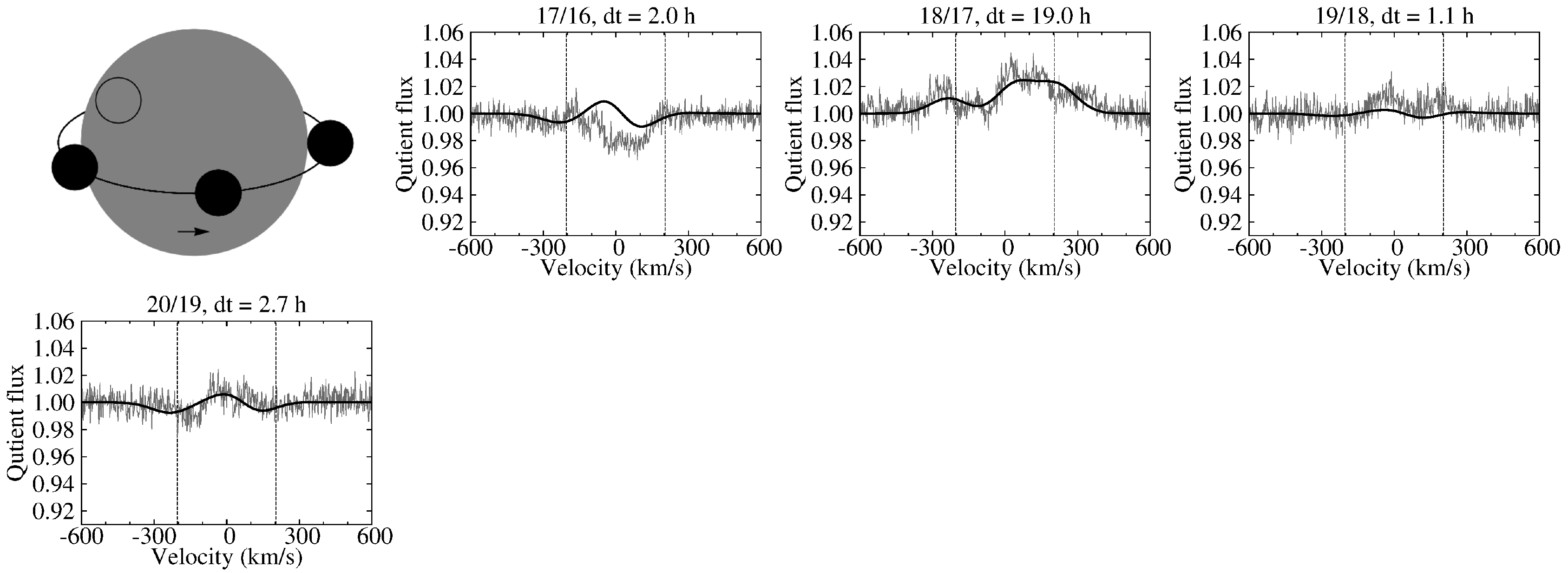}
\caption{Same as Fig.\,\ref{fig:qfit2012a}, but for  spectra 13--20.}
\label{fig:qfit2012c}
\end{figure*}


\clearpage
\subsection{The 2013 dataset}
\label{A2013}

\begin{figure*}[ht!]
\includegraphics[width=0.95\columnwidth]{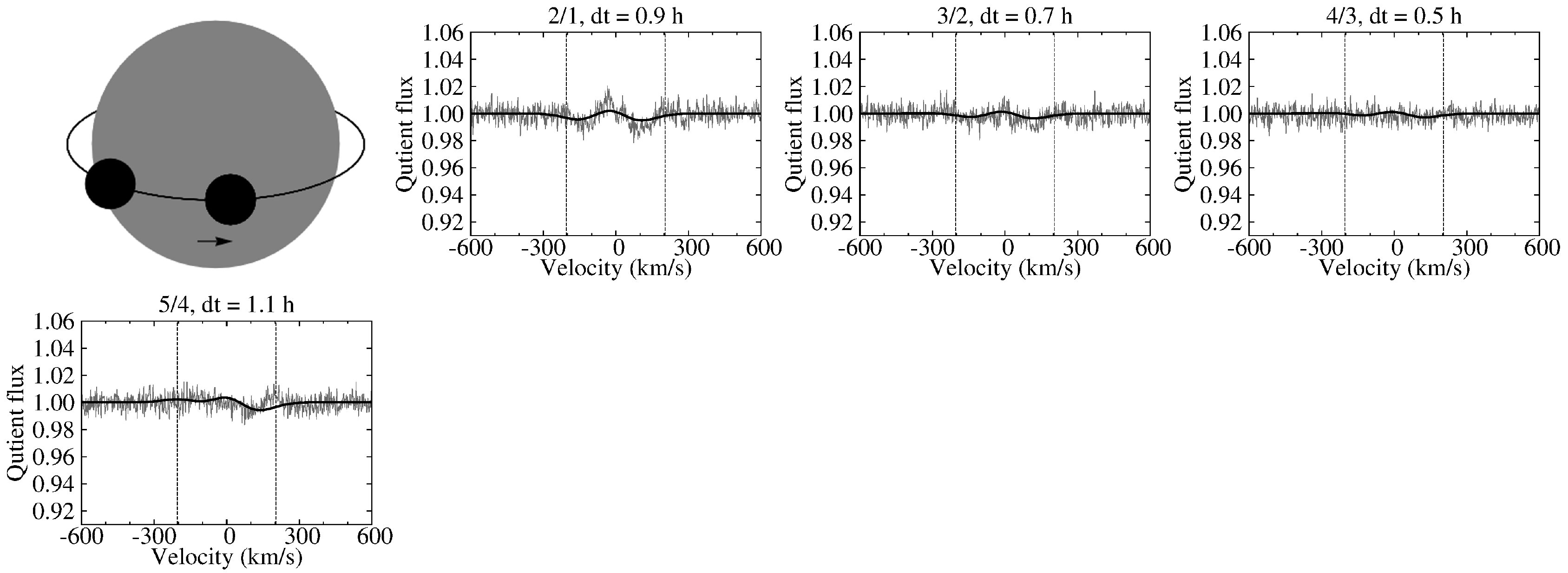}

\includegraphics[width=0.95\columnwidth]{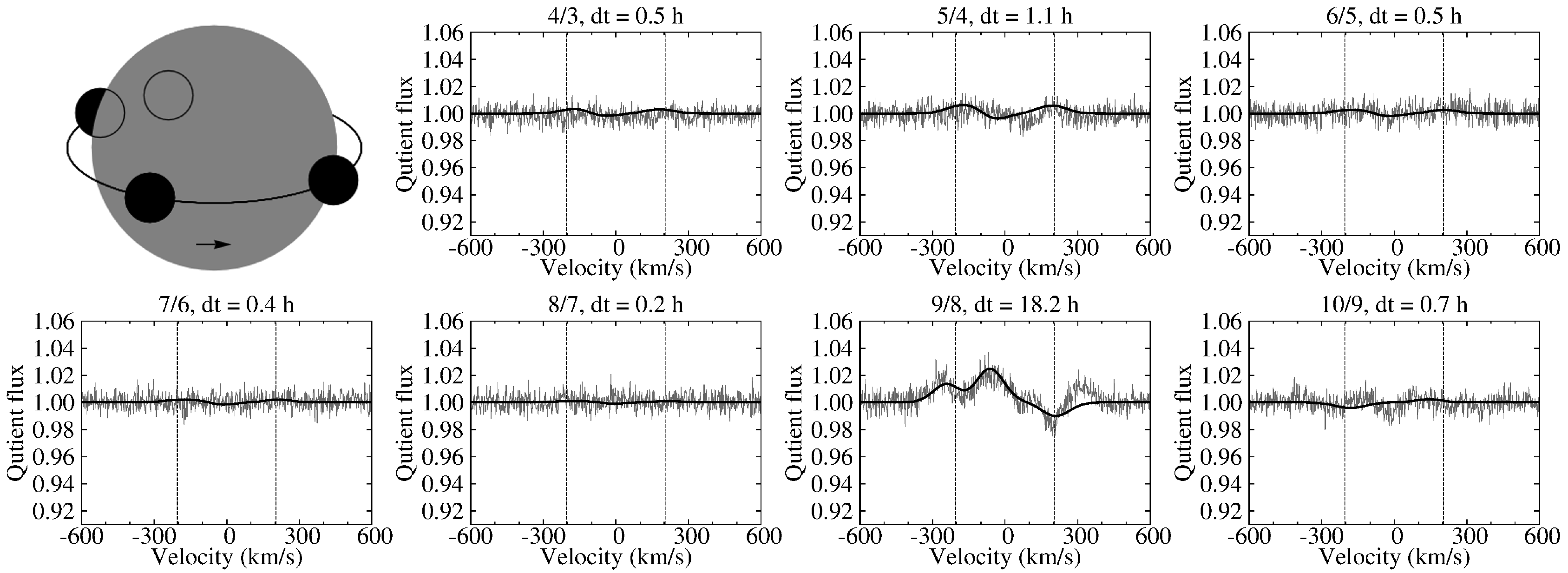}

\includegraphics[width=0.95\columnwidth]{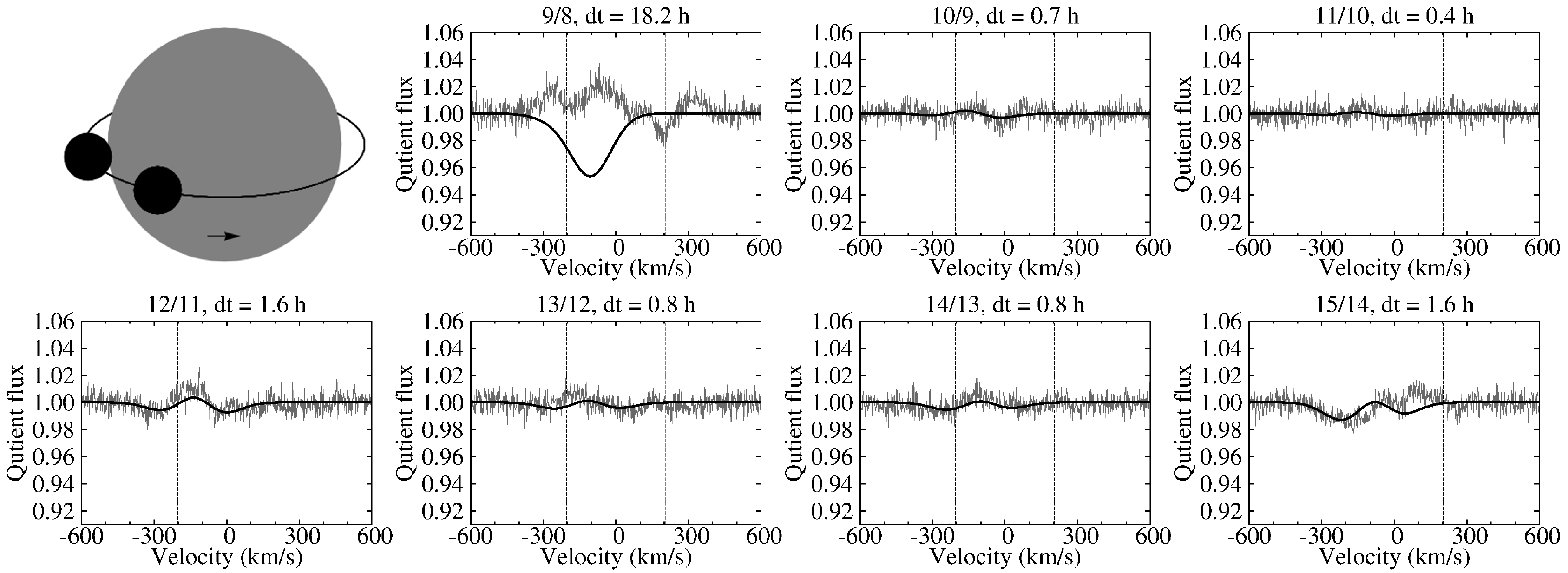}
\caption{Same as Fig.\,\ref{fig:qfit1989a}, but for the  2013 dataset, spectra 1--15.}
\label{fig:qfit2013a}
\end{figure*}

\begin{figure*}[ht!]
\includegraphics[width=0.95\columnwidth]{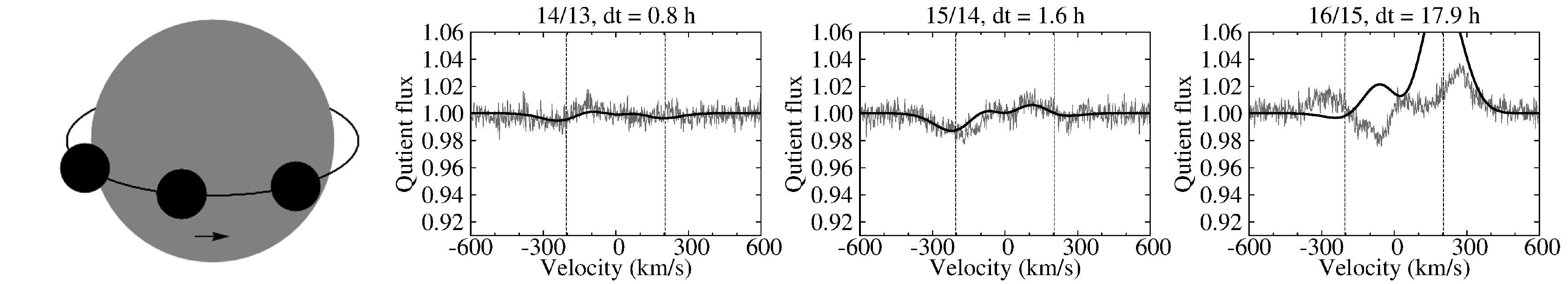}

\includegraphics[width=0.95\columnwidth]{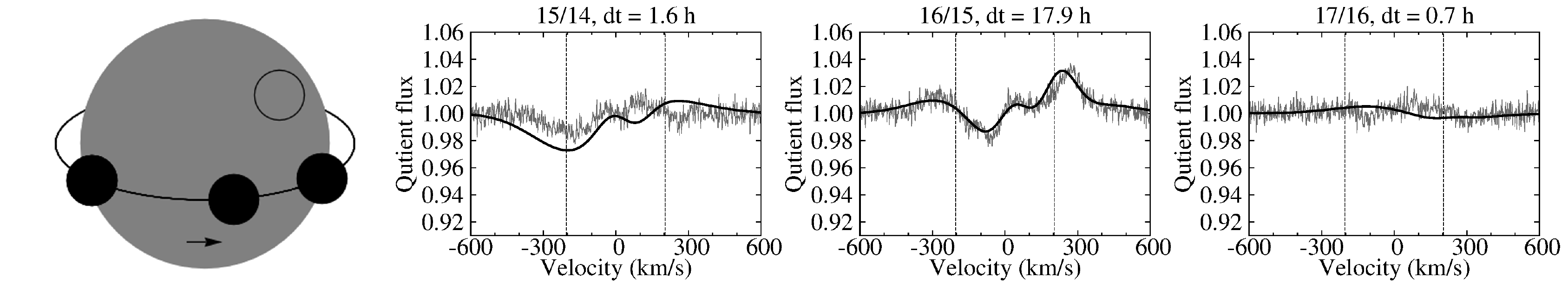}

\includegraphics[width=0.822\columnwidth]{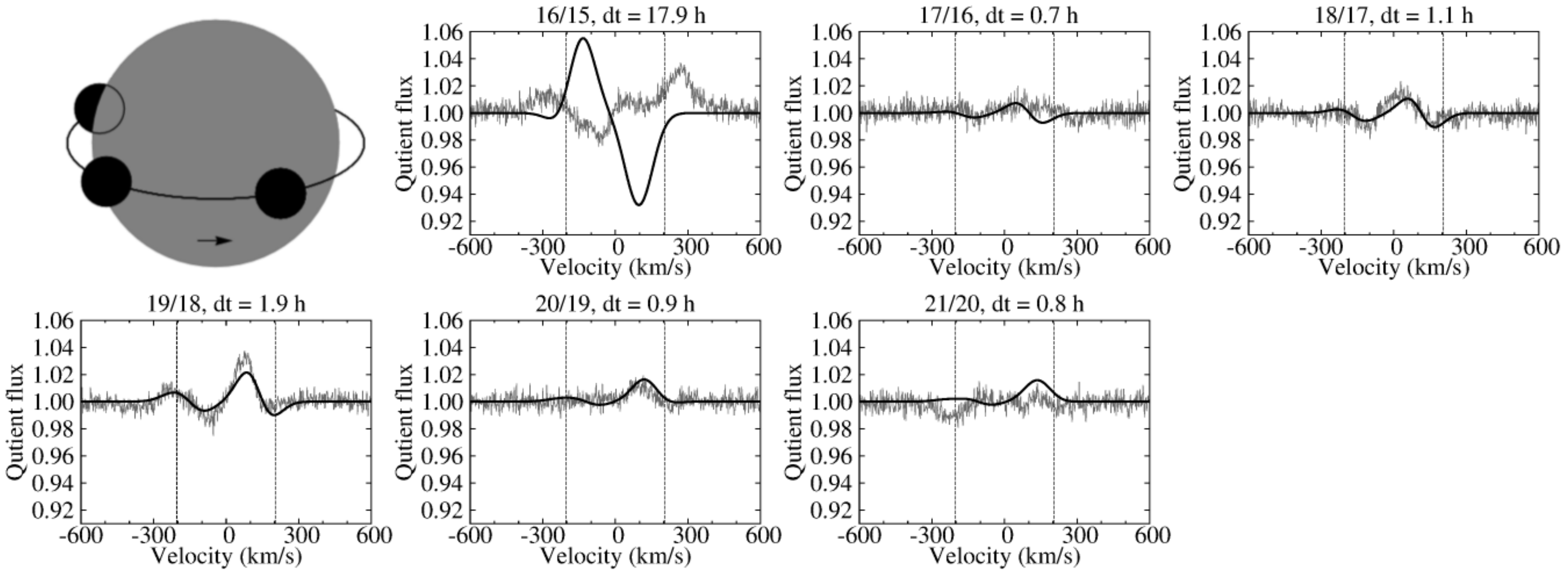}

\includegraphics[width=0.95\columnwidth]{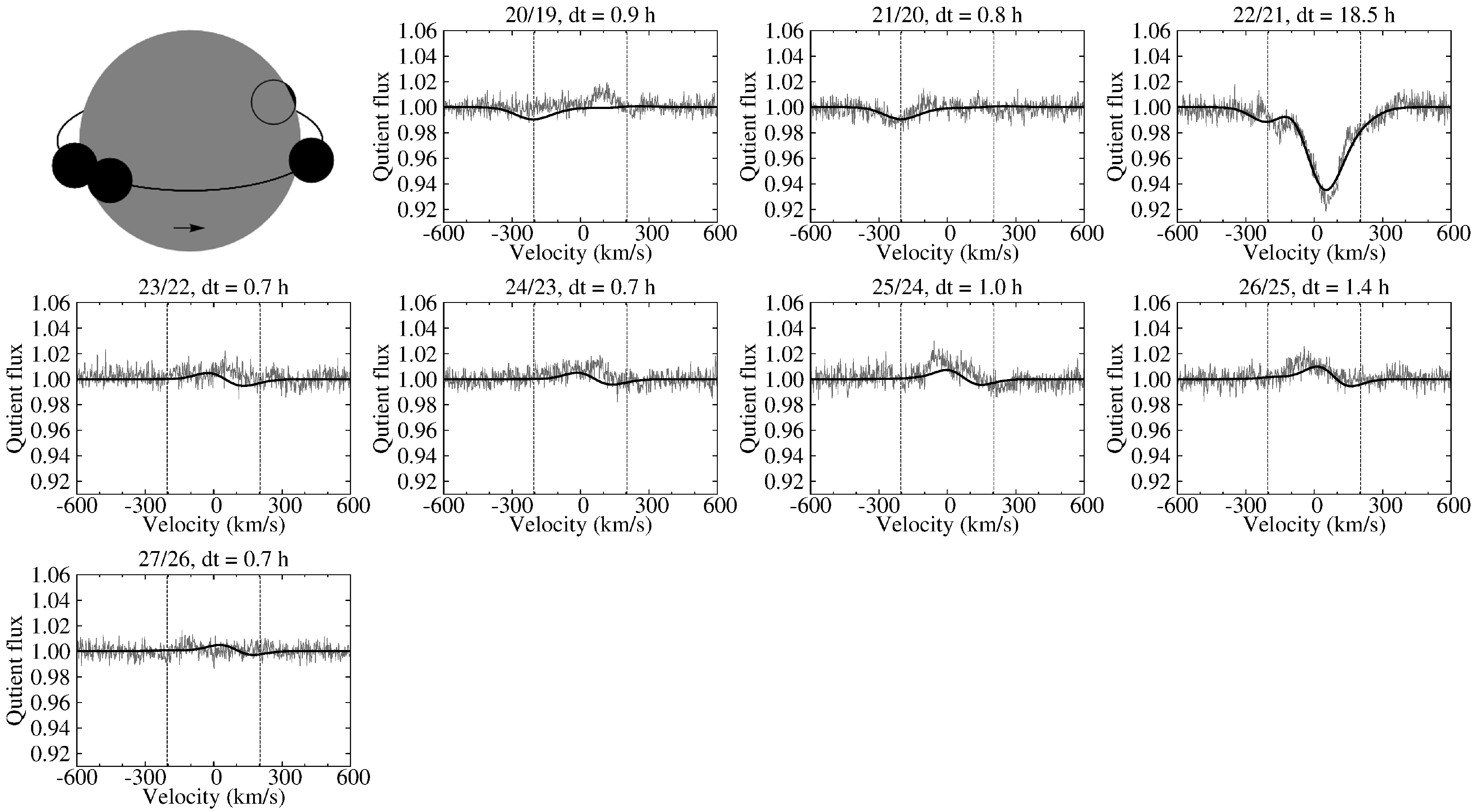}
\caption{Same as Fig.\,\ref{fig:qfit2013a}, but for  spectra 13--27.}
\label{fig:qfit2013b}
\end{figure*}

\begin{figure*}[ht!]
\includegraphics[width=0.822\columnwidth]{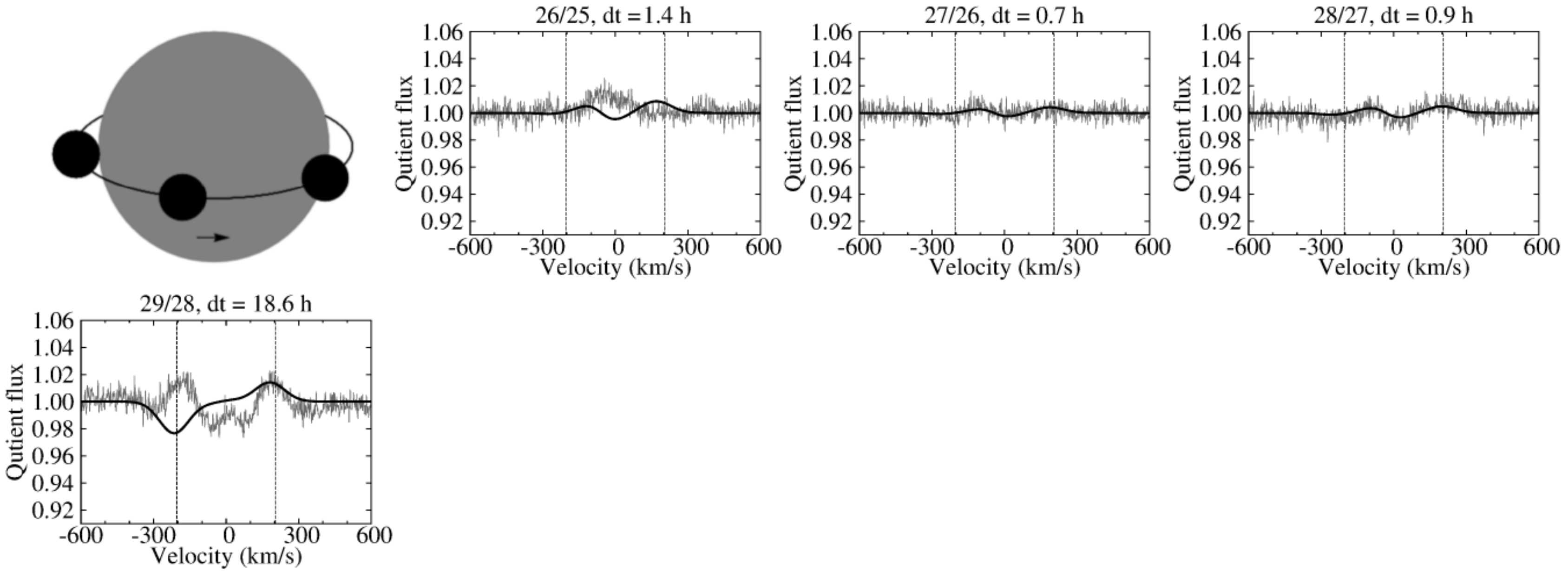}

\includegraphics[width=0.95\columnwidth]{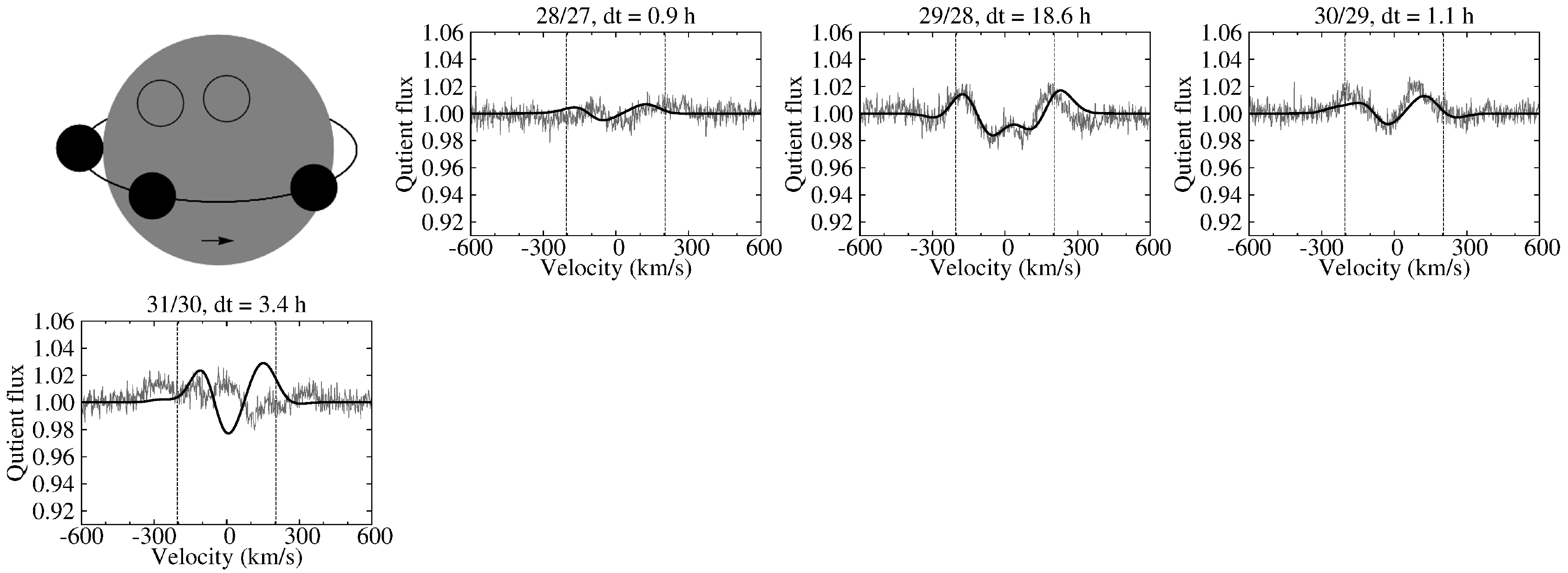}

\includegraphics[width=0.95\columnwidth]{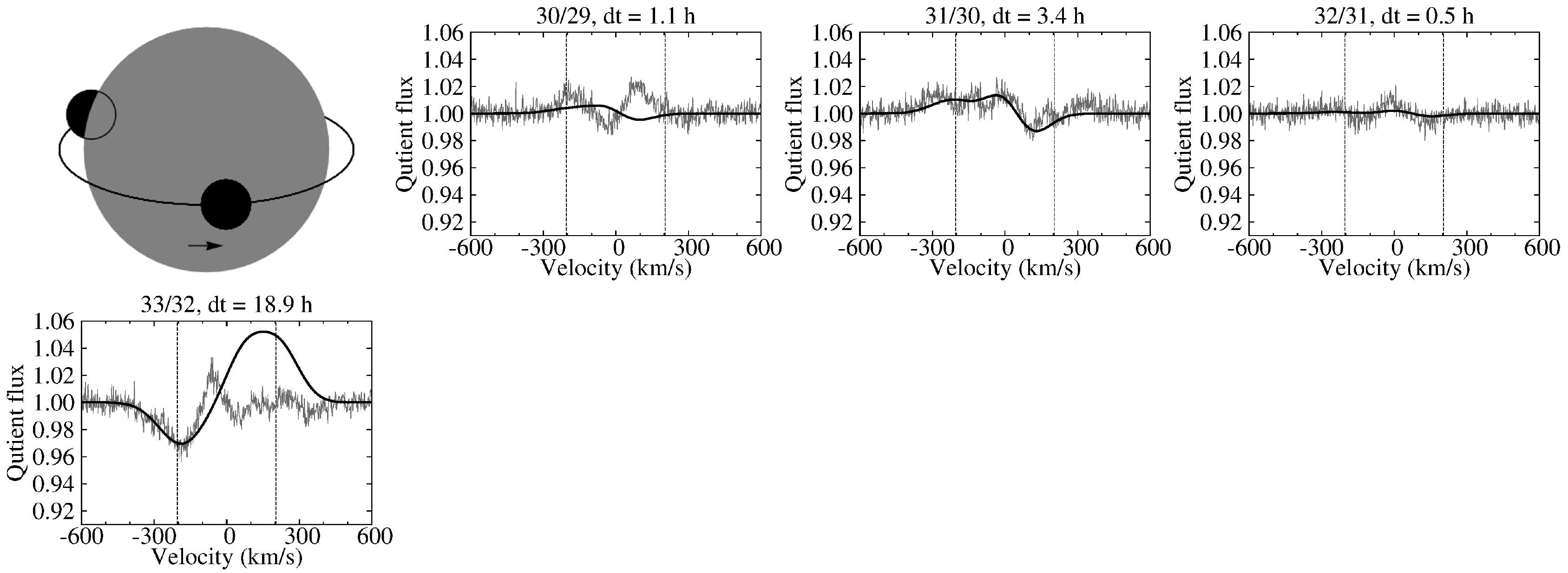}

\caption{Same as Fig.\,\ref{fig:qfit2013a}, but for  spectra 25--33.}
\label{fig:qfit2013c}
\end{figure*}

\begin{figure*}[ht!]
\includegraphics[width=0.822\columnwidth]{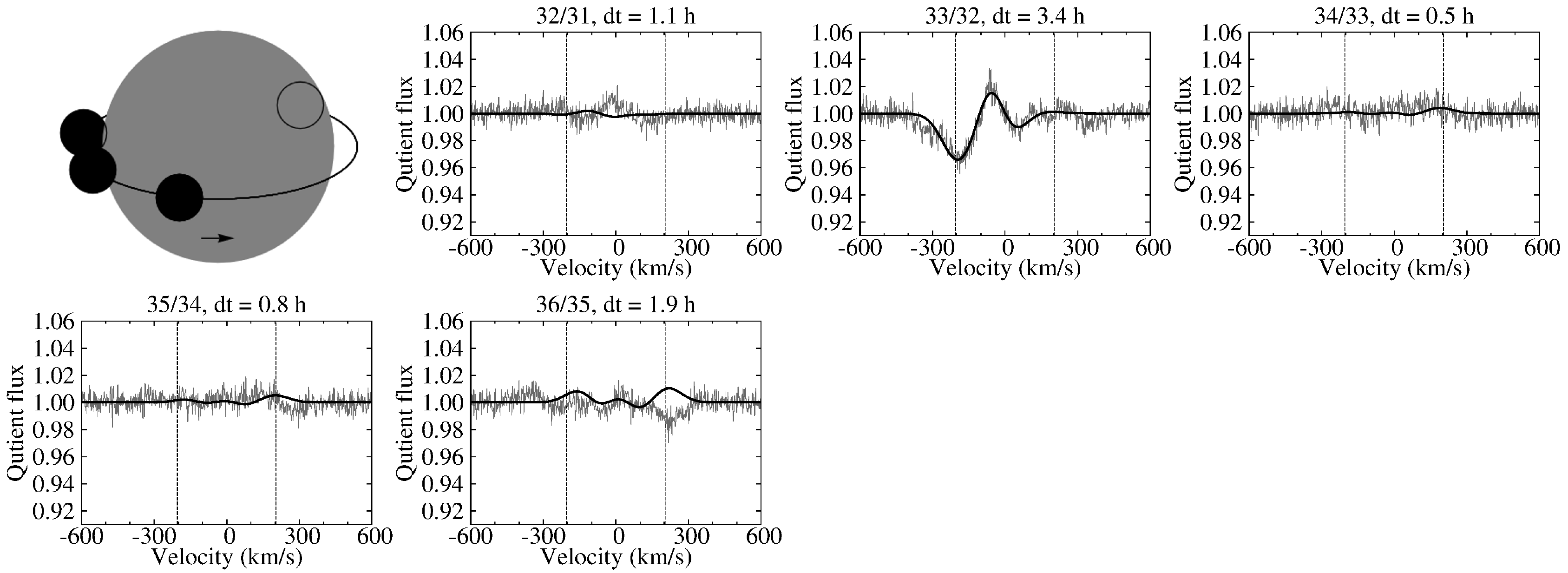}

\includegraphics[width=0.95\columnwidth]{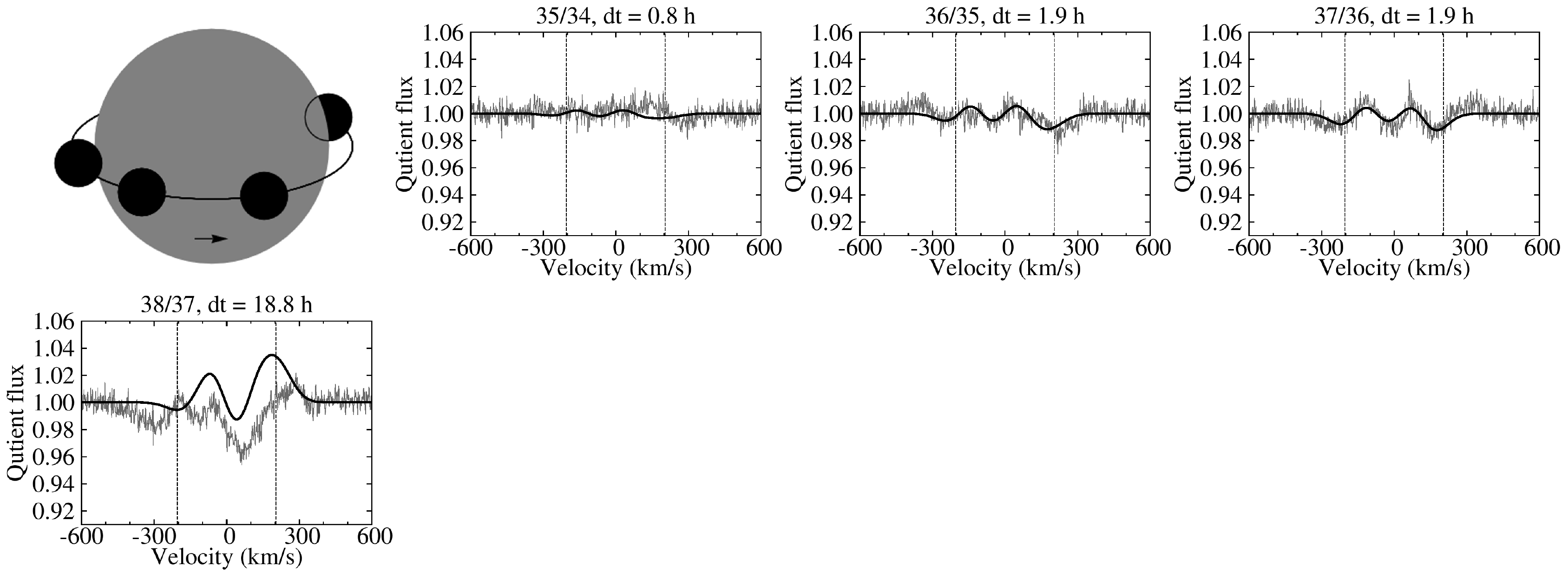}

\includegraphics[width=0.95\columnwidth]{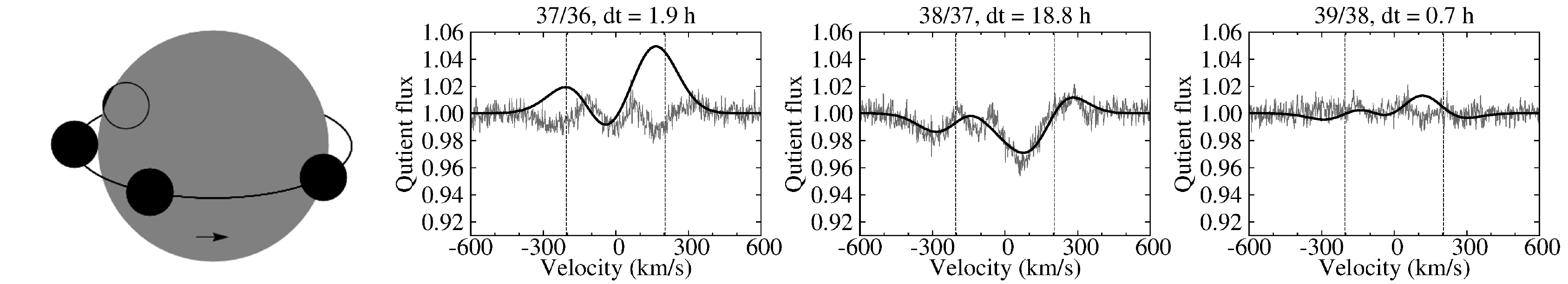}

\includegraphics[width=0.95\columnwidth]{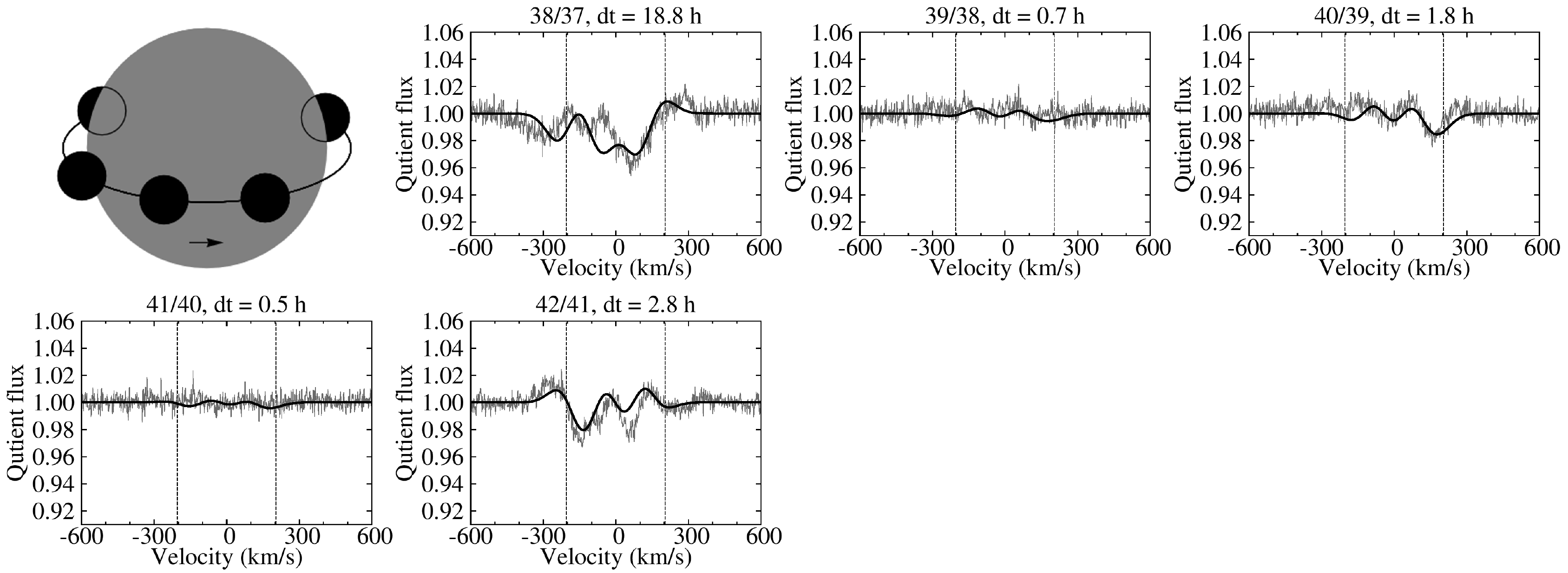}
\caption{Same as Fig.\,\ref{fig:qfit2013a}, but for  spectra 31--42.}
\label{fig:qfit2013d}
\end{figure*}

\end{appendix}

\end{document}